\documentclass[apj]{emulateapj}

\usepackage{amsmath,amssymb,latexsym, graphicx}
\usepackage{epsfig}
\usepackage{booktabs}
\usepackage[dvips]{color}

\newcommand{\tn}[1]{\textnormal{#1}}

\newcommand{\kref}[1]{{\S} \ref{#1}}

\newcommand{\EBV}{E(B-V)}

\newcommand{\AV}{\tn{\emph{A}}_{\tn{V}}}

\newcommand{\tb}{\tn{t}_{\tn{b}}}

\newcommand{\MB}{\tn{M}_{\tn{B}}}
\newcommand{\TNT}{\tn{T}_{90}}
\newcommand{\n}{\tn{n}}

\def\MgII{\hbox{{\rm Mg~}\kern 0.1em{\sc ii}}}
\def\OII{\hbox{{\rm [O~}\kern 0.1em{\sc ii}{\rm ]}}}
\def\CIV{\hbox{{\rm C~}\kern 0.1em{\sc iv}}}

\slugcomment{Submitted to ApJ, 16th of July 2010}

\shorttitle{\emph{Swift} Type II GRB Afterglows}
\shortauthors{Kann et al.}

\begin{document}

\title{The Afterglows of \emph{Swift}-era Gamma-Ray Bursts. I. Comparing pre-\emph{Swift} and \emph{Swift} era Long/Soft (Type II) GRB Optical Afterglows
\footnote{Based in part on observations obtained with the Very Large Telescope under ESO program 075.D-0787, PI Tagliaferri. Also based partly on observations made with the Italian Telescopio Nazionale Galileo (TNG) operated on the island of La Palma by the Fundaci\'{o}n Galileo Galilei of the INAF (Istituto Nazionale di Astrofisica) at the Spanish Observatorio del Roque de los Muchachos of the Instituto de Astrofisica de Canarias under program TAC 1238.}
}

\author{
D.~A.~Kann,\altaffilmark{1}
S.~Klose,\altaffilmark{1}
B.~Zhang,\altaffilmark{2}
D.~Malesani,\altaffilmark{3}
E.~Nakar,\altaffilmark{4,5}
A.~Pozanenko,\altaffilmark{6}
A.~C.~Wilson,\altaffilmark{7}
N.~R.~Butler,\altaffilmark{8,9}
P.~Jakobsson,\altaffilmark{10,11}
S.~Schulze,\altaffilmark{1,11}
M.~Andreev,\altaffilmark{12,13}
L.~A.~Antonelli,\altaffilmark{14}
I.~F.~Bikmaev,\altaffilmark{15}
V.~Biryukov,\altaffilmark{16,17}
M.~B\"ottcher,\altaffilmark{18}
R.~A.~Burenin,\altaffilmark{6}
J.~M.~Castro Cer\'on,\altaffilmark{3,19}
A.~J.~Castro-Tirado,\altaffilmark{20}
G.~Chincarini,\altaffilmark{21,22}
B.~E.~Cobb,\altaffilmark{23,24}
S.~Covino,\altaffilmark{21}
P.~D'Avanzo,\altaffilmark{22,25}
V.~D'Elia,\altaffilmark{14,26}
M.~Della Valle,\altaffilmark{27,28,29}
A.~de Ugarte Postigo,\altaffilmark{21}
Yu.~Efimov,\altaffilmark{16}
P.~Ferrero,\altaffilmark{1,30}
D.~Fugazza,\altaffilmark{21}
J.~P.~U.~Fynbo,\altaffilmark{3}
M.~G{\aa}lfalk,\altaffilmark{31}
F.~Grundahl,\altaffilmark{32}
J.~Gorosabel,\altaffilmark{20}
S.~Gupta,\altaffilmark{18}
S.~Guziy,\altaffilmark{20}
B.~Hafizov,\altaffilmark{33}
J.~Hjorth,\altaffilmark{3}
K.~Holhjem,\altaffilmark{34,35}
M.~Ibrahimov,\altaffilmark{33}
M.~Im,\altaffilmark{36}
G.~L.~Israel,\altaffilmark{14}
M.~Je\'{l}inek,\altaffilmark{20}
B.~L.~Jensen,\altaffilmark{3}
R.~Karimov,\altaffilmark{33}
I.~M.~Khamitov,\altaffilmark{37}
\"{U}. K{\i}z{\i}lo\v{g}lu,\altaffilmark{38}
E.~Klunko,\altaffilmark{39}
P.~Kub\'anek,\altaffilmark{19}
A.~S.~Kutyrev,\altaffilmark{40}
P.~Laursen,\altaffilmark{3}
A.~J.~Levan,\altaffilmark{41}
F.~Mannucci,\altaffilmark{42}
C.~M.~Martin,\altaffilmark{43}
A.~Mescheryakov,\altaffilmark{6}
N.~Mirabal,\altaffilmark{44}
J.~P.~Norris,\altaffilmark{45}
J.-E.~Ovaldsen,\altaffilmark{46}
D.~Paraficz,\altaffilmark{3}
E.~Pavlenko,\altaffilmark{17}
S.~Piranomonte,\altaffilmark{14}
A.~Rossi,\altaffilmark{1}
V.~Rumyantsev,\altaffilmark{17}
R.~Salinas,\altaffilmark{47}
A.~Sergeev,\altaffilmark{12,13}
D.~Sharapov,\altaffilmark{33}
J.~Sollerman,\altaffilmark{3,31}
B.~Stecklum,\altaffilmark{1}
L.~Stella,\altaffilmark{14}
G.~Tagliaferri,\altaffilmark{21}
N.~R.~Tanvir,\altaffilmark{48}
J.~Telting,\altaffilmark{33}
V.~Testa,\altaffilmark{14}
A.~C.~Updike,\altaffilmark{49}
A.~Volnova,\altaffilmark{50}
D.~Watson,\altaffilmark{3}
K.~Wiersema,\altaffilmark{48,51}
D.~Xu\altaffilmark{3}
}

\altaffiltext{1}{Th\"uringer Landessternwarte Tautenburg,  Sternwarte 5, D--07778 Tautenburg, Germany}
\altaffiltext{2}{Department of Physics and Astronomy, University of Nevada, Las Vegas, NV 89154}
\altaffiltext{3}{Dark Cosmology Centre, Niels Bohr Institute, University of Copenhagen, Juliane Maries Vej 30, DK-2100 K{\o}benhavn {\O}, Denmark}
\altaffiltext{4}{Division of Physics, Mathematics, and Astronomy, California Institute of Technology, Pasadena, CA 91125, USA}
\altaffiltext{5}{Raymond and Beverly Sackler School of Physics \& Astronomy, Tel Aviv University, Tel Aviv 69978, Israel}
\altaffiltext{6}{Space Research Institute (IKI), 84/32 Profsoyuznaya Str, Moscow 117997, Russia}
\altaffiltext{7}{Department of Astronomy, University of Texas, Austin, TX 78712, USA}
\altaffiltext{8}{Townes Fellow, Space Sciences Laboratory, University of California, Berkeley, CA, 94720-7450, USA}
\altaffiltext{9}{Astronomy Department, University of California, 445 Campbell Hall, Berkeley, CA 94720-3411, USA}
\altaffiltext{10}{Centre for Astrophysics Research, University of Hertfordshire, College Lane, Hatfield, Herts, AL10 9AB, UK}
\altaffiltext{11}{Centre for Astrophysics and Cosmology, Science Institute, University of Iceland, Dunhagi 5, IS 107 Reykjavik, Iceland}
\altaffiltext{12}{Terskol Branch of Institute of Astronomy of RAS, Kabardino-Balkaria Republic 361605 Russian Federation}
\altaffiltext{13}{International Centre of Astronomical and Medico-Ecological Research of NASU, 27 Akademika Zabolotnoho St. 03680 Kyiv, Ukraine}
\altaffiltext{14}{INAF, Osservatorio Astronomico di Roma, via Frascati 33, 00040, Monteporzio Catone (RM), Italy}
\altaffiltext{15}{Kazan State University and Academy of Sciences of Tatarstan, Kazan, Russia}
\altaffiltext{16}{Crimean Laboratory of the Sternberg Astronomical Institute, Nauchny, Crimea, 98409, Ukraine}
\altaffiltext{17}{SRI ``Crimean Astrophysical Observatory'' (CrAO), Nauchny, Crimea, 98409, Ukraine}
\altaffiltext{18}{Astrophysical Institute, Department of Physics and Astronomy, Clippinger 339, Ohio University, Athens, OH 45701, USA}
\altaffiltext{19}{European Space Agency (ESA), European Space Astronomy Centre (ESAC), P.O. Box - Apdo. de correos 78, 28691 Villanueva de la Ca\~nada, Madrid, Spain}
\altaffiltext{20}{Instituto de Astrof\'{\i}sica de Andaluc\'{\i}a (IAA-CSIC), Apartado de Correos, 3004, E-18080 Granada, Spain}
\altaffiltext{21}{INAF, Osservatorio Astronomico di Brera, via E. Bianchi 46, 23807 Merate (LC), Italy}
\altaffiltext{22}{Universit\`{a} degli studi di Milano-Bicocca, Dipartimento di Fisica, piazza delle Scienze 3, 20126 Milano, Italy}
\altaffiltext{23}{Department of Astronomy, Yale University, P.O. Box 208101, New Haven, CT 06520, USA}
\altaffiltext{24}{Department of Astronomy, 601 Campbell Hall, University of California, Berkeley, CA 94720--3411}
\altaffiltext{25}{Dipartimento di Fisica e Matematica, Universit\`{a} dell'Insubria, via Valleggio 11, 22100 Como, Italy}
\altaffiltext{26}{ASI-Science Data Centre, Via Galileo Galilei, I-00044 Frascati, Italy}
\altaffiltext{27}{INAF, Osservatorio Astronomico di Capodimonte, Salita Moiariello, 16 80131, Napoli, Italy}
\altaffiltext{28}{European Southern Observatory, Karl Schwarschild Strasse 2, D-85748 Garching bei M\"unchen, Germany}
\altaffiltext{29}{International Centre for Relativistic Astrophysics Network, Piazzale della Republica 2, Pescara, Abruzzo, Italy}
\altaffiltext{30}{Instituto de Astrof\'isica de Canarias, C/ V\'ia L\'actea, s/n E38205 - La Laguna (Tenerife). Espa\~na}
\altaffiltext{31}{Stockholm Observatory, Department of Astronomy, Stockholm University, AlbaNova University Centre, 106 91 Stockholm, Sweden}
\altaffiltext{32}{Department of Physics and Astronomy, University of Aarhus, Ny Munkegade, 8000 {\AA}rhus C, Denmark}
\altaffiltext{33}{Ulugh Beg Astronomical Institute, Tashkent 700052, Uzbekistan}
\altaffiltext{34}{Nordic Optical Telescope, Apartado 474, Santa Cruz de La Palma, Spain}
\altaffiltext{35}{Argelander-Institut f\"ur Astronomie, Universit\"at Bonn, Auf dem H\"ugel 71, D-53121 Bonn, Germany}
\altaffiltext{36}{Center for the Exploration of the Origin of the Universe (CEOU), Astronomy Program, Dept. of Physics \& Astronomy, Seoul National University, 56-1 San, Shillim-dong, Kwanak-gu, Seoul, South Korea}
\altaffiltext{37}{TUBITAK National Observatory, Antalya, Turkey}
\altaffiltext{38}{Middle East Technical University, Ankara, Turkey}
\altaffiltext{39}{Institute of Solar-Terrestrial Physics,  Lermontov st., 126a, Irkutsk, 664033 Russia}
\altaffiltext{40}{Observational Cosmology Laboratory, NASA/GSFC, 8800 Greenbelt Rd, Greenbelt, MD 20771-2400, USA}
\altaffiltext{41}{Department of Physics, University of Warwick, Coventry, CV4 7AL, UK}
\altaffiltext{42}{INAF, Osservatorio Astrofisico di Arcetri, largo E. Fermi 5, I-50125 Firenze, Italy}
\altaffiltext{43}{Loyola College, 4501 N. Charles Street, Baltimore, MD 21210, USA}
\altaffiltext{44}{Ram\'on y Cajal Fellow; Dpto. de F\'isica At\'omica, Molecular y Nuclear, Universidad Complutense de Madrid, Spain}
\altaffiltext{45}{Department of Physics and Astronomy, University of Denver, Denver, CO 80208, USA}
\altaffiltext{46}{Institute of Theoretical Astrophysics, University of Oslo, P. O. Box 1029 Blindern, N-0315 Oslo, Norway}
\altaffiltext{47}{Grupo de Astronom\'ia, Facultad de Ciencias F\'isicas y Matem\'aticas, Universidad de Concepci\'on, Concepci\'on, Chile}
\altaffiltext{48}{Department of Physics and Astronomy, University of Leicester, University Road, Leicester LE1 7RH, United Kingdom}
\altaffiltext{49}{Department of Physics and Astronomy, Clemson University 118 Kinard Laboratory, Clemson, SC 29634, USA}
\altaffiltext{50}{Sternberg Astronomical Institute, Moscow State University, Universitetsky pr., 13, Moscow 119992, Russia}
\altaffiltext{51}{Astronomical Institute ``Anton Pannekoek'', University of Amsterdam, Kruislaan 403, 1098 SJ Amsterdam, The Netherlands}

\begin{abstract}
We have gathered optical photometry data from the literature on a large sample of \emph{Swift}-era gamma-ray burst (GRB) afterglows including GRBs up to September 2009, for a total of 76 GRBs, and present an additional three pre-\emph{Swift} GRBs not included in an earlier sample. Furthermore, we publish 840 additional new photometry data points on a total of 42 GRB afterglows, including large data sets for GRBs 050319, 050408, 050802, 050820A, 050922C, 060418, 080413A and 080810. We analyzed the light curves of all GRBs in the sample and derived spectral energy distributions for the sample with the best data quality, allowing us to estimate the host galaxy extinction. We transformed the afterglow light curves into an extinction-corrected $z=1$ system and compared their luminosities with a sample of pre-\emph{Swift} afterglows. The results of a former study, which showed that GRB afterglows clustered and exhibited a bimodal distribution in luminosity space, is weakened by the larger sample. We found that the luminosity distribution of the two afterglow samples (\emph{Swift}-era and pre-\emph{Swift}) are very similar, and that a subsample for which we were not able to estimate the extinction, which is fainter than the main sample, can be explained by assuming a moderate amount of line-of-sight host extinction. We derived bolometric isotropic energies for all GRBs in our sample, and found only a tentative correlation between the prompt energy release and the optical afterglow luminosity at one day after the GRB in the $z=1$ system. A comparative study of the optical luminosities of GRB afterglows with echelle spectra (which show a high number of foreground absorbing systems) and those without reveals no indication that the former are statistically significantly more luminous. Furthermore, we propose the existence of an upper ceiling on afterglow luminosities and study the luminosity distribution at early times, which was not accessible before the advent of the \emph{Swift} satellite. Most GRBs feature afterglows that are dominated by the forward shock from early times on. Finally, we present the first indications of a class of long GRBs which form a bridge between the typical high-luminosity, high-redshift events and nearby low-luminosity events (which are also associated with spectroscopic SNe) in terms of energetics and observed redshift distribution, indicating a continuous distribution overall.
\end{abstract}


\section{Introduction}
\setcounter{footnote}{0}
The study of the optical afterglows of gamma-ray bursts (GRBs), first discovered over a decade ago \citep{vanParadijs1997}, has taken a great leap forward with the launch of the \emph{Swift} satellite \citep{Gehrels2004}. Its high $\gamma$-ray sensitivity and rapid repointing capabilities have ushered in an era of dense early afterglow observations. In the optical regime, one sobering result is that early afterglows are not as bright as expected, and early optical faintness seems to be the norm rather than the exception \citep{Roming2006}. Furthermore, \emph{Swift}-era optical afterglows are usually observed to have fainter magnitudes than those of the pre-\emph{Swift} era \citep[e.g.,][]{BergerFirstSwift, BergerSecondSwift, Fiore2007} (mainly due to even faint afterglows often being discovered thanks to the rapid XRT positions), and also lie at higher redshifts \citep{Jakobsson050814, Bagoly2006}, with the most distant up to now at $z=8.2$ \citep{Tanvir090423, Salvaterra090423}. For recent reviews of the impact of \emph{Swift} on GRB research, see \cite{ZhangSwiftReview}, \cite{MeszarosSwiftReview} and \cite{GehrelsARAA}.

The pre-\emph{Swift} afterglows have been studied extensively, both in terms of their light curve behavior \citep[e.g.][and references therein]{PK2002, ZKK}, and via their spectral energy distributions \citep[e.g.,][]{Stratta2004, PaperIII, StarlingPaperI}, which allow conclusions to be drawn concerning the rest frame line-of-sight extinction and even the dust type in some cases. In our previous study \citep[][henceforth K06]{PaperIII}, we found that the afterglows that met our selection criteria typically had little line-of-sight extinction, and that the dust properties were best described by Small Magellanic Cloud (SMC) dust, which shows no UV bump and strong FUV extinction \citep[e.g.,][]{Pei1992}. These results were confirmed by \cite{StarlingPaperI}, who studied a smaller sample but also incorporated X-ray afterglow data, as well as \cite{Schady2007, Schady2010}, who also employed joint optical-to-X-ray fits as well as UVOT (and ground-based) data.

Almost six years after the launch of \emph{Swift}, the amount of published data on optical/NIR GRB afterglows have become sufficient to compile a large sample comparable to the pre-\emph{Swift} sample studied by K06, and to determine if some of the afterglows detected by \emph{Swift} are truly fundamentally different to those of the pre-\emph{Swift} era. K06, as well as \cite{LZ2006} and \cite{Nardini2006}, found a clustering in the optical luminosities after correcting the afterglows for line-of-sight extinction (Galactic and intrinsic) and host contribution, followed by a transformation to a common redshift \citep[such a clustering has also been reported in the host-frame near-infrared bands by][]{GendreNIR}. Therefore, one could speculate that the observationally fainter \emph{Swift} afterglows might be intrinsically fainter too, implying that the clustering of luminosities may be due to a sample selection effect.

Another triumph of \emph{Swift} \citep[and \emph{HETE II},][see \citealt{LambHETE} for more HETE results]{Villasenor050709} was the discovery of optical counterparts to short GRBs \citep{Hjorth050709, Fox050709, Covino050709}, which, along with X-ray \citep{Gehrels050509B} and radio \citep{Berger050724} afterglows, have placed these GRBs into a cosmological context too, via accurate localization and host galaxy spectroscopy. Further observations, though, have blurred the ``classical'' short/hard vs. long/soft GRB dichotomy \citep{Kouveliotou1993, Ford1995}. The first was the discovery of GRBs with light curves consisting of short, hard spikes followed by extended soft emission components which led to $\TNT\approx100$ s, such as in the case of GRB 050724, which is unambiguously associated with an early-type galaxy \citep{Barthelmy050724, Berger050724, Grupe050724, Gorosabel050724, Malesani050724}. Further complications arose with the discovery of the temporally long, nearby events GRB 060505 and GRB 060614, which lacked SN emission down to very deep levels \citep{FynboNature, DellavalleNature, GalyamNature, Ofek060505}. The absence of SN emission is a hallmark of short GRBs\footnote{GRB 060614 and GRB 060505 are discussed in detail in \cite{KannShortII}. Still, we wish to point out here that alternative explanations tying these GRBs to the deaths of massive stars have been proposed \citep{FynboNature, DellavalleNature, GalyamNature, Thoene060505, McBreen060505}, with negligible radioactivity-driven emission being either due to fallback black holes \citep{Fryer2006, Fryer2007} or low energy deposition when the jet penetrates the star \citep{Nomoto2007, Tominaga2007}. At this time, the available observational information points to GRB 060614, despite its long duration, not being associated with the death of a young, massive star \citep{GehrelsNature, Zhang060614, Zhang080913}, while GRB 060505 is, both through analysis of its environment \citep{Thoene060505} and due to non-negligible spectral lag \citep{McBreen060505}.} \citep[e.g.,][]{Hjorth050509B, Fox050709, Ferrero050813}, which are thought to derive from the merger of compact objects such as neutron stars and black holes \citep{Blinnikov1984a, Blinnikov1984b, Paczynski1986, Goodman1986, Eichler1989}. Long GRBs, on the other hand, have been conclusively linked to the explosions of massive stars \citep{Galama1998, Hjorth030329, Stanek030329, Malesani2004, Pian060218, Chornock100316D, Starling100316D}, showing a photometric SN signature in case the redshift was $z\lesssim0.7$ and deep searches were carried out \citep{Zeh2004}. The existence of temporally long events which otherwise show signatures of ``short'' GRBs \citep[e.g., beginning with short, spiky emission and having negligible spectral lag,][]{NorrisBonnell} led \citet[][see also \citealt{ZhangNature, GehrelsNature}]{Zhang060614} to introduce an alternative definition. GRBs that have compact-object mergers as progenitors\footnote{To be more general, those that are not associated with the core collapse of young, massive stars.} (independent of their duration) are labeled Type I GRBs, while those with massive star progenitors (including X-Ray flashes, XRFs) are Type II GRBs. While this nomenclature has its disadvantages\footnote{Not only does it need multiple observational aspects, which are often not accessible, to associate a GRB with one type or the other \citep{Zhang080913}, but it also links the greater part of all detected GRBs through rather diverse properties (duration and/or spectral lag, for example) to the small sample of events where clear distinctions can be made, such as GRB 050724 (not associated with massive star formation) and GRB 030329 (associated directly with the explosion of a massive star).}, we will adopt it in the following.

In this study, we compile a large amount of optical/NIR photometric data on \emph{Swift}-era Type II GRB afterglows (detected by \emph{Swift}, \emph {INTEGRAL}, \emph{HETE II}, \emph{Fermi} and the \emph{IPN}). We create three samples (see \kref{data} for details). The ``Golden Sample'' (\kref{Golden}) follows the quality criteria given in K06. The ``Silver Sample'' (\kref{Silver}) consists of GRB afterglows with good light curve coverage, but where certain assumptions have to be made to create the SED. The ``Bronze Sample'' (\kref{Bronze}) comprises GRBs with good light curve coverage but where we are not able to derive an SED. For these afterglows, we assume no dust extinction, thus deriving a lower limit on the afterglow luminosity only. Furthermore, we compile the prompt emission parameters of all bursts (including the K06 GRBs) and derive the cosmological corrections and bolometric isotropic energies (\kref{Shift}). We then undertake a comparison of the GRB afterglows of the pre-\emph{Swift} and the \emph{Swift} era (\kref{RaD}). A comparison of the complete Type II GRB afterglow sample with the afterglows of Type I GRBs is presented in a companion paper \citep[][henceforth Paper II]{KannShortII}. There, we also detail the criteria for separating GRBs into the Type I and Type II samples. The GRBs presented in this paper are those that do not show any indication of being a Type I event, more specifically, following Fig. 8 of \cite{Zhang080913}, all GRBs presented in this paper are either classified as Type II GRBs or Type II GRB candidates. This is especially true for the two highest redshift events, GRB 080913 \citep{Greiner080913} and GRB 090423 \citep{Tanvir090423, Salvaterra090423}, which are intrinsically of short-duration, but are very probably due to collapsars both from their observational properties \citep{Zhang080913} as well as the basic observability of high-$z$ events with current detectors \citep{Belczynski080913}.

In our calculations we assume a flat universe with a matter density $\Omega_M = 0.27$, a cosmological constant $\Omega_\Lambda=0.73$, and a Hubble constant $H_0=71$ km s$^{-1}$ Mpc$^{-1}$ \citep{Spergel2003}. Errors are given at the $1\sigma$, and upper limits at the $3\sigma$ level for a parameter of interest.

\section{Data Collection and Analysis Methods}
\label{data}

\subsection{Additional photometric observations}
\label{ObservationsChap}

In addition to collecting all available data from the literature, we present further photometric observations of GRB afterglows included in our samples in this paper, many of them published here for the first time, with the rest being revised from preliminary values originally published in the GCN Circulars\footnote{http://gcn.gsfc.nasa.gov/gcn3\_archive.html}. Some are identical to GCN magnitudes reported earlier, when these have been deemed to be the final values.

In Appendix \ref{Observations}, we present information on observations as well as a table (Table \ref{tabPHOT}) containing 840 data points of 42 GRBs contained in our samples (including the last two pre-\emph{Swift} GRBs of the K06 sample, GRBs 040924 and GRB 041006). Observations have been reduced and analyzed with standard procedures under IRAF\footnote{http://iraf.noao.edu/ IRAF is distributed by the National Optical Astronomy Observatories, which are operated by the Association of Universities for Research in Astronomy, Inc., under cooperative agreement with the National Science Foundation.} and MIDAS\footnote{http://www.eso.org/sci/data-processing/software/esomidas/}, magnitudes being derived by aperature and PSF photometry against calibrator stars. See Appendix \ref{Observations} for more details on specific calibrators and special analysis techniques in some cases.

\subsection{The Samples}

Further to our own photometry, we compiled optical/NIR afterglow data from public sources (all references can be found in the Appendices) on a total of 79 GRBs with redshifts (three of them photometric, all others spectroscopic) and good light curve coverage (extending to about 1 day in the observer frame if the GRB were at $z=1$, with a few exceptions) from the \emph{Swift} era (from the end of 2004 to September 2009), a few of which have been localized by missions other than \emph{Swift}, such as \emph{HETE II}, \emph{INTEGRAL}, the \emph{Third Interplanetary Network} (\emph{IPN}) and most recently \emph{Fermi}. This is to be compared with a total of 251 Type II GRB afterglows in the \emph{Swift}-era as of the end of September 2009, of these, 146 have redshifts (122 spectroscopic, 6 photometric, 18 host galaxy-derived)\footnote{http://www.mpe.mpg.de/$\sim$jcg/grbgen.html}. All the remaining GRBs did not have redshifts and/or sufficient light curve coverage to be included in the sample. Depending on the data quality, we sort the 79 GRB afterglows into three different samples (with one further split temporally). All afterglow data are corrected for Galactic extinction using the maps of \cite{SFD}.

\subsubsection{The pre-\emph{Swift} era ``Golden Sample'' -- An update}
\label{PreGolden}

This sample comprises three GRBs. These were all included in the complete sample of K06, but not in their Golden Sample due to the SEDs not conforming to the sample selection criteria. Additional data (GRB 990510) and more diligent analysis (GRB 011211, GRB 030323) have led us to include them in the pre-\emph{Swift} Golden Sample. See Appendix \ref{App0} for more details.

\subsubsection{The \emph{Swift}-era ``Golden Sample''}
\label{Golden}

This sample comprises 48 GRBs. These GRBs fulfill the criteria of the Golden Sample of K06 and are thus directly comparable\footnote{Note that in our case, the derived extinction is always model-dependent, see, e.g., \cite{Watson050401} for a discussion on model-dependent and model-independent extinction estimations.}. The criteria are (K06):
\begin{enumerate}
\item The 1$\sigma$ error in $\beta$ ($\Delta\beta$) and the 1$\sigma$ error in
      $\AV$ ($\Delta\AV$) should both be $\leq0.5$.
\item $\AV+\Delta\AV\geq0$.
\item We do not consider GRBs where all fits (MW, LMC, and SMC) find $\AV<0$,
      even if the previous criterium is fulfilled\footnote{Such negative extinction is usually found when scatter in the SED results in data points in the blue region being too bright, or an overbright data point creating a ``2175 {\AA} emission bump''. In such cases, the fitting program determines that the best fit is then ``emissive dust'' which creates an upward curvature, $\AV<0$ and $\beta>\beta_0$.}. 
\item $\beta>0$ (although we do not reject cases with 
      $\beta-\Delta\beta\leq0$).
\item A known redshift (derived from absorption line spectroscopy in most cases, with some GRBs having redshifts from host galaxy spectroscopy (e.g., XRF 050416A, GRB 061126), and some being photometric redshifts (e.g., GRB 050801, GRB 080916C).
\end{enumerate}
Details on the GRB afterglows can be found in Appendix \ref{AppA}.

\subsubsection{The \emph{Swift}-era ``Silver Sample''}
\label{Silver}

This sample comprises 14 GRBs. These GRBs have well-detected multi-color afterglow light curves but the derived SEDs do not conform to the quality standards of the ``Golden Sample''. Certain reasonable assumptions are made to derive $\beta$ and thus $\AV$ using the theoretical relations derived from the fireball model \citep[e.g.,][]{ZhangMeszaros2004}. We treated different cases individually, the assumptions are listed for each GRB in Appendix \ref{AppB}. In some cases, $\beta$ is derived from the measured pre-break afterglow decay slope $\alpha$, in other cases, we us the X-ray spectral slope $\beta_X$ as given on the XRT repository webpage \citep{EvansXRT1, EvansXRT2}, with the assumed optical spectral slope being either $\beta=\beta_X$ or $\beta=\beta_X-0.5$. Further details on the GRB afterglows can be found in Appendix \ref{AppB}, see there especially for the case of GRB 071025. Note that all GRBs presented in K06 which were not included in their Golden Sample (an additional 11 GRBs, their Table 1) fit these ``Silver Sample'' selection criteria (three of these 11 GRBs are now presented in this work and have been added to the pre-\emph{Swift} ``Golden Sample'').

\subsubsection{The \emph{Swift}-era ``Bronze Sample''}
\label{Bronze}

The sample selection criteria used to define the Golden and Silver Samples includes a significant selection bias against dust obscured systems \citep{FynboSpectra}. In limiting ourselves to afterglows that have good multicolor observations (which is usually only the case for observationally bright afterglows), we may be missing out a population of fainter afterglows that would increase the spread of luminosities, in principle possibly bringing Type II GRB afterglows closer to those of Type I GRBs (Paper II). Reducing this selection bias is a complicated task, though. As detailed in Paper II, we expect no significant rest frame dust extinction along the line of sight to Type I GRB afterglows, and only a small spread in redshift, ranging (for now) from $z=0.1$ to roughly $z=1$. Both assumptions are invalid for Type II GRBs, which have been detected up to $z=8.2$ \citep{Tanvir090423, Salvaterra090423}, and can be strongly extinguished by dust in their host galaxies \citep[e.g., GRB 051022:][GRB 060923A: \citealt{Tanvir060923A}, GRB 061222A, GRB 070521: \citealt{PerleyHosts2009}, GRB 070306: \citealt{Jaunsen070306}, GRB 080607: \citealt{Prochaska080607} and GRB 090417B: \citealt{Holland090417B}]{Rol051022, CastroTirado051022, Nakagawa051022}. An unknown redshift can have a strong influence on the magnitude shift (denoted $dRc$, see \kref{Shift}), that is applied when transforming an afterglow from its observed redshift to a common redshift of $z=1$. An unknown rest frame extinction can only make the afterglow more luminous if it were corrected for\footnote{Note that an unknown redshift also implies an additional uncertainty in the host galaxy extinction, but this effect will usually be minor compared to the pure distance effect.}. Therefore we create a third Type II GRB afterglow sample which we denote the ``Bronze Sample''. The selection criteria are the following: Firstly, the GRB must have a well-constrained redshift. In all selected GRBs, this is a spectroscopic redshift, but we would not exclude GRBs with well-constrained photometric redshifts. We will not use GRBs that only have pseudo-redshifts \citep{pseudoz}. The knowledge of the redshift removes the strongest uncertainty in the luminosity derivation. Secondly, the redshift must be $z\leq4$, to keep the $R_C$ band (which is usually the band with the best measurements) unaffected by host galaxy Lyman $\alpha$ absorption or intergalactic Lyman $\alpha$ blanketing. Thirdly, the GRB must have sufficient afterglow data in the $R_C$ band (in some cases, we create composite light curves by shifting other colors to the $R_C$ zero point if sufficient overlap exists) to allow at least a confident extrapolation of the light curve to 0.5 rest frame days after the corresponding burst, where we determine the optical luminosity (\kref{Shift}). This final criterion removes many GRBs that only have very early detections published, usually \emph{Swift} UVOT observations. In total, we find 14 additional GRBs that fulfill these selection criteria. Clearly a selection bias against very faint afterglows still applies, but if one sets the sample selection threshold even lower (e.g., including also optically dark GRBs with no afterglow detection and no redshift), hardly any useful information can be gleaned. Furthermore, an analysis of all \emph{Swift} afterglow upper limits in the optical/NIR bands is beyond the scope of this paper.

In Appendix \ref{AppC}, we list the GRBs of the ``Bronze Sample''. We took data from the references given, constructed the afterglow light curve, and shifted it to a common redshift of $z=1$ assuming, identical to the Type I GRB sample (Paper II), a spectral index in the optical/NIR bands of $\beta=0.6$ and a host galaxy extinction of $\AV=0$ \citep[a single exception is GRB 060605, where an extrapolation of the X-ray slope finds that the optical slope must be identical, and also that the $R_C$ band is unaffected by Lyman $\alpha$ absorption,][]{Ferrero060605}. A value of $\beta=0.6$ was found to be the mean value for pre-\emph{Swift} afterglows (see, e.g., K06), and we find a similar value for the \emph{Swift}-era Golden Sample (\kref{AVres}). Most likely, in many cases, these afterglows suffer from host frame extinction as well, even if it is likely to be only a small amount (K06 found a mean extinction of $\AV=0.2$ in their sample, see also \kref{AVres}), so the derived luminosities are lower limits only. If the host extinction were known, and we would correct for it, the afterglows would always become more luminous. Therefore, the mean absolute magnitude of this sample is a lower limit only.

\subsection{Methods}
\label{Shift}
With knowledge of the redshift $z$, the extinction-corrected spectral slope $\beta$ and the host galaxy rest frame extinction $\AV$, we can use the method described in K06 to shift all afterglows to a common redshift of $z=1$, corrected for extinction along the line of sight. The $dRc$ values of the ``Silver'' (Appendix \ref{AppB}) and ``Bronze'' (Appendix \ref{AppC}) samples are only estimations and lower limits, respectively. After shifting, we derive for all afterglows the apparent magnitude at one day after the GRB (0.5 days in the source frame). In addition, for those GRBs where the light curves extend far enough, we also derive the magnitude at four days. Furthermore, we derive the magnitude in all possible cases at $10^{-3}$ days (43 seconds in the source frame) to compare the luminosity distribution at very early times. Such an analysis was not possible in the pre-\emph{Swift} era, as only a few GRBs were detected at such early times.

Our unique sample of afterglow luminosities allows us to look for correlations between the prompt emission and the optical afterglow parameters (see also Paper II). To achieve a comparison with the energies of the GRBs, we compiled the fluences and Band function (or cutoff power law) parameters for all afterglows in K06 and in this paper (Table \ref{tabTypeIISample}). We denote the low-energy and high-energy power-law indices of the Band function with $\alpha_B$ and $\beta_B$ respectively, to prevent them from being confused with the afterglow decay slopes $\alpha_{1,2}$ and the optical/NIR spectral index $\beta$. Using the given spectral parameters and the redshifts, we derive cosmological corrections for the rest-frame bolometric bandpass from 0.1 to 10000 keV following the method of \cite{BloomBol}. Using the correction, the fluences and the luminosity distances, we then derive the bolometric isotropic energy $E_{\rm iso,bol}$ for all GRBs. Note that if no high energy index $\beta_B$ is given, we used a cutoff power law instead of a Band function. In the cases of X-Ray flashes (XRFs, no $\alpha_B$ given), we assumed $\alpha_B=-1\pm0.3$. As only a small part of the spectral range lies beneath the peak energy in these cases, the choice of the value does not strongly influence the correction. For GRB 000301C, the Band function parameters could not be determined from Konus-Wind data (V. Pal'shin, priv. comm.), and we used mean values \citep{Preece2000}. In rare cases (e.g., GRB 050408) the errors of the correction were not constrained due to missing errors on the input data, and we assumed conservative errors.

\section{Results and Discussion}
\label{RaD}

The energetics, including the bolometric isotropic energies, for the complete sample (including the 19 GRBs from K06) can be found in Table \ref{tabTypeIISample}. The results of our SED fits with a Milky Way (MW), Large Magellanic Cloud (LMC) and Small Magellanic Cloud (SMC) extinction curve \citep{Pei1992} are given in Table \ref{tabALL} for the Golden Sample. For the Silver Sample, approximative results can be found in the individual GRB descriptions in Appendix \ref{AppB}. Apparent and absolute magnitudes at 1 and 4 days after the GRB can be found in Table \ref{tabMB}. The magnitudes of a selected sample of GRBs with early observations and/or well-defined peaks in the afterglow evolution can be found in Table \ref{tabPeak}.

\subsection{Observed Light Curves of \emph{Swift}-era GRB Afterglows}

\begin{figure*}[!t]
\epsfig{file=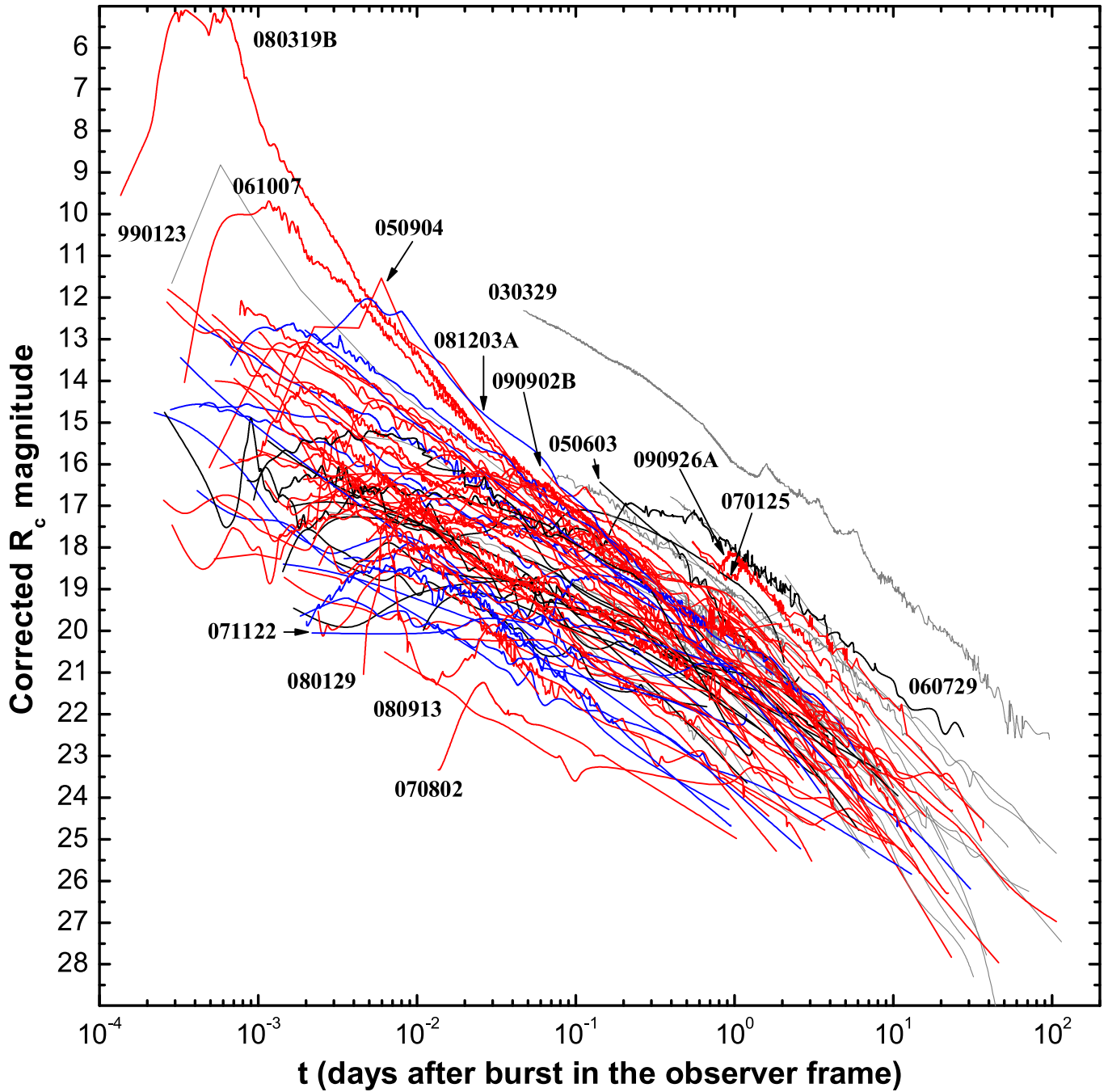,width=1\textwidth}
\caption[]{The afterglows of Type II GRBs in the observer frame. All data have been corrected for Galactic extinction \citep{SFD} and, where possible, the contribution of the host galaxy has been subtracted. Thin gray lines are Type II GRBs from the pre-\emph{Swift} era, taken from K06 (three further GRBs are presented in this paper but are added to the pre-\emph{Swift} sample). Thick red lines are the \emph{Swift}-era Golden Sample. The Silver Sample is blue, and the Bronze Sample is black. The large number of early afterglow detections is evident. Clearly, there are several afterglows that are significantly fainter than the pre-\emph{Swift} sample. At late times, the non-breaking afterglow of GRB 060729 \citep{Grupe060729} is brighter than any other except for GRB 030329. At very early times, the prompt flash of the ``naked-eye'' GRB 080319B \citep{Racusin080319B, Wozniak080319B, Beskin080319B} reaches four magnitudes brighter than the previous record-holder, GRB 990123. GRB 061007 comes close to the magnitude of the optical flash of GRB 990123, making it the third-brightest afterglow ever detected. At late times, the afterglow of the nearby GRB 030329 still remains brighter than any other afterglow discovered since.}
\label{Bigfig1}
\end{figure*}

In Fig. \ref{Bigfig1} we show, analogous to Fig. 7 in K06, the observed light curves of afterglows of \emph{Swift}-era GRBs (after correcting for Galactic extinction and also host galaxy contribution, if the latter is possible), in comparison with the Golden Sample of K06. The most immediate result is that the rapid and precise localization capabilities of \emph{Swift}, as well as the proliferation of rapid-slewing autonomous robotic telescopes, have strongly increased the number of afterglows that are detected at early times, typically starting within the first minutes after the GRB trigger. A strong spread in early magnitudes is also evident. Only a few GRBs of the \emph{Swift} era are observationally as bright as the brightest pre-\emph{Swift} afterglows. At early times, the prompt optical flash of the ``naked-eye'' GRB 080319B \citep{Racusin080319B, Bloom080319B, Wozniak080319B, Beskin080319B} lies several magnitudes above all other afterglows. Otherwise, only the afterglow of GRB 061007 is comparable\footnote{We note that GRB 060117, the burst with the second highest peak photon flux in the complete \emph{Swift} sample (after GRB 090424, which is included in our sample), had an early afterglow light curve \citep{Jelinek060117} which is almost identical in magnitude and evolution to the afterglow of GRB 061007. It was very close to the Sun at trigger time and could not be observed spectroscopically, therefore, it is not included in our sample.} to the optical flash of GRB 990123 \citep{Akerlof1999}. Several further early afterglows reach magnitude $\approx12$. At later times, beyond 0.1 days, the afterglows of GRB 050603, GRB 090926A, GRB 070125 and, at very late times, GRB 060729, are among the brightest observed, the latter due to a long plateau phase and a slow, unbreaking decline \citep{Grupe060729}. At early times, the faintest afterglows are GRB 071122 \citep[which has a long plateau phase,][]{CenkoDark}, GRB 080129 \citep[which was undetected down to 23rd magnitude in the beginning before rising to a huge optical flare,][]{Greiner080129}, GRB 080913 \citep[at very high redshift, assuming a fully transparent universe,][]{Greiner080913} and GRB 070802 \citep[which exhibited a late rise and a highly extinguished afterglow,][]{Kruehler070802, Eliasdottir070802}. After 0.1 days, several further afterglows (GRB 050401, XRF 050416A, GRB 060927, GRB 070419A, GRB 050502A) are considerably fainter than the faintest afterglow in the pre-\emph{Swift} sample, GRB 040924 (K06). This confirms the early reports about the faintness of the afterglows of \emph{Swift} GRBs \citep{BergerFirstSwift, BergerSecondSwift, ZhengDark}.

\subsection{Results from SED Fitting -- Low Host Extinctions at High Redshifts}
\label{AVres}

\begin{figure}[!t]
\epsfig{file=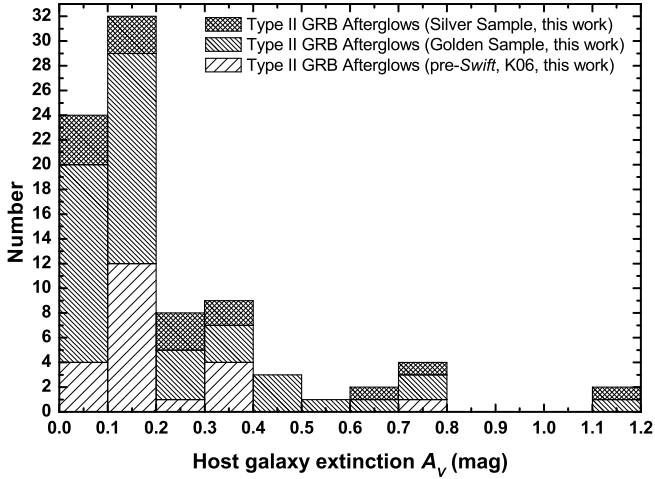,width=1\columnwidth}
\caption[]{Distribution of the derived host galaxy visual extinction $\AV$ in the source frame for the bursts of the Golden Sample of K06 (plus three additional ones presented in this paper) and the values derived in this work (Golden Sample, Table \ref{tabALL} and Silver Sample, Appendix \ref{AppB}), updating Fig. 2 in K06. In contrast to the sample presented in K06, we find two bursts with $\AV\geq0.8$, GRB 070802 with $\AV=1.18\pm0.19$ and GRB 060210 with $\AV=1.18\pm0.10$, the latter case being unsure, though. Most bursts have $\AV\leq0.2$, just as in the pre-\emph{Swift} sample.}
\label{AVHisto}
\end{figure}

\begin{figure}[!t]
\epsfig{file=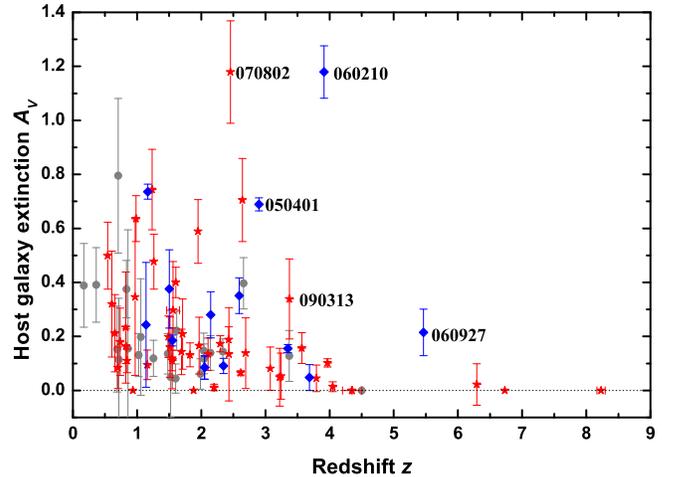,width=1\columnwidth}
\caption[]{Derived host galaxy visual extinction $\AV$ in the source frame for the Golden Sample of K06 bursts (grey circles, plus three additional pre-\emph{Swift} GRBs presented here) and the values derived for the GRBs in this work (Golden Sample, Table \ref{tabALL}, red stars, and Silver Sample, Appendix \ref{AppB}, blue diamonds) plotted as a function of the redshift $z$. A trend of lower extinctions toward higher redshifts is visible but only weakly supported (Kendall's $\tau=-0.34$, Spearman's $\rho=-0.42$). It is unclear if this is due to a selection effect (highly extinct bursts at high redshift are too faint to yield good data and are thus not included in our sample) or to the evolution of dust properties or quantities. Note that for most Silver Sample GRBs, the errors of the derived extinction are underestimated due to parameter fixing in the fitting process. Several exceptional GRBs are indicated.}
\label{AVz}
\end{figure}

SMC dust is preferred for most GRBs in our Golden Sample, which strengthens the results of K06. Clear exceptions are GRB 060124 (which features a small 2175 {\AA} bump, Kann et al., in preparation) and GRB 070802 \citep[][see Appendix \ref{AppA} for details]{Kruehler070802, Eliasdottir070802}, and several other GRBs are better fit, though not with high statistical significance, with LMC or MW dust (see Appendix \ref{AppA} for more details on each event). In several cases, especially for MW dust, the fitting process finds unphysical ``negative extinction'' which we interpret as strong evidence that the corresponding dust extinction curve is ruled out. We find a total of 17 GRBs for which the preference for SMC dust is strong, whereas the preference for SMC dust is weak for a further 17 events (cf. K06, which discuss in more detail the dependence of dust model preference on redshift and dust amount). For a total of six afterglows, we find no evidence at all for dust and use the spectral slope without extinction correction (in further cases, the amount of dust is negligible within errors), most of these events (four out of six) lie at high redshift, $z>4$ (see below). We find a total of seven GRBs with $\AV>0.5$ within $1\sigma$ errors, two of these, GRB 070802 with $\AV=1.18\pm0.19$ and GRB 060210 with $\AV=1.18\pm0.10$ (see Appendix \ref{AppB} for caveats on this event, though), lie significantly above the highest value in K06, that of GRB 991208 ($\AV=0.80\pm0.29$). The further GRBs are GRB 050408 ($\AV=0.74\pm0.15$), XRF 071010A ($\AV=0.64\pm0.09$), and GRB 080210 ($\AV=0.71\pm0.15$, though the data are sparse on this event) in the Golden Sample, and GRB 050401 ($\AV=0.69\pm0.02$) and GRB 070208 ($\AV=0.74\pm0.03$) in the Silver Sample, note here that the errors are underestimated due to the spectral slope in the fit not being a variable. Furthermore, evidence for high extinction of an uncommon type is found for GRB 071025, we use the value derived in the work of \cite{Perley071025}. Otherwise, all afterglows show very low extinction (Fig. \ref{AVHisto}). The mean extinction value for the 48 GRB afterglows of the Golden Sample is $\overline{\AV}=0.21\pm0.03$ (FWHM 0.24), identical to the pre-\emph{Swift} sample value of $\overline{\AV}=0.20\pm0.04$ (K06, with the additional three GRBs presented here). Similarly, the mean value for the extinction-corrected spectral slope, $\overline{\beta}=0.66\pm0.04$ (FWHM 0.25), is also in decent agreement within errors with the value from K06, $\overline{\beta}=0.54\pm0.05$. The reason for the larger value in this sample (offset by $1.9\sigma$, which is not yet statistically relevant) is not obvious, though. A possible explanation may be that a higher number of \emph{Swift} GRB afterglows have SEDs which include NIR data, which, being less affected by extinction, allow less flexibility and perhaps overestimation of the extinction. To test this conjecture, we create a subsample of the Golden Sample which contains only SEDs which fulfill the following criteria: 1.) They include NIR data ($YJHK$ as well as \emph{Spitzer} data). 2.) The best fit included a dust model (i.e., no straight power-law fits). 3.) There are at least four SED data points left after removing the NIR points to assure a free fit is possible. A total of 32 afterglow SEDs fulfill these criteria. For the original sample, we find $\overline{\beta}=0.69\pm0.04$ (FWHM 0.24), in full agreement with the complete Golden Sample (48 SEDs). Removing all the NIR data and refitting, we find that more than half the fits are not physically reasonable any more (mostly due to $\beta<0$), and it is $\overline{\beta}=0.02\pm0.39$ (FWHM 2.19 due to some strong outliers). Generally, this indicates that our conjecture is correct, and points out the need for NIR data to achieve good SED results.

For the Silver Sample, we find a strong preference for SMC dust in some cases (e.g., GRB 051109A and GRB 051111), but even for SMC dust, these fits are still not good and formally rejected. Such strongly curved SEDs were also found for some pre-\emph{Swift} GRB afterglows (e.g., GRB 971214, K06). A free fit to such an SED results in very high extinction and a negative spectral slope $\beta$. We find a mean host extinction which is slightly higher than that of both our Golden Samples ($\overline{\AV}=0.32\pm0.08$), but we caution that the derived extinctions depend upon fixed spectral index values derived from theoretical relations only.

One big difference between the \emph{Swift-}era sample and the sample of K06 is that there was only one burst \citep[GRB 000131,][]{Andersen000131} in the pre-\emph{Swift} sample which had $z\geq3$ (as GRB 030323 [\citealt{Vreeswijk030323}] has only been included in the pre-\emph{Swift} sample in this work), while in the present Golden Sample, 27\% (13 out of 48) of the GRBs lie at such high redshifts (an additional 36\%, five out of 14, in the Silver Sample). Like GRB 000131, almost all these high-$z$ GRBs show very small host extinction (Fig. \ref{AVz}). This seems to confirm the initial suspicion in K06 that host extinction declines with higher redshift. Exceptions are GRB 060210 \citep[see above, and][]{CenkoDark}, GRB 071025 \citep[see][]{Perley071025}, GRB 090313 with a moderate extinction of $\AV=0.34\pm0.15$, and possibly GRB 060927, but the result here is unsure. Also, GRB 050401, at $z\approx3$, clearly shows signs of moderate line-of-sight reddening \citep[see][]{Watson050401}. To check the significance of this possible result, we use two rank correlation tests, Kendall's $\tau$ and Spearman's $\rho$, on the combined Golden Sample bursts of this work and K06. We find $\tau=-0.34$ and $\rho=-0.42$, both results indicate that while there is a (negative, as expected) correlation, it is only weakly significant at best. To estimate the influence of the errors, we do the same tests on the maximum ($\AV+\Delta\AV$) and minimum possible ($\AV-\Delta\AV$) values. We find $\tau=-0.14\cdots-0.39$ and $\rho=-0.25\cdots-0.52$ (with the minimum extinction yielding the lowest rank correlation coefficient, and vice versa). Furthermore, a Kolmogorov-Smirnov (KS) test on two samples (taken from the Golden Samples only), divided by $z<2$ and $z>2$, shows that they are likely to be taken from the same distribution ($P=0.039$). While this low-extinction result might be expected from several evolutionary factors, such as the metallicity evolution of the universe \citep[][and references therein]{LapiHosts, Li2007}, different dust depletion patterns at high redshift \citep[see][for an overview]{Savaglio2006} or the lack of dust-producing AGB stars \citep{Fiore2007, Stratta050904}, we caution that several biases might be involved \citep[as already discussed in K06, see also][]{Fiore2007} and the evidence is thus not conclusive at all. A higher redshift implies what we see in the optical (and measure with a spectrograph) lies further and further into the rest frame ultraviolet, which is much more affected by dust (especially if it is similar to SMC dust, which is usually found). Therefore, unless rapidly observed, highly extinct high redshift afterglows are much more likely to not be observed successfully spectroscopically, thus excluding them from our sample. 

\subsection{Rest-frame Light Curves of \emph{Swift}-era GRB Afterglows}

\subsubsection{On the Rest-Frame Clustering and Redshift-Dependent Bimodality of the Afterglow Distribution in the \emph{Swift}-Era}
\label{Bimodal}

\begin{figure*}[!t]
\epsfig{file=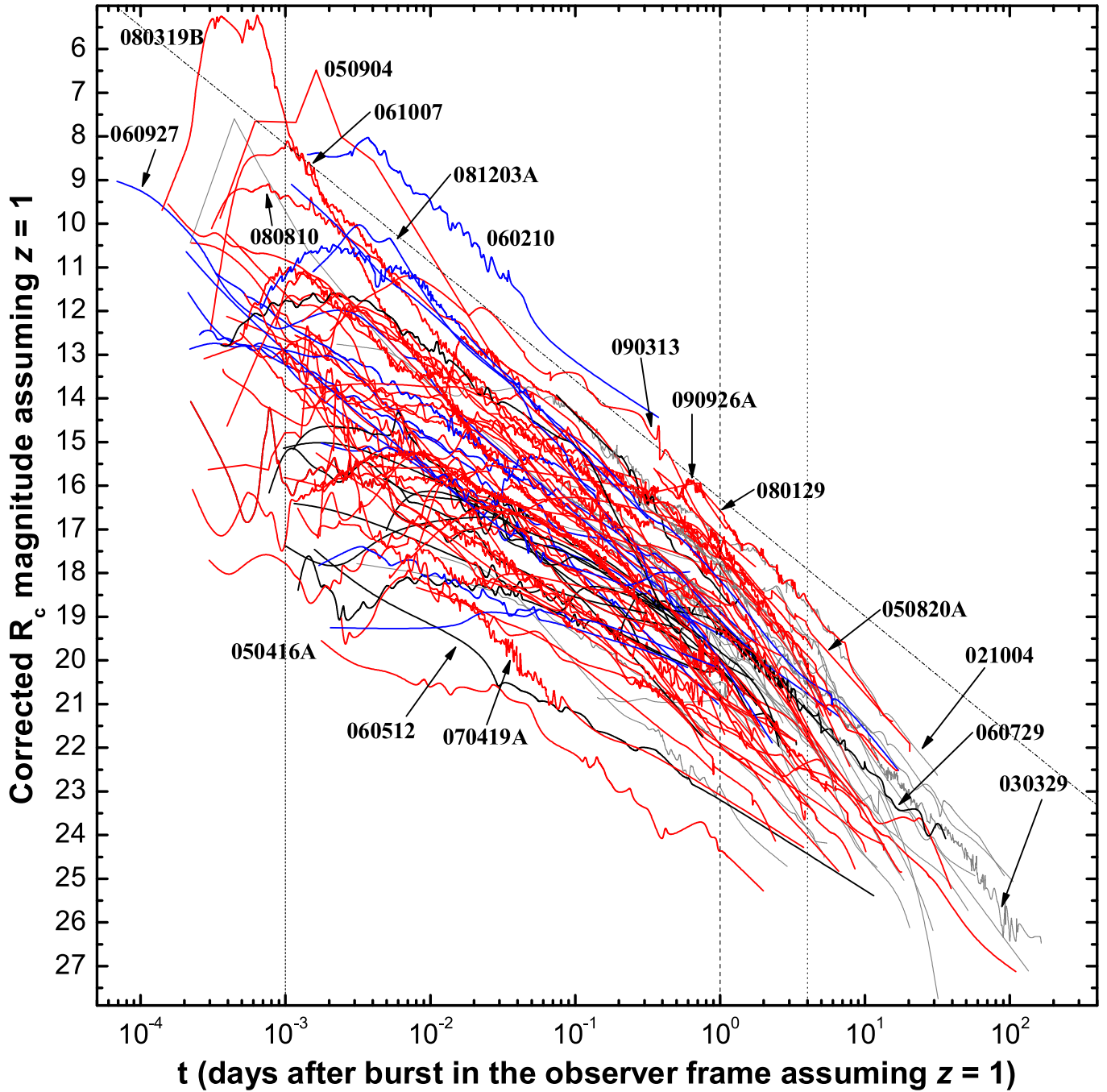,width=1\textwidth}
\caption[]{The afterglows of Type II GRBs in the observer frame after transforming them to a common redshift of $z=1$. The labeling is identical to Fig. \ref{Bigfig1}. Note that the Bronze Sample afterglows are lower limits only, as no dust extinction correction could be applied. The vertical lines denote $10^{-3}$, 1 and 4 days, times at which the luminosities were determined. All additional \emph{Swift}-era afterglows fit into the tight clustering reported by K06, \cite{LZ2006} and \cite{Nardini2006}. The width of the distribution expands slightly, with XRF 050416A, XRF 060512 and GRB 070419A being fainter than the faintest afterglow of the pre-\emph{Swift} era, GRB 040924. Note that XRF 060512 is in the Bronze Sample, its luminosity may be underestimated. At very early times, a large spread is visible, as well as several cases of strong variability beyond a simple decay. The brightest bursts at early times are (as in the observer frame) GRBs 080319B, 050904, 061007 and 060210 (the latter an unsure case, the extinction may be overestimated), and several more GRBs (060927, 080413A, 080810, 080721 and 081203A) exceed tenth magnitude. The dot-dashed, slanted line ($\alpha\approx1$) indicates what may be an upper ceiling for the afterglow luminosity at later times (\kref{UpperLimit}). Similar to the afterglow of GRB 030329 (K06), the afterglow of GRB 060729 is now seen to be of average luminosity at one day, and quite subluminous at early times. Exceptional afterglows, both at the bright and at the faint ends of the distribution, are indicated.}
\label{Bigfig2}
\end{figure*}

It was independently found by three groups \citep[K06;][]{LZ2006,Nardini2006} studying pre-\emph{Swift} afterglows that the magnitude distribution becomes tighter (clusters) compared to the observed distribution if the afterglows are corrected for host-frame extinction and transformed to a common redshift ($z=1$ was used). Closer study revealed that this clustering was best described by two populations (a bimodality) which were separated in redshift. Nearby afterglows were, in the mean, less luminous than more distant ones (K06).

In Fig. \ref{Bigfig2} we show (analogous to figure 8 in K06) the light curves of all optical afterglows shifted to $z=1$ (for the method, see K06, their Appendix A). The additional 76 afterglows of the \emph{Swift}-era samples seem to confirm the clustering of intrinsic afterglow luminosities. Only three afterglows, those of XRF 050416A, XRF 060512 and GRB 070419A, are fainter than the one of GRB 040924 at one day, but the difference is only large for XRF 050416A, $\approx1.3$ magnitudes. Also, note that XRF 060512 is from the Bronze Sample and thus the luminosity is probably higher, and the low redshift, derived from a galaxy spatially coincident with the optical afterglow, is in dispute, the GRB may be a background event \citep{OatesUVOT}. Furthermore, also only three afterglows exceed the previously brightest one, GRB 021004, these being GRB 090313, GRB 090926A and especially GRB 080129, which is 0.7 magnitudes brighter. We point out that this was an extremely peculiar afterglow which exhibited a long plateau phase and multiple rebrightenings \citep{Greiner080129}, and something similar was seen for GRB 090926A \citep{CenkoLAT, Rau090926A}. In K06, a bimodality in the afterglow luminosities was found after dividing the samples into two redshift bins, with $z=1.4$ as a dividing line. The new afterglows further bolster this finding, with the faintest afterglows at early times (GRB 060729, GRB 060904B, GRB 071122) and at later times (XRF 050416A, XRF 060512, GRB 070419A) all lying at $z\lesssim1$. A quantitative analysis leads us to be cautious about this result, though. While we find that the total spread of pre-\emph{Swift} afterglows is indeed reduced (7 to 5.7 magnitudes for those detected at one day), the FWHM of the two samples is identical, though (1.51 vs. 1.54 mags). For the \emph{Swift} era sample, we even find that both the spread (6.9 vs. 7.8 mags) as well as the FWHM (1.48 vs. 1.63 mags) actually \emph{increases}. Only for a complete sample of all afterglows, the range is still reduced (8.9 vs. 7.8 mags) while the FWHM is similar (1.56 vs. 1.61 mags). The effect of the reduced spread is mostly due to a single afterglow, that of the very nearby GRB 030329, whereas the increase in spread in the \emph{Swift}-era data is due to the very faint afterglow of XRF 050416A.

\begin{figure}[!t]
\epsfig{file=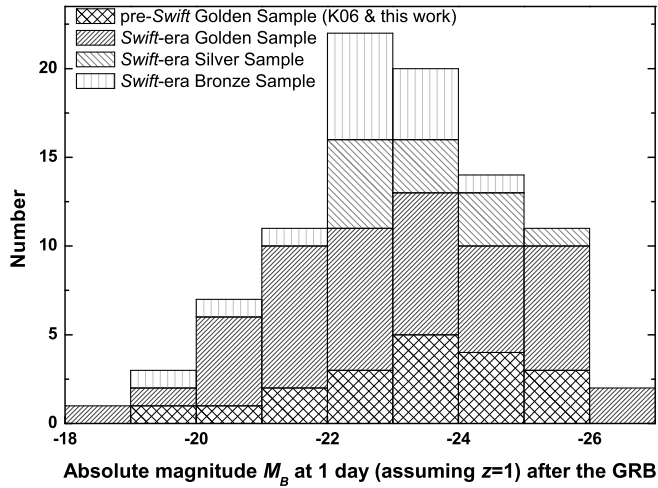,width=1\columnwidth}
\caption[]{The absolute magnitudes of Type II GRB afterglows at one day after the burst assuming $z=1$. The \emph{Swift}-era Type II GRB afterglows in the Golden Sample analyzed in this paper (mean magnitude $\overline{\MB}=-23.02\pm0.27$, FWHM 1.82 mags) are very similar to those of the K06 (plus three additional GRBs presented in this work) sample (mean magnitude $\overline{\MB}=-23.44\pm0.36$, FWHM 1.59 mags). The Silver Sample afterglows are identical in luminosity with the pre-\emph{Swift} sample (mean magnitude $\overline{\MB}=-23.69\pm0.32$, FWHM 1.11 mags). The Bronze Sample afterglows represent the least luminous afterglows (mean magnitude $\overline{\MB}=-22.53\pm0.34$, FWHM 1.25 mags), and the assumption of a small amount of rest frame extinction makes them as bright as the Golden Sample afterglows.}
\label{Histo}
\end{figure}

While \emph{Swift} has clearly allowed us to detect afterglows that are observationally fainter (we find $\overline{R_C}=20.08\pm0.36$ for the pre-\emph{Swift} GRBs, and $\overline{R_C}=21.30\pm0.18$ for the \emph{Swift}-era GRBs), are these afterglows also less luminous? In Fig. \ref{Histo} we show the distribution of afterglow magnitudes measured in the host frame at one day after the GRB assuming a common redshift of $z=1$ (Table \ref{tabMB}). Evidence for the bimodality is not directly evident. Indeed, several recent works \citep{MelandriROBONET, CenkoDark, OatesUVOT}, working on small, homogeneous samples derived from single instruments, do not report finding a bimodality. As a whole, the clearer bimodality of the pre-\emph{Swift} sample has disappeared (if one does not do the redshift separation), though see \cite{Nardini2008}. Indeed, while the magnitude distribution is not fit very well with a unimodal distribution (both Gaussian or Lorentzian distributions yield $\chi^2$/d.o.f. $>2$ for both the \emph{Swift}-era data set as well as the complete data set [see below]), we were also unable to find a bimodal distribution which significantly improved the fit. We thus, working on a larger sample, find agreement with above-mentioned works \citep{MelandriROBONET, CenkoDark, OatesUVOT}, in contrast to the clear bimodality seen by \cite{Nardini2008}.

It is also evident that our four samples are not significantly different from each other. K06 found a mean absolute magnitude of $\overline{\MB}=-23.3\pm0.4$ for their Golden Sample, to which we now add three further GRBs (for a total of 19, as three events do not have data at one day), and find $\overline{\MB}=-23.44\pm0.36$ (FWHM 1.59 mags). This value is almost identical to our new Golden Sample (a total of 46 GRBs, two not having data at one day), where we find $\overline{\MB}=-23.02\pm0.27$ (FWHM 1.82 mags). A KS test shows that both data sets are consistent with being drawn from the same distribution ($P=0.78$). The Silver Sample (14 GRBs, two do not have data at one day) is slightly more luminous on average, with $\overline{\MB}=-23.69\pm0.32$ (FWHM 1.11 mags), whereas the Bronze Sample (14 GRBs) is slightly less luminous on average, with $\overline{\MB}=-22.53\pm0.33$ (FWHM 1.25 mags), respectively. The difference is not statistically significant, though, a KS test shows that they are taken from the same distribution as the \emph{Swift}-era Golden Sample ($P=0.15$ and $P=0.43$ for the Silver and Bronze Samples, respectively). Furthermore, as no extinction correction has been applied to the Bronze Sample, its mean absolute magnitude is a lower limit only (\kref{Bronze}). Assuming $\AV=0.3$ for all afterglows of this sample, a value just slightly above the mean extinction of the Golden Samples (\kref{AVres}), we find a mean magnitude $\overline{\MB}=-23.45\pm0.38$ (FWHM 1.42 mag), identical with to the Golden Samples of both papers. Therefore, a small amount of host extinction is sufficient to explain the slightly fainter mean magnitude. This is also confirmed by the recent analysis of \cite{Schady2010}, who derive $\AV$ values for eight of our Bronze Sample GRB afterglows, and find $\AV\approx0.3$ or even (significantly) higher for seven of these GRBs, GRB 060729 being the only exception (Appendix \ref{AppC}).

As the Golden Samples of K06 and this work can be readily compared (\citealt{ZhengDark} and \citealt{Gehrels2008} also find that the distribution of optical-to-X-ray spectral indices, $\beta_{OX}$, is identical in pre-\emph{Swift} and \emph{Swift} GRB afterglows) we create one total Golden Sample and split it along the $z=1.4$ division used by K06. Here, we find clear evidence for bimodality, it is $\overline{\MB}=-21.89\pm0.32$ (FWHM 1.52 mags) for the low-$z$ (22 GRB afterglows) and $\overline{\MB}=-23.78\pm0.23$ (FWHM 1.51 mags) for the high-$z$ (43 GRB afterglows) sample. The difference is statistically significant, a KS-test shows that they are probably not taken from the same distribution ($P=2.9\,\times\,10^{-4}$). Using the \emph{Swift}-era sample only, and dividing it along the $z=1.4$ line, we find $P=8.4\,\times\,10^{-4}$, further strengthening the result. Non-parametric rank correlation tests find further, albeit relatively weak, evidence for the magnitude increase toward higher redshifts; it is Kendall's $\tau=-0.41$ and Spearman's $\rho=-0.53$ for the \emph{Swift}-era-only sample, and Kendall's $\tau=-0.31$ and Spearman's $\rho=-0.51$ for the complete sample, these very similar values also show that mixing the two samples is not problematic.

While \emph{Swift} has increased the recovery rate of afterglows, and also the percentage of afterglows with successful spectroscopy in follow-up observations, \emph{Swift} GRBs that actually did meet our selection criteria, especially those of the Golden Sample, are quite rare events. These bursts usually not only have a lot of optical follow-up, but are also interesting in such a manner that publications with data on these bursts are preferred over the many others \emph{Swift} has delivered. For example, GRB 050904 held the record for highest redshift ever discovered for a burst for several years, the afterglow of GRB 060206 showed a very powerful rebrightening, that of GRB 060526 showed a complex optical light curve, and GRB 061007, GRB 070125 and especially GRB 080319B were exceptionally bright, both in gamma-rays and in the optical. GRB 050408, one of the observationally faintest afterglows in our new Golden Sample, was very well observable from both hemispheres, leading to a lot of observations. In other words, our Golden Sample contains mostly GRBs that are not typical of the faint \emph{Swift} era bursts, but more typical of the \emph{Beppo-SAX} era. While the  selection criteria of the Silver and especially the Bronze Sample are less stringent, the amount of derived information is also reduced.

Still, it seems clear that for the GRBs in our combined \emph{Swift} sample (i.e., Golden, Silver and Bronze), the larger amount of faint afterglows is an effect based mostly on the increased mean ensemble redshift \citep{Jakobsson050814, Bagoly2006}. This is mainly a result of \emph{Swift} BAT's low-energy sensitivity and novel triggering methods, such as image triggers, which find GRBs whose light curves are strongly stretched due to redshift \citep[e.g.,][]{CampanaRT, Salvaterra2007a, Salvaterra2007b, Ukwatta2009}. Another factor is the rapid localization capability of \emph{Swift} combined with rapid ground-based follow-up, which is crucial for long-slit spectroscopy of faint high-$z$ targets. But the need for a spectroscopic redshift and decent light curve coverage is, of course, still a strong restriction for inclusion into our sample \citep[see][for a discussion on these selection effects]{Fiore2007}. There are afterglows which are clearly strongly extinguished by host extinction, such as GRB 051022 \citep{Nakagawa051022, Rol051022, CastroTirado051022}, GRB 060923A \citep{Tanvir060923A}, GRB 070306 \citep{Jaunsen070306}, GRB 061222A, GRB 070521 \citep{PerleyHosts2009}, GRB 080607 \citep{Prochaska080607} and GRB 090417B \citep{Holland090417B}. In such cases, we are unable to derive the afterglow luminosity \citep[which was probably very high in the case of GRB 051022, since it was an highly energetic burst with a very bright X-ray afterglow;][]{CastroTirado051022}. More highly extinguished or intrinsically faint afterglows very probably can be found among those \emph{Swift} afterglows that did not match our selection criteria, even for the Bronze Sample. Therefore, the question if ``dark'' GRBs are usually optically undetected due to strong host extinction or intrinsic faintness remains unsolved as yet, though several recent works find evidence for dust attenuation being the main factor \citep{Gehrels2008, ZhengDark, CenkoDark, PerleyHosts2009}. It therefore remains possible that a population of afterglows would remain that are significantly less luminous than all in our complete sample. In this case, the clustering of afterglow luminosities itself, as inferred by K06, \cite{LZ2006} and \cite{Nardini2006} (evidence for which has already been reduced with our larger sample) may be an observational bias, both due to optical sampling criteria \citep[good multicolor light curves and redshift, see, e.g.,][for how spectroscopic response time can induce further bias to the redshift distribution, and \citealt{FynboSpectra} for other selection effects involving redshift determination]{CowardSpeed} and gamma-ray detection criteria, similar to what has been proposed for the existence of high-energy correlations \citep[e.g.,][]{NakarPiran, BandPreece, ButlerBAT}. On the other hand, a recent study of pre-\emph{Swift} ``dark'' GRBs has shown that the bimodal clustering persists even after the inclusion of these events \citep{Nardini2007}, but we caution that this sample is built upon satellites that were less capable of detecting subluminous GRBs than \emph{Swift}. On the whole, a combination of factors makes the \emph{Swift} afterglow sample less biased than that of the pre-\emph{Swift} era, and thus more representative of the (unknown) true luminosity distribution. That we find only weak evidence for clustering with our less biased sample may indicate that an unknown observational bias has played a role in the pre-\emph{Swift} data. Recent research indicates that a luminosity evolution in the prompt emission of GRBs does exist \citep[e.g.,][]{SalvaterraLF}, and as we find (\kref{Berger}) that bursts with high isotropic energy release are usually also associated with brighter afterglows, this connection, while still involving many insecurities, implies that a true luminosity evolution in the afterglows is also plausible. Recently, \cite{ImeritoSpeed} have also reported that the result of \cite{CowardSpeed} implies a true evolution in the luminosity function, with afterglows at higher redshifts being more luminous, though their result does not allow a physical interpretation to be derived, nor the true shape of the luminosity function to be discerned.

\subsubsection{Does the high number of {\MgII} foreground absorbers depend on afterglow flux?}

\cite{ProchterMgII}, studying medium- and high-resolution spectra of bright GRB afterglows, found a high number of strong ($W_r(2796)>1$\,\AA) intervening {\MgII} absorption systems, $4\pm2$ times more than along the lines of sight to quasars studied in the SDSS. Such a discrepancy was not found in intervening {\CIV} systems \citep{SudilovskyCIV, TejosCIV}, and multiple explanations have been proposed \citep[see, e.g.,][for an overview]{PorcianiMgII}. Differing beam sizes of quasars and GRBs \citep{FrankMgII} as an explanation can not account for the case of GRB 080319B, where high-S/N multi-epoch data show no temporal variations over several hours \citep[][see also \citealt{PontzenMgII, VerganiUVES}]{D'Elia080319B}\footnote{\cite{Hao060206} claimed variable equivalent widths of foreground absorber {\MgII} lines seen in multi-epoch spectra of the GRB 060206 afterglow, but this was later refuted by \cite{Thoene060206} and \cite{Aoki060206} using high-S/N Subaru and WHT data.}. \cite{SudilovskyMgII} simulate the effect of dust in the foreground absorbers on quasar detection efficiency and rule out a strong contribution from this factor \citep[see also][]{CucchiaraMgII}. \cite{CucchiaraMgII} compare properties of foreground {\MgII} systems along quasar and GRB sightlines and find no significant differences, concluding that the GRB systems are probably not associated with material ejected near the GRB at relativistic velocities (intrinsic origin). \cite{TejosMgII} and \cite{VerganiUVES} also study the incidence of weak {\MgII} systems, and both come to the conclusion that the incidence of weak systems is similar along quasar and GRB afterglow sightlines, implying that the best explanation is that the GRB afterglows of the echelle sample have been amplified by gravitational lensing \citep[see also \citealt{Wyithe2010}, but see][]{CucchiaraMgII}. Both studies also find that the excess is smaller than originally deduced from the original small sample by \cite{ProchterMgII}, but the significance that the excess is real has increased with increasing sample size and redshift path.

All GRBs in the UVES sample of \cite{VerganiUVES} are included in our Golden Sample (with GRB 021004 being part of the pre-\emph{Swift} Golden Sample, and GRB 060607A not having any data at 0.5 rest-frame days, so we will not include it in this discussion), and two further GRBs with echelle spectra (Keck HIRES) from the sample of \cite{TejosMgII} are also part of our Silver Sample. Furthermore, a GRB with published UVES spectroscopy not included in the sample of \cite{VerganiUVES} is XRF 080330, which also shows a very strong {\MgII} foreground absorber \citep{D'Elia080330}, and another strong foreground system is seen in the afterglow of GRB 090313, as measured by X-Shooter \citep{deUgarte090313}. That all of these GRBs are in our samples is not surprising, only very bright afterglows can be successfully observed with echelle spectrographs, and will therefore very likely also have extensive photometric follow-up (and a redshift, of course), allowing inclusion in our sample \citep[though this situation is now changing with X-Shooter, see][]{deUgarte090313}. While the telescopes capable of deriving echelle spectra of GRBs (VLT/UVES+X-Shooter, Magellan/MIKE, and Keck/HIRES) are all concentrated in one hemisphere (Chile and Hawaii), and thus some GRBs with bright afterglows are missed because they have become too faint once they are observable (GRB 061007 being a good example), the isotropic distribution of GRBs should ensure that the echelle sample is mostly unbiased.

We first create two samples. The first one, the ``UVES'' sample, contains the nine GRBs from the sample of \cite{VerganiUVES} (excluding, as mentioned, GRB 060607A), furthermore GRB 051111 and GRB 080810 \citep[][while these are from the Silver Sample, the extinction correction is small in both cases, so the insecurity in the luminosity is not large]{TejosMgII} XRF 080330 \citep{D'Elia080330} and GRB 090313 \citep{deUgarte090313}. The ``non-echelle'' sample contains all other GRBs from the Golden Samples, for a total of 54 GRBs. Note that this sample still includes several GRBs with echelle observations. But in the cases of GRB 020813 \citep{Fiore2005}, GRB 050502A \citep{ProchaskaGCN050502A} and especially GRB 071003 \citep{Perley071003}, the resulting spectra had very low S/N. GRB 081008 was also observed by UVES \citep{D'AvanzoGCN081008}, but no information has been published on whether a foreground system exists or not. In the case of GRB 030329 \citep{Thoene030329}, the GRB is so close that there are no intervening absorbers, and gravitational lensing is very unlikely. We find the mean absolute $B$ magnitude of the UVES sample to be $\overline{\MB}=-23.72\pm0.41$, while the other sample has $\overline{\MB}=-23.05\pm0.25$, implying the the UVES sample is brighter only at the $1.4\sigma$ level, which is not statistically relevant ($P=0.28$). On the other hand, the UVES sample can be brighter by, at the $2\sigma$ level, up to 1.63 mags, which is a factor of $4.5\times$, which lies above the amplification of $1.7\times$ inferred by \cite{PorcianiMgII}, therefore we are not able to rule at such a rather subtle amplification with any significance either.

A second point is that the sample selection as we are using it now just gives us information about the afterglows which were, or were not, observed with high-resolution spectrographs, which can be due to nothing but luck (declination and explosion time). A more precise analysis needs to compare afterglows with strong foreground absorbing system with those that definitely do not have any. We therefore create two subsamples of the UVES sample. The ``strong sample'' contains GRBs 021004, 050820A, 051111, 060418, 080319B, 080330 and 090313, while the ``weak sample'' contains GRBs 050730\footnote{\cite{TejosMgII} find one ``Very Strong'' foreground absorber for this event from Magellan MIKE spectroscopy, but \cite{VerganiUVES} give an upper limit for this system below the $W_r(2796)>1$\,{\AA} cutoff after correcting for sky contamination using UVES data.}, 050922C, 071031, 080310, 080413A and 080810\footnote{There is tentative evidence for a very weak ($W_r(2796)<0.07$\,\AA) {\MgII} foreground system in the spectrum which was too weak to even be included in the sample of \cite{TejosMgII} (N. Tejos, priv. comm.). Of course, this does not influence the fact that GRB 080810 belongs to the weak sample.}. We find $\overline{\MB}=-23.65\pm0.66$ for the strong sample, and $\overline{\MB}=-23.80\pm0.49$ for the weak sample, implying they are identical ($P=0.95$), with the weak sample actually being marginally brighter (we caution that we are in the realm of low-number statistics here). We conclude that if the {\MgII} statistics are influenced by lensing, the effect is not statistically relevant, on the other hand, we can also not rule out a small amplification factor with any significance either. Clearly, the sample of high-S/N, high-res afterglow spectra must be increased before further conclusions can be drawn, X-Shooter will make an important contribution here \citep{deUgarte090313}.

We also note that on the issue of dust reddening by foreground systems, we find no evidence for large absorption in any of the GRBs in the UVES sample, with the highest values being found for GRB 060418 ($\AV=0.20\pm0.08$), where one foreground absorbing system may contribute significantly \citep{Ellison060418, VerganiUVES}, and GRB 090313 ($\AV=0.34\pm0.15$). We caution that the latter value is based on GCN data only so far, but there is corroborating evidence for dust found in the spectrum \citep{deUgarte090313}.

\subsubsection{Does an Upper Limit on the Forward Shock Luminosity Exist?}
\label{UpperLimit}
Compared to K06, the luminosity range of our afterglow sample has slightly expanded, both to lower and higher luminosities, but this must be seen in the context of a much larger sample. In the K06 sample, the afterglow of GRB 021004 dominated the luminosity distribution over a long period of time. In the present sample, several more GRBs are added which parallel the evolution of the afterglow of GRB 021004. The large early luminosity of the afterglow of GRB 050904 has been discussed in \cite{Kann050904}. Its light curve evolution is clearly anomalous, featuring an early rise, a plateau and a superposed sharp peak. Multiple papers \citep[e.g.,][]{Racusin080319B, Wozniak080319B, Beskin080319B, Kumar080319BA, Kumar080319BB} discuss the extreme prompt optical flash of GRB 080319B. Finally, the derived very high extinction for GRB 060210 is unsure (see Appendix \ref{AppB} for more details), implying that the afterglow, which seems to show a standard (not rapid, like GRB 050904 and GRB 080319B) decay after a short plateau and peak, may be much less luminous. Excluding these special events, the early afterglow of GRB 061007 \citep{Mundell061007, Schady061007} is the most luminous in the sample, although it decays rapidly\footnote{At one day, it has become so faint that it falls into the faint bin of the bimodal distribution, which could be ``expected'' from its redshift $z=1.261\leq1.4$.}. Between 0.01 and 0.5 days, the afterglow of GRB 090313 is the most luminous, though we caution that so far, we have only an extensive GCN data set. It is then exceeded by the last strong rebrightening of GRB 080129, which is then followed after about 1.5 days by the afterglow of GRB 090926A, which shows a very similar evolution to that of GRB 021004.

In Fig. \ref{Bigfig2} we plot as a boundary a power-law decay and attach it to the brightest afterglow detections at times from hours to days (we find $\alpha\approx1$). At early times, this slope is exceeded, and at least for GRB 990123, GRB 050904 and GRB 080319B, additional emission components dominate over the forward shock afterglow \citep[e.g.,][]{Akerlof1999, NakarPiran990123, Boer2006, Wei2006, Zou2006, Beskin080319B, Kumar080319BA, Kumar080319BB}. This may also be the case for GRB 061007, although this burst's afterglow showed a remarkable, unbroken broadband (from gamma-rays to optical) power-law decay from very early times onward \citep{Mundell061007, Schady061007}. Beyond $\approx2$ days, the light curves usually become steeper due to jet breaks. This upper boundary may imply that there exists an upper limit for the luminosity of forward-shock generated afterglow emission in the optical bands. \cite{Johanesson2007} have studied a large sample of synthetic afterglows created by using the standard fireball model and find that the luminosity function of afterglows (in wavebands from the X-rays to the radio) can be described by a lognormal distribution with an exponential cutoff at high luminosities, which may be considered a theoretical prediction of our result, although they do not explicitly state that. Determining the actual luminosity distribution from the data is clearly non-trivial, especially trying to discern between, e.g., a regular power-law distribution and one that needs an exponential cutoff at high luminosities (as a power-law distribution itself will trend toward zero, just not as sharply as the exponential cutoff). Furthermore, determining the slope of the power-law is complicated by selection effects such as Eddington bias at low luminosities (see, e.g., \citealt{Teerikorpi2004} for a discussion in terms of galaxy counts), as well as all the selection effects we have pointed out concerning our optically selected sample.

In the standard afterglow theory \citep[see, e.g.,][for the equations]{PK2000}, the optical flux generally depends on the isotropic kinetic energy $E_{\rm k,iso}$, the ambient density ($n$ for an ISM or $A^{*}$ for a wind), and the shock microphysics parameters $p$ (electron spectral index), $\varepsilon_e$ (fraction of energy in electrons) and $\varepsilon_B$ (fraction of energy in magnetic fields). This upper limit therefore is relevant to a combination of these parameters and cannot be used to pose a limit for each individual parameter. On the other hand, if one makes the assumption that the microphysics parameters do not vary significantly among bursts, this upper limit may suggest that bursts do not have an exceptionally large $E_{\rm k,iso}$ and the fireball is usually not expanding into an ambient medium of very high density. \cite{Johanesson2007} also find that variation of the initial energy release is one of the main drivers of the luminosity distribution (the others are the microphysical parameters, but we argue that they should not vary overly much from burst to burst). It may be possible that a very high circumburst density, as one would find within a molecular cloud, is connected to very large gas and dust column densities, and thus to a large line-of-sight extinction, which prevents us from detecting the afterglow or at least adding it to our sample. We note that \cite{Johanesson2007} find that a range of circumburst densities has little influence on the afterglow luminosity, but they only vary the density between 0.1 and 10 cm$^{-3}$. We also note that several of the GRBs that populate the region of the upper limit only reach it due to additional injections of energy into the external shock, e.g. GRB 021004 \citep{deUgarte021004}, GRB 060206 \citep{Wozniak060206, Monfardini060206}, GRB 070125 \citep{Updike070125, Chandra070125}, GRB 080129 \citep{Greiner080129} and GRB 090926A \citep{CenkoLAT, Rau090926A}. For GRB 050603 and especially GRB 991208 (cf. K06), the lack of early afterglow data makes the situation less clear. The afterglow of GRB 050820A has a relatively slow decay and a very late break. Therefore, several factors may account for the potential existence of this upper luminosity limit, and the afterglow sample will have to increase strongly to reach further conclusions, as these bright events are very rare.

\subsubsection{The Luminosity Distribution at Early Times -- Diversity and Clustering}
\label{LumEarly}

\begin{figure}[!t]
\epsfig{file=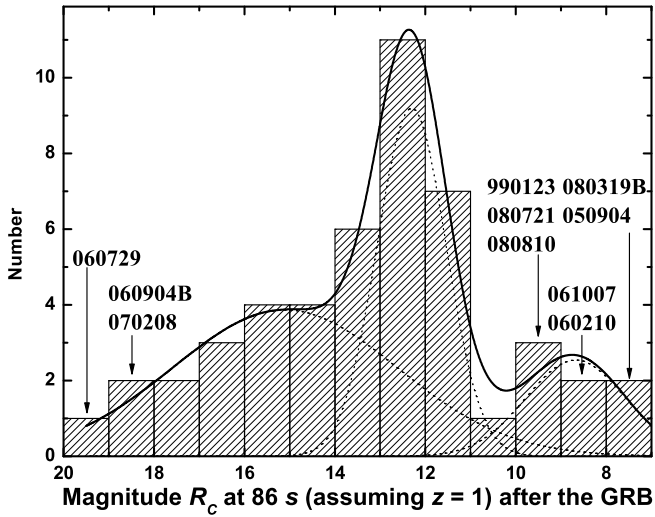,width=1\columnwidth}
\caption[]{The distribution of optical transient magnitudes at 86 seconds ($10^{-3}$ days) after the GRB trigger in the observer frame, after shifting all afterglows to $z=1$. While the complete spread is very wide (12 magnitudes), there is a strong clustering around $13^{\tn{th}}$ magnitude. We interpret these GRB afterglows as those where the forward shock emission dominates already at early times. In some cases, an additional component dominates strongly, making the afterglow even brighter, while many other afterglows suffer from early suppression. The complete distribution is trimodal and well-fit by three overlapping Gaussians. The brightest and faintest afterglows are indicated.}
\label{Early}
\end{figure}

As mentioned before, many of the \emph{Swift}-era GRBs in our sample have afterglows that have been detected at very early times, when they were for the most part still bright. This allows us to derive the luminosity distribution at early times, an exercise that was not possible in the pre-\emph{Swift} era. We choose $10^{-3}$ days (86.4 seconds) at $z=1$, which is equivalent to only 43.2 seconds after the GRB trigger in the source frame (in several cases, GRB prompt high-energy emission is still ongoing at this time). The sample comprises 48 afterglows, with GRB 990123 as the only burst from the pre-\emph{Swift} era\footnote{There is only one other pre-\emph{Swift} afterglow that is detected at such early times, GRB 021211. It is not included here as it did not yield a usable SED (K06). The flat, blue SED and the redshift close to $z=1$ imply a minimal $dRc-$shift, though, and it would be $R_C\approx14$ at 43 seconds in the rest frame.}. The distribution is presented in Fig. \ref{Early}. It is, on the one hand, very broad, which was already apparent from Fig. \ref{Bigfig2}. The total width is 11.5 magnitudes, almost twice as wide as the luminosity distribution at one day. On the other hand, 50\% of all afterglows (24 out of 48) cluster within only three magnitudes (a similar tight clustering has been found by \citealt{OatesUVOT} at 100 seconds after the GRB onset in the rest-frame). Eight afterglows (GRBs 080319B, 050904, 061007, 060210, 080810, 080721, 990123 and 080413B) are brighter than this cluster (albeit significantly, in some cases). Most of these are probably dominated by additional emission components at early times (see below), although the unbroken decay from very early times on in the case of GRB 061007 may speak against an additional component \citep{Mundell061007, Schady061007}. GRB 080721 is a similar case \citep{Starling080721}. The strongly clustered afterglows would then be those that are dominated by the forward shock emission already at early times, while the fainter afterglows may suffer from optical supression \citep{Roming2006} or a late afterglow onset \citep[e.g.][]{Molinari060418, Nysewander060607A}. In some cases \citep[e.g., GRB 060729,][]{Grupe060729}, there are also indications that significant long-term energy injection similar to what may cause the shallow decay/plateau phase of the ``canonical'' X-ray afterglow \citep{Nousek2006, Zhang2006, Panaitescu2006a} occurs, although in most cases the plateau phase in X-rays and the following break to a ``classical'' afterglow decay is not mirrored in the optical \citep{Panaitescu2006b}. This highlights the possibility that there are afterglows that start at a similar faintness to, e.g., GRB 060729, and then follow a straightforward decay instead of remaining roughly constant. These optical afterglows would be too faint to be included in our sample due to the selection criteria, and might also be much less luminous at 0.5 rest-frame days than the afterglows presented here.

Fitting the complete distribution with a single Gaussian does not yield an acceptable result, it is $\chi^2$/d.o.f.$=1.58$, and, more importantly, the fit finds that a constant $y_0\approx2$ has to be added, which is unphysical. Instead, following the idea that we are seeing three different types of early behavior, we are able to fit the distribution with three overlapping Gaussians (see Fig. \ref{Early}). We find a significantly improved fit, it is $\chi^2$/d.o.f.$=0.58$, no constant term is needed ($y_0=0$), and the three Gaussians are centered at $8.67\pm0.48$ (FWHM 2.20) mags, $12.31\pm0.09$ (FWHM 1.52) mags, and $15.11\pm1.23$ (FWHM 4.95) mags, for the ``overluminous'', ``standard'' and ``subluminous'' types, respectively.

Several caveats apply, however, and the picture is not so simple. \cite{Kann050904} discussed the possibility of different spectral slopes at early times, in application to the prompt optical emission of GRB 050904. They found that assuming achromaticity (and thus the spectral slope derived from the late-time, forward-shock dominated afterglow), the luminosity of the prompt flash was higher than in the case that spectral slopes more appropriate for early emission were considered (e.g., fast cooling phase, or injection frequency still above the optical band). Therefore, such color evolution may also apply to other afterglows in our early sample, possibly widening the clustering in one photometric band. For some GRBs, early multicolor afterglow data are available, but these yield an inconclusive picture. For example, the prompt optical flare of GRB 061121 \citep{Page061121} is more pronounced in the $V$ band (\emph{Swift} UVOT) than in unfiltered observations (ROTSE). On the other hand, the color evolution of the afterglow of GRB 061126 \citep{Perley061126} goes from redder to bluer, similar to the case of the very-well sampled early afterglow of GRB 080319B \citep{Bloom080319B, Racusin080319B, Wozniak080319B}. Several other afterglows show no early color changes at all, e.g. those of GRB 060418 and GRB 060607A \citep{Molinari060418, Nysewander060607A} and GRB 061007 \citep{Mundell061007, Schady061007}.

Furthermore, several cases in the ``cluster'' exist where a detailed study has shown additional emission components. An early reverse shock component has been proposed for GRB 050525A \citep{Klotz050525A, Shao050525A}, this is also an interpretation for the early steep decay of GRB 061126 \citep{Perley061126} and GRB 060908 \citep{Covino060908}. In the case of GRB 051111, an early steep decay is associated with the tail of the prompt emission \citep{Yost0511,Butler051111}. Once again, there are counterexamples, e.g. for GRB 060418, early upper limits on the polarization of the optical afterglow point to a weak (or even negligible) reverse shock component \citep{Mundell060418, Jin060418}, in agreement with a dominating forward shock at very early times.

\begin{figure}[!t]
\epsfig{file=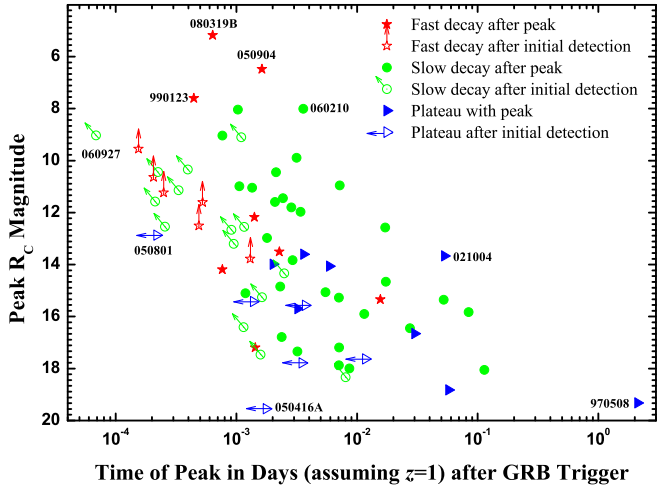,width=1\columnwidth}
\caption[]{The peak magnitudes of 70 afterglows from our samples, as derived from the extinction-corrected $z=1$ light curves (Fig. \ref{Bigfig2}). We discern between six groups: Those with early peaks followed by rapid decays (possibly of reverse-shock origin, filled red stars), those where the peak is before the earliest detection, but the early decay is also steep (empty red stars with upward-pointing arrows), those with early peaks followed by slower decays (probably of forward-shock origin, filled green discs), those where the peak is before the earliest detection, but the early decay is also slower (green rings with slanted arrows), those with early plateau phases which also show magnitude peaks (very slow rise and decay, filled blue triangles), and those where the early decay is very slow, but the peak lies before the first detection (empty blue triangles with left-pointing arrows). While there is clearly an envelope seen, the scatter is very large. Several outstanding events have been labeled. See text for more details on special cases.}
\label{PV}
\end{figure}

Recently, \cite{PanaitescuVestrand} have presented a study of early afterglow behavior, investigating different classes and finding a possible correlation between peak luminosity and peak time for afterglows with fast rises (which can be both reverse-shock flashes and forward-shock peaks; from their sample, \citealt{OatesUVOT} find the rises are consistent with forward-shock evolution), which they claim might even be used as a redshift indicator. Our large sample allows us to further study this possible correlation. In total, we find 72 afterglows (including several more from the pre-\emph{Swift} era) which have either very early detections, or show later peaks. We have gathered these afterglows in Table \ref{tabPeak}, where we give the relevant time (peak or earliest detection) and the $R_C$ magnitude in the extinction-corrected $z=1$ frame (errors are statistical only). We discern between six classes, and indicate additional noteworthy features in the comments to Table \ref{tabPeak}:
\begin{itemize}
\item{\bf Afterglow peak followed by a fast decay:} These afterglows show a fast rise to a peak, followed by a fast decay ($\alpha\approx1.5-2$), which usually becomes flatter later. This behavior is interpreted as an additional component superposed on the forward-shock afterglow, which, due to its rapid decay, quickly becomes less luminous than the forward-shock afterglow, leading to the steep-to-shallow transition. Often, this component is attributed to a reverse-shock flash, with the classical example being GRB 990123 \citep{MeszarosRees1997, MeszarosRees1999, SariPiran}. In other cases, it is probably tied to optically emissive internal shocks, that is, direct central engine activity, as for GRB 080319B\footnote{\cite{Racusin080319B} interpret the intermediately rapid decay in the early light curve of this afterglow as a reverse shock flash component that becomes dominant over the very rapidly fading prompt optical emission.} \citep[][but see \citealt{Kumar080319BB}]{Racusin080319B, Beskin080319B}, GRB 060526 \citep{Thoene060526}, GRB 061121 \citep{Page061121}, and GRB 080129 \citep{Greiner080129}, making this a diverse class. These afterglows (or, more correctly, optical transients), are the most luminous among GRB or any other phenomena \citep{Kann050904, Bloom080319B}. We find seven afterglows (10\%) in this category.
\item{\bf Initial fast decay:} These afterglows show a similar steep-to-shallow transition as described above, but the observations did not begin until after the peak, implying it must be very early. An example is GRB 090102 \citep[\citealt{Gendre090102}, see also][]{Stratta060111B}. To our knowledge, an early steep decay for GRB 090424 is reported here for the first time. This category contains six afterglows (8\%), and the combined fast decay categories make for 18\% of all afterglows, in agreement with \cite{KlotzTAROT}.\footnote{A reverse shock origin has also been implied for the very early, steeply decaying emission of GRB 060117 \citep[mentioned before,][]{Jelinek060117} and GRB 060111B \citep{Stratta060111B}, but the latter GRB, which incidentally shows evidence for very high host extinction as well, has no redshift beyond estimates and is also missing from our sample.}
\item{\bf Afterglow peak followed by a slow decay:} In these cases, after a usually fast rise and a turnover, the decay index is typical for a forward-shock afterglow with constant blastwave energy (aside from the radiative losses, and opposed to a forward shock with energy injection), and there is no further transition between different decay indices. This has been interpreted as the rise of the forward-shock afterglow at deceleration time, with classical examples being GRBs 060418 and 060607A \citep{Molinari060418, Nysewander060607A}. \cite{PanaitescuVestrand}, from afterglow modeling, also favor this explanation, with a second valid interpretation being non-uniform jets beamed off-axis with respect to the observer. Such an interpretation is favored for late peaks (if the initial Lorentz factor is also a function of angle), as in the case of GRB 080710 \citep{Kruehler080710}. Some special cases also exist, like the extreme rebrightening (following a standard forward-shock decay) of GRB 060206, which has been interpreted as an extreme energy injection event \citep{Wozniak060206, Monfardini060206}. This group contains the most afterglows, 30 (41\%).
\item{\bf Initial slow decay:} In these cases, the decay index is typical for a forward-shock dominated afterglow, and no peak is seen. All afterglows in our total sample which we do not discuss here would fit into this category, but have been detected at such late times (e.g., almost all afterglows of the pre-\emph{Swift} era) that no real conclusions can be gathered about their early behavior. Intriguingly, some afterglows with very early detections already feature a typical forward-shock decay from the first detection on. While once seen as the most typical behavior, most forward-shock dominated afterglows peak late enough that their peaks are detected (see above), putting less afterglows in this category, a total of 15 (21\%). In total, the early dominance of the forward shock is found to be the most common case, with 45 afterglows (62\%).
\item{\bf A plateau with a discernible peak magnitude:} In these cases, a rising-to-decaying transition is seen as well, but the rise and initial decay are very shallow, leading to a plateau phase where the afterglow luminosity barely changes over long times. Such a behavior has often been seen in connection with the spectrally soft X-Ray flashes\footnote{Note that some X-ray flashes show a normal, forward-shock-like decay from very early on, e.g., XRF 050406 \citep{Schady050406} and XRF 050824 \citep{Sollerman050824}.}, and may indicate a jet viewed off-axis (e.g., XRF 080310 and XRF 080330, \citealt{Guidorzi080330} analyze the latter in detail). Several special cases are included in this category, like GRB 021004, which is dominated by multiple energy injections at early times \citep{deUgarte021004}, and the highly peculiar afterglow of GRB 970508, which begins with a very faint plateau followed by a very late, strong rebrightening. This category contains nine afterglows (12\%).
\item{\bf An early very shallow decay:} Here, the afterglow decays from the first observation onwards, but the decay index is very shallow, less than is expected from a classical forward shock ($\alpha\approx0-0.4$), creating a plateau phase. A classical example of such behavior is GRB 050801 \citep{Rykoff050801, DePasquale050801}, and it has also been seen in the highest redshift GRB 090423 \citep{Tanvir090423}. These very slow decays are quite rare, with only six afterglows in the sample showing them (8\%).
\end{itemize}

We plot all 73 data points (GRB 060729 has resulted in two measurements) in Fig. \ref{PV}, discerning between the six classes. Clearly, an envelope is seen which traces the correlation found by \cite{PanaitescuVestrand}, but the scatter is much larger than what they find in their small sample, indicating that the significance of the correlation is much smaller than assumed. Applying rank correlation tests, we find Kendall's $\tau=0.43$, and Spearman's $\rho=0.62$, indicating the existence of a correlation with only moderate significance. A similar result was found by \cite{KlotzTAROT} using observations of the TAROT robotic telescopes. In magnitudes, the width of the scatter is around ten magnitudes even if we only choose those afterglows which exhibit a fast rise (and fast or slow decay). Intriguingly, those afterglows which are already decaying at first detection cluster more strongly than those with detected peaks (especially those with rapid decays), on the left hand side of the correlation, indicating an extension in this direction and even larger scatter, were the optical follow-up to be even more rapid. All afterglows not included in our sample of 72 would be found in the lower right hand corner, usually beyond 0.1 days and fainter than 16th magnitude.

\cite{OatesUVOT} find a statistically significant correlation between the early decay index and the magnitude of the afterglow, with bright afterglows decaying more rapidly. This hints both at possible additional early components in the afterglow, as well as pointing to the late clustering (a fast-decaying, bright afterglow will be at similar magnitude at one day as a fainter afterglow that decays more slowly, though note the transition time plays an important role as well). Further studies of the early diversity, especially the optically subluminous cases, which must decay slowly or even rebrighten to fit into the late clustering (and are therefore usually found in the two plateau groups mentioned above), are beyond the reach of this work.

\subsubsection{Existence of a Correlation between Optical Luminosity and Isotropic Energy?}
\label{Berger}

\begin{figure}[!t]
\epsfig{file=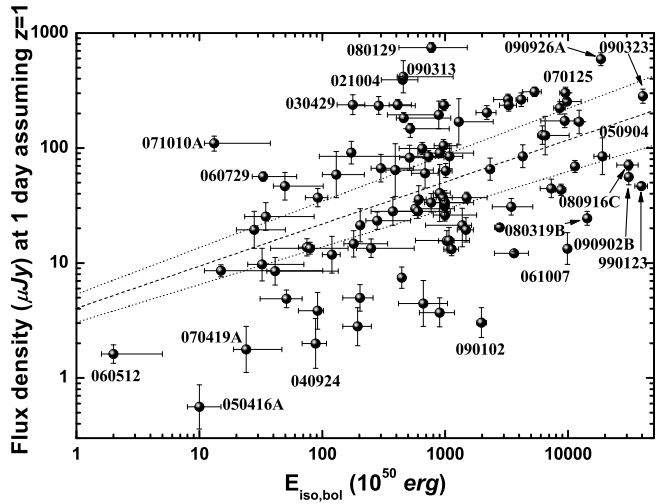,width=1\columnwidth}
\caption[]{The flux density in the $R_C$ band at one day (in the host frame assuming $z=1$) plotted against the bolometric isotropic energy of the prompt emission for all GRBs in the optically selected sample (except GRB 991208, which was only discovered after several days, and GRBs 060210, 060607A, 060906 and 080319C, where the follow-up does not extend to one day). While no tight correlation is visible, there is a trend of increasing optical luminosity with increasing prompt energy release. This is confirmed by a linear fit (in log-log space), using a Monte Carlo analysis to account for the asymmetric errors. The dashed line shows the best fit, while the dotted line marks the $3\sigma$ error region. Several special GRBs are marked.}
\label{Amati}
\end{figure}

In Fig. \ref{Amati} we show the flux density in the $R_C$ band at one day in the host frame assuming $z=1$ (Fig. \ref{Bigfig2}, Table \ref{tabMB}) plotted against the bolometric isotropic energy of the prompt emission (Table \ref{tabTypeIISample}). This plot is similar to that of \citet[][see also \citealt{FW2001, LZ2006, Amati060614, Kaneko2007, Gehrels2008}, and \citealt{NakarReview, Berger2007} for Type I GRBs]{Kouveliotou2004}, who used the X-ray luminosity at 10 hours \citep[for a detailed discussion, see][]{Granot2006, FanPiran}, as well as \cite{Nysewander2009}, who also studied the $R-$band luminosity (as well as the X-ray luminosity) at 11 hours. Similar to the correlations found by the aforementioned authors, a trend is visible in Fig. \ref{Amati}: The optical luminosity increases with increasing prompt energy release. But the scatter is very large, especially in contrast to the often very well constrained flux densities (i.e., the offset from the best fit in units of the individual flux density errors $\sigma_F$ is much larger than one in many cases, $\frac{|F-F_{Fit}|}{\sigma_F}\gg1$, with $F_{Fit}$ being the flux density expected from the correlation). This can be clearly seen both in flux density and in isotropic energy. GRB 061007 and GRB 070125 have almost identical isotropic energy releases, but the flux densities of their optical afterglows differ by a factor of $23^{+12}_{-8}$. The span between GRB 080129 and GRB 050502A is even larger, over two orders of magnitude. GRB 990123 has an isotropic energy release roughly 1000 times higher than GRB 060729, but its optical afterglow has a slightly fainter luminosity at one day. The trend is almost non-existent except for three faint bursts: XRF 060512, XRF 050416A and GRB 070419A have been mentioned in \kref{Bimodal}, and here we see that these events are also subenergetic in their prompt emission. The faintest optical afterglow of the K06 sample, GRB 040924, is seen to be among the least energetic GRBs too, but it is still part of the ``cloud''. In log-log space, we use a linear fit, accounting for the asymmetric error bars with a Monte Carlo simulation. In 30000 runs, we find the following correlation:

\begin{eqnarray}
\frac{F_{\rm opt}\;({\rm at}\; t=1\; {\rm day})}{1 \mu\rm Jy} = \qquad\qquad\qquad\qquad\qquad\qquad\nonumber\\
10^{(0.607\pm0.041)}\times \left(\frac{E_{iso,bol}}{10^{50}\;erg}\right)^{(0.366\pm0.013)}.
\end{eqnarray}

Using an unweighted fit, we find exactly the same slope and a slightly smaller (though identical within $1\sigma$ errors) normalization, indicating that the intrinsic scatter dominates over the errors of the data points. Using Kendall's rank correlation coefficient $\tau$, we check the significance of the correlation. We find $\tau=0.29$ (significance $4.1\sigma$) for the complete data set. Therefore, the correlation is only of low significance. As would be expected, removing the three subenergetic events reduces the significance even more, it is $\tau=0.24$ (significance $3.3\sigma$). We conservatively estimate the errors on $\tau$ by creating maximally tight and maximally scattered data sets. In the first case, we shift data beneath the best fit closer by $-\delta E_{iso,bol}$ and $+\delta F_{\rm opt}$, and data above the best fit closer by $+\delta E_{iso,bol}$ and $-\delta F_{\rm opt}$. In the latter case, the data are shifted away from the correlation in the reverse way. For the maximally tight data set, we find $\tau=0.44$ (significance $6.1\sigma$) for the complete data set and $\tau=0.40$ (significance $5.5\sigma$) when we remove the three subenergetic events. For the maximally scattered data set, the values are $\tau=0.18$ (significance $2.6\sigma$) and $\tau=0.13$ (significance $1.8\sigma$), respectively.

\cite{NakarReview} and \cite{Berger2007} argue that as the cooling frequency is usually beneath the X-ray range \cite[but see][who find that 30\% (9 of 31) of the X-ray afterglows they studied to still have $\nu_c>\nu_X$ at up to ten hours after the GRB]{Zhang2007}, the X-ray luminosity is independent of the circumburst density and it thus represents an acceptable proxy for the kinetic energy, $L_X\propto\epsilon_eE_K$ (with $\epsilon_e$ being the fraction of energy in relativistic electrons). Clearly this is not the case here, as the cooling break lies above the optical bands in most cases \citep[e.g, K06;][]{Panaitescu2006a, Panaitescu2006b, StarlingPaperI, Schady2007, Curran_p}. Therefore, the strong spread in optical luminosities may be explained by the effect of the spread in the circumburst density, which, while typically lying at $1 - 10\;\rm cm^{-3}$ \cite[cf.][]{FB2005}, can reach several hundred cm$^{-3}$, e.g. in the case of GRB 050904 \citep{Frail050904}. Still, the existence of this trend is intriguing, and further observations will hopefully reveal more subluminous GRBs. One extension possibility is to do a similar analysis for Type I GRBs, these results are presented in the companion Paper II. Furthermore, using measured jet break times, one could correct for the jet collimation, and step from an ``Amati-like'' (using $E_{\rm iso}$) to a ``Ghirlanda-like'' (using $E_\Gamma$) plot, albeit with less events.

\subsubsection{A New Population of Low-Luminosity GRBs at Low Redshifts?}

There is clear evidence for one Type II sub-population that probably extends the $L_{\rm opt}- E_{\rm iso}$ correlation to significantly lower energy values. These are the so-called low-luminosity SN bursts, GRB 980425, GRB 031203 and XRF 060218 \citep{Pian060218, Soderberg060218, Liang2007a, Guetta2007}\footnote{The recently discovered XRF 100316D \citep{Chornock100316D, Starling100316D} does not yet have enough analyses published to be further included here.}. In all these three cases, while luminous, basically unreddened SN emission was detected, there were no or only marginal indications of a ``classical'' optical afterglow \citep[e.g.,][]{Galama1998, Malesani2004, Campana060218, Pian060218, Ferrero060218, Cobb060218, Mirabal060218, Modjaz060218, Sollerman060218, Kocevski060218}. On the other hand, the prompt emission energy release of these GRBs is orders of magnitude beneath typical Type II events and thus, they cannot be readily compared with each other. The SN emission and, in the case of GRB 031203, the bright host galaxy \citep{Prochaska031203, Mazzali031203, Margutti031203} prevent us from setting definite limits on afterglow emission, thus, they can not be included in our study.

Recently, systematic photometric and spectroscopic observations of GRB host galaxies\footnote{We note that, as many of these events were optically dark, their host galaxies are selected via \emph{Swift} X-ray error boxes. This presents problems similar to Type I GRBs, which up to now have host-galaxy determined redshifts exclusively (see Paper II). Indeed, several host galaxy candidates have been ruled out after a refined X-ray analysis \citep[e.g.][]{PerleyHosts2009}. Also, there is the increased possibility of a chance superposition \citep{CobbBailyn2007, CampisiLi}.} (see, e.g., \citealt{JakobssonVLTLP} for a short introduction to the VLT survey, and \citealt{PerleyHosts2009} for first results from the Keck survey) have started to reveal a population of GRBs that are intermediate in luminosity, both in terms of prompt emission and afterglow, lying between most of the GRBs in our optically selected sample and the local universe low-luminosity events mentioned above. These GRBs are defined by low fluence, usually soft spectra (several are XRFs), usually a simple prompt emission light curve, faint or non-existent afterglows and low redshifts ($z\lesssim1$)\footnote{Note that our sample only overlaps in XRF 060218 with the \emph{Swift} long-lag sample presented by \cite{XiaoSchaefer2009}, which they find is not associated with the Local Supercluster (whereas several of our optically selected GRBs, namely GRB 051111, GRB 060502A and GRB 060607A, are in the sample of those authors). Therefore, a long spectral lag is not a common feature of the sample we present here.}. Recently, the existence of this population was inferred theoretically by comparing the distribution of measured redshifts with what is expected if the GRB rate follows the star formation history of the universe \citep{Coward2007}. Several examples are included in our sample (XRF 050416A, XRF 060512, GRB 070419A) and have been mentioned above, although these still have afterglows that are relatively bright observationally. Another example is XRF 050824, although this event has an even brighter optical afterglow. We have searched the literature for further examples of these low-redshift events. Similar to our main Type II sample, we compiled their energetics, which are presented in Table \ref{tabTypeIIlowzSample}. This contains ten events from the \emph{Swift} era and three pre-\emph{Swift} events. Next to GRB 980425 and GRB 031203 we have added XRF 020903. The latter burst did have a faint afterglow and showed a spectroscopic (albeit of low significance) and photometric SN signature \citep{Soderberg020903A, Soderberg020903B, Bersier020903}. Due to limited publicly available photometry (or no afterglow detection at all), these GRBs can not be included in our main sample either.

\citet[][see also \citealt{Kistler2007}]{Fiore2007} speculate that many bright \emph{Swift} GRBs without optical afterglows could be low-$z$, dust obscured events. Such events clearly exist, GRB 051022 was mentioned above, and other examples are GRB 060814A and GRB 070508. Both were very bright GRBs with bright X-ray afterglows, but they were optically dark (060814A) or faint (070508) and their afterglows were discovered only in the NIR \citep[060814A,][]{LevanGCN060814} or were very red \citep[070508,][]{PiranomonteGCN070508}. GRB 060814A lies at $z=0.84$ \citep{ThoeneGCN060814}, GRB 070508 (probably) at $z=0.82$ \citep[][but see \citealt{FynboSpectra}]{JakobssonGCN070508}. But the GRBs listed in Table \ref{tabTypeIIlowzSample} are clearly a different population, although here too, evidence for dust obscuration does exist, e.g. GRB 060202 (at $z=0.783$), which had a bright X-ray afterglow, was dark even in the $K$ band \citep{WangGCN060202}, while the X-ray faint GRB 050223 was situated in a dusty, red galaxy \citep{Pellizza050223}. Intrinsic faintness or dust obscuration is undecided in the other cases, but we point out again that this population has low-luminosity prompt emission. This population may partly be responsible for an excess of dark bursts at low gamma-ray peak fluxes, reported by \cite{DaiLF}.

\begin{figure}[!t]
\epsfig{file=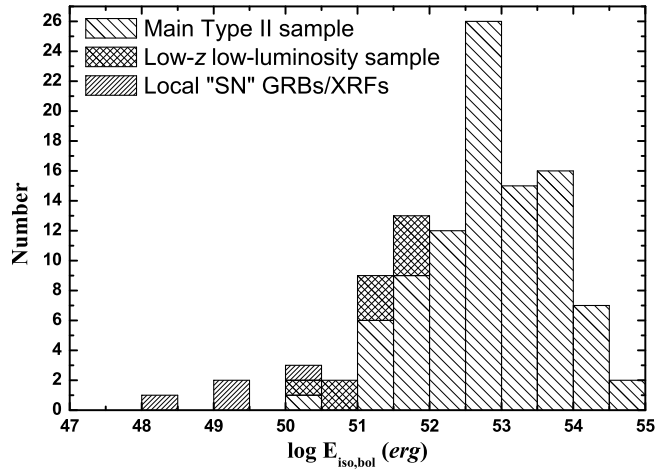,width=1\columnwidth}
\caption[]{The distribution of bolometric isotropic energies for all the GRBs of the ``optically selected'' main sample of this paper (Table \ref{tabTypeIISample}) and the low-$z$ low-luminosity events (Table \ref{tabTypeIIlowzSample}) for which we have no optical afterglow information. Here, we also differentiate between the four local ``SN'' GRBs/XRFs and the new sample which is being uncovered mostly by host galaxy observations. These form an intermediate population between the optically selected sample and the local events with spectroscopic SN signatures.}
\label{lowz}
\end{figure}

In Fig. \ref{lowz} we show the distribution of the bolometric isotropic energies for the ``optically selected'' main sample of this paper (which also includes the K06 bursts), the four SN GRBs/XRFs and the low-$z$ low-luminosity sample. All in all, the isotropic energy releases are distributed over six orders of magnitude without gaps, with only GRB 980425 being over one order of magnitude less energetic than the next faintest event (XRF 020903). Clearly, the three samples are quite distinct from each other. The mean (logarithmic) bolometric isotropic energy of the optically selected sample is $\overline{E_{\rm iso,bol}}=52.88\pm0.09$, it is $\overline{E_{\rm iso,bol}}=51.30\pm0.16$ for the low-$z$ sample and $\overline{E_{\rm iso,bol}}=49.22\pm0.43$ for the ``SN'' sample. A KS-test finds that the optically selected GRBs and the low-$z$ GRBs do not stem from the same sample with high significance ($P=2.7\,\times\,10^{-6}$). This low probability is hardly surprising, as the low-$z$ sample was selected according to the criteria of low fluence and low redshift, necessarily implying a low isotropic energy release. We note that the five faintest GRBs (in terms of prompt energy release) in the optical sample all lie at $z<1$. Next to XRF 060512, XRF 050416A, GRB 070419A and XRF 050824 (all mentioned above) this includes XRF 071010A. The latter event, though, has an observationally bright optical afterglow. If we move the first four into the low-$z$ sample, we find $\overline{E_{\rm iso,bol}}=52.96\pm0.08$ for the rest of the optically selected sample and $\overline{E_{\rm iso,bol}}=51.21\pm0.14$ for the larger low-$z$ sample, and the difference becomes even more significant ($P=7.4\,\times\,10^{-9}$). On the other hand, the almost continuous distribution of energy releases may indicate that all events are part of a single population that can be described by a single power-law luminosity function. A deeper analysis is beyond the scope of this paper, but it would require an estimation of the true rate of these events, similar to what has been done for the local universe SN events \citep{Pian060218, Soderberg060218, Liang2007a, Guetta2007, Virgili2009}.

As pointed out above, most of these low-$z$ events are only now being identified as such due to host galaxy spectroscopy campaigns. These are, of course, biased to low redshift events, both due to host galaxies being observationally brighter and due to effects like the ``redshift desert'' when the {\OII} line moves into the airglow region of the spectrum at $z\gtrsim1$. Still, it is intriguing that many of the GRBs with new redshift information are not the bright, dust-enshrouded events \cite{Fiore2007} predicted, but a population that falls beneath the ``standard energy reservoir'' as already pointed out by \cite{Kocevski061028}. These events, being optically dim or even dark, are only observable due to the X-ray localization capabilities of \emph{Swift} (with the flux sensitivity playing a lesser role) which allows discovery of the host galaxy in many cases (although we caution again that the significance some the associations may be questionable). With more host galaxy observation results likely to be published in the future, it is expected that this sample will continue to grow. As most of these redshifts were not found until months after the event, searches for SN signatures have not been undertaken, so it is as yet unclear if these events are also subluminous (or perhaps superluminous) in terms of their SN explosions (assuming that they truly are related to the deaths of massive stars). But the clear association of events that are even fainter with powerful broad-lined Type Ic SNe indicates that the basic collapsar mechanism will probably also underlie this new population. Future observatories, such as \emph{SVOM} and especially \emph{EXIST}, are predicted to yield much higher detection rates of these subluminous GRBs \citep{ImeritoEXIST}, though we point out that \emph{Fermi} GBM, despite its wide FOV, is not well equipped for this task, as the low localization precision precludes a high identification rate.

\section{Summary and Conclusions}

We have compiled a large amount of optical/NIR photometry of \emph{Swift}-era GRB afterglows, creating a total sample of 76 GRBs (as well as three more pre-\emph{Swift} events), considering events up to the end of September 2009. Following \cite{ZKK} and K06, we analyzed the light curves and spectral energy distributions. We also collected data on the energetics of the GRBs. We used this sample to compare the \emph{Swift}-era afterglows to the pre-\emph{Swift} sample taken from K06, and looked for correlations between the optical afterglow luminosity and parameters of the prompt emission. To summarize, we come to the following results.

\begin{itemize}
\item As has been found before, observed optical afterglows in the \emph{Swift} era are typically fainter than those of the pre-\emph{Swift} era. The rapid localization and follow-up capabilities available today give us access to this fainter population.
\item In terms of luminosity, we find no statistically significant difference between the pre-\emph{Swift} and the \emph{Swift}-era afterglows, the relative faintness of the \emph{Swift}-era afterglows can typically be attributed to a larger mean redshift. But we caution that several selection biases still apply.
\item We still find (see also K06) that SMC-like dust is usually preferred and that the line-of-sight extinctions through the GRB host galaxies are usually low. Still, at least one clear case (GRB 070802) of high $\AV$ exists.
\item The trend seen in K06 of lower extinction at higher redshifts is confirmed in our new sample, which increases the number of $z>3$ GRBs from one to 17. We caution though, that the correlation is only weak, and it is still unclear if this is due to a true evolution or to a selection bias.
\item The clustering of optical afterglow luminosities at one day reported by K06, \cite{LZ2006} and \cite{Nardini2006} is found to be less significant than before, indeed, the spread of magnitudes actually increases in luminosity space due to the discovery of exceptionally over- as well as underluminous events. As the \emph{Swift} sample is less biased than earlier samples, this indicates that the clustering found in pre-\emph{Swift} data may be the result of selection effects only. The bimodal distribution when splitting the afterglows into two redshift bins is confirmed though, but the total sample itself is not found to be bimodally distributed anymore, in agreement with several other recent results.
\item Our samples contain all GRBs which have been observed with high-resolution echelle spectroscopy (and have had these results published beyond GCN circulars). \cite{ProchterMgII} found a high incidence of strong foreground absorption systems in comparison to a QSO sightline sample. One possible explanation, that these foreground galaxies are gravitationally lensing the afterglow (making them brighter and consequently more accessible to echelle spectroscopy) would imply that the echelle sample is significantly more luminous than other afterglows. We find no evidence for such an increased luminosity, both in comparison between the echelle-observed afterglows and those with low-res spectroscopy only, as well as between those with and without strong foreground absorption systems within the echelle sample. We caution, though, that we are also not able to rule out a low amplification factor.
\item We find that an upper boundary on the optical luminosity of a forward-shock driven afterglow seems to exist.
\item At very early times, the apparent magnitude spread is much larger than at later times but, intriguingly, half the afterglows strongly cluster within three magnitudes. Basically, there seem to be three classes: optical afterglows with additional early emission components, afterglows dominated by the constant-blastwave-energy forward shock already at early times, and optically faint afterglows that show plateau phases or later rebrightenings (possibly due to energy injections into the forward shock or off-axis viewing geometry). The forward-shock dominated afterglows make up 60\% of the sample that had early detections (or late, definite peaks), and the afterglows with additional emission components, which are the most luminous ones, are also the most rare. While there is a trend between the peak time and the peak luminosity of afterglows with fast initial rises, a strong correlation \citep{PanaitescuVestrand} is not observed in our larger sample.
\item A trend is visible between the isotropic energy release in gamma-rays and the optical luminosity at a fixed late time in the rest frame. The scatter is large, probably due to circumburst density variations, but low-luminosity events support the reality of this trend.
\item We propose the existence of a population of low-redshift low-luminosity events that bridge the gap between the Type II GRBs in our main sample (selected due to their optical afterglows) and the ``SN'' GRBs/XRFs that have been detected in the local universe. These events are optically dim or dark and are being revealed mostly by systematic host galaxy observations which are able to determine that their redshifts are low.
\end{itemize}

At the time the results of \cite{ZKK} and K06 were published, the pre-\emph{Swift} era was a closed chapter and a clear overview of the properties of pre-\emph{Swift} GRB afterglow could be given. The \emph{Swift} era, on the other hand, continues, and with time many more results, larger samples and probably more surprises await us. Therefore, this work and its companion paper (Paper II) can only give a first overview, by nature incomplete, of the less biased sample \emph{Swift} is delivering. And the studies of the properties of dust and the chemical evolution of galaxies in the reionization era, thanks to the high rate of GRBs in the early universe \citep{YukselSFR, Kistler090423, Salvaterra090423, Tanvir090423}, has only just begun, and awaits the development of more powerful instruments to study GRB afterglows in the NIR/MIR wavelength regions.

\acknowledgments
We thank the anonymous referee for helpful comments that improved this paper.
D.A.K. thanks C. Guidorzi for helpful comments as well as the GRB 061007 calibration, D. A. Perley for ``better-late-than-never'' comments, and A. Zeh for the fitting scripts. D.A.K., S.K. and P.F. acknowledge financial support by DFG grant Kl 766/13-2. B.Z. acknowledges NASA NNG 05GC22G and NNG06GH62G for support. The research activity of J.G. is supported by Spanish research programs ESP2005-07714-C03-03 and AYA2004-01515. D.M. thanks the Instrument Center for Danish Astrophysics for support. The Dark Cosmology Centre is funded by the Danish National Science Fundation. A.U. acknowledges travel grant Sigma Xi Grant G2007101421517916. We are grateful to K. Antoniuk (CrAO) for observation of the GRB 060927. A.P. acknowledges the CRDF grant RP1-2394-MO-02 which supported observations in CrAO in 2003-2005. S.S. acknowledges support by a Grant of Excellence from the Icelandic Research Fund. I.B., R.B., I.K. and A.G. give thanks to TUBITAK, IKI and KSU for partial support in using RTT150 (Russian-Turkish 1.5-m telescope in Antalya) with project number 998,999. I.B. and A.G. are grateful for partial support by grants ``NSh-4224.2008.2'' and ``RFBR-09-02-97013-p-povolzh'e-a''. M.I. acknowledges the support from the Creative Research Initiative program, No. 2010-0000712,  of the Korea Science and Engineering Foundation (KOSEF) funded by the Korea government (MEST). Furthermore, we wish to thank Scott Barthelmy, NASA, for the upkeep of the GCN Circulars, Jochen Greiner, Garching, for the ``GRB Big List'', Robert Quimby et al. for GRBlog and D. A. Perley et al. for GRBOX. This work made use of data supplied by the UK Swift Science Data Centre at the University of Leicester.

\newpage\clearpage
\appendix
\label{App}

\section{Observations}
\label{Observations}
\begin{itemize}
\item{{\bf GRB 040924}: Observations were obtained with the Small and Moderate Aperture Research Telescope System (SMARTS) 1.3m telescope, equipped with the ANDICAM (A Novel Double-Imaging CAMera) detector (only upper limits could be obtained). Detections in $R_C$ were obtained with the 1.5m Russian-Turkish Telescope (RTT150) in Turkey. Calibration was done against Landolt fields.}
\item{{\bf GRB 041006}: Observations were calibrated against comparison stars obtained by Arne Henden\footnote{The calibration files of Arne Henden have been downloaded at ftp://ftp.aavso.org/public/calib/grb. The files have the names ``grb******.dat'', where ****** is the date of the GRB, e.g., ``grb041006.dat''}. Data were obtained by the 1.5m AZT-22 telescope of the Maidanak observatory in Uzbekistan, equipped with the CCD direct camera, the 0.64m AT-64 telescope of the Crimean Astrophysical Observatory, Ukraine, the 1.34m Schmidt telescope of the Th\"uringer Landessternwarte Tautenburg (Thuringia State Observatory), Germany, and the RTT150. The results of AT-64 telescope (CrAO) observations were originally presented in \cite{PozanenkoCRAO}.}
\item{{\bf GRB 050315}: Observations were obtained by the SMARTS 1.3m telescope. Imaging was obtained in all possible filters ($BVR_CI_CJHK$), but the afterglow was only detected in $R_C$ and $I_C$. The observations were calibrated against Landolt fields.}
\item{{\bf GRB 050319}: We obtained late and deep observations of this afterglow with multiple telescopes. The 2.56m Nordic Optical Telescope, NOT, equipped with the Andalucia Faint Object Spectrograph and Camera (ALFOSC), the Standby CCD Camera (StanCam) and the NOTCam near-IR camera/spectrograph; the 3.6m Telescope Nazionale Galileo (TNG) equipped with the Low Resolution Spectrograph (LRS) (both telescopes on La Palma in the Canary Islands, Spain); the 1.5m Maidanak and the RTT150. Photometry was obtained using a Henden calibration in the optical. In the $K$ band, we found only a single object (a probable elliptical galaxy) in the small NOTcam FOV which was also significantly detected in the 2MASS $K_S$ band image, but it is too faint to be in the 2MASS catalog. We obtained a zero point of the 2MASS image (standard deviation 0.01 magnitudes) from multiple bright 2MASS sources, measured the single source to be $K_S=16.07\pm0.18$, and used it to derive the zero point of the NOTcam image and determine the afterglow magnitude (adding the statistical errors of the 2MASS and NOTcam magnitudes in quadrature), which is in full agreement with what is expected from the optical afterglow colors.}
\item{{\bf GRB 050401}: Observations (only upper limits could be obtained) were obtained with the SMARTS 1.3m telescope. Calibration was done against Landolt fields.}
\item{{\bf GRB 050408}: Observations were obtained at multiple epochs and in five colors ($UBVR_CI_C$) with the TNG, and a single observation each with the CrAO AT-64 and the Maidanak 1.5m. TNG observations were calibrated against converted magnitudes from the SDSS, while the AT-64 and Maidanak data points used the calibration by Arne Henden.}
\item{{\bf XRF 050416A}: Observations were obtained with the Maidanak 1.5m telescope in $BR_C$, and have been calibrated against a Henden calibration. Further observations, yielding detections in $I_C$ but only upper limits in $J$, were obtained with the SMARTS 1.3m, including a late detection of the host galaxy. They were calibrated against Landolt standards.}
\item{{\bf GRB 050502A}: Observations were obtained with the 0.8m IAC80 telescope at Observatorio del Teide, Canary Islands, Spain, the Maidanak 1.5m (shallow upper limit only due to clouds), as well as the 2.5m Isaac Newton Telescope equipped with the Wide Field Camera (WFC) on La Palma. Photometry was performed against a Henden calibration.}
\item{{\bf GRB 050525A}: Observations were obtained with the Southeastern Association for Research in Astronomy (SARA) 0.9m telescope at Kitt Peak National Observatory (KPNO), Arizona, USA, the 0.4m telescope of Ussuriysk Astrophysical Observatory (UAPhO) in the far east of Russia, the CrAO AT-64, the Maidanak 1.5m, the SMARTS 1.3m and the RTT150. Observations were calibrated against Landolt standards (RTT150) and against a calibration by Arne Henden. We detect the afterglow in SMARTS, SARA and RTT150 observations, and obtain upper limits otherwise, with a very deep upper limit from the Maidanak telescope. Some of the results of Maidanak 1.5m and CrAO observations were previously presented in \cite{PozanenkoCRAO}.}
\item{{\bf GRB 050730}: Observations were obtained by the SMARTS 1.3m \citep[only upper limits, detections are presented in][]{Pandey050730} and the RTT150, from which we obtained detections in the $R_C$ and $I_C$ bands. Observations were calibrated against Landolt standards.}
\item{{\bf GRB 050801}: Observations were obtained by the Danish 1.54m telescope at La Silla Observatory, Chile, equipped with the Danish Faint Object Spectrograph and Camera (DFOSC), and with the SMARTS 1.3m. Calibrations in $BVI_C$ were done using five stars each from \cite{OvaldsenHosts}, calibration in $R_C$ was done against Landolt field stars observed near the GRB time with the D1.54m, and calibration in $JK$ was done against five 2MASS sources.}
\item{{\bf GRB 050802}: We obtained a large data set on this GRB, the only (beyond the GCN) ground-based observations published so far. Observations were obtained with the 1.5m telescope of the Sierra Nevada Observatory (OSN), Spain, the 2.6m Shajn telescope of the Crimean Astrophysical Observatory, the NOT, the TNG, and the Maidanak 1.5m. Photometry was performed against a Henden calibration.}
\item{{\bf GRB 050820A}: Optical observations were obtained with the UAPhO 0.4m, the Maidanak 1.5m, the Shajn 2.6m and the RTT150. NIR observations were obtained with the TNG and the NICS (Near-Infrared Camera and Spectrometer) detector as well as with the 3.8m United Kingdom Infra-Red Telescope (UKIRT) on Mauna Kea, Hawaii, USA and the University of Wyoming's 2.3m Wyoming Infrared Observatory telescope. Optical data were calibrated against a Henden calibration, while the NIR data were calibrated against 2MASS. WIRO data were strongly affected by weather and detector systematics, while the statistical error are small ($\leq0.06$ magnitudes), we added an 0.1 magnitude error in quadrature to account for this; still, some scatter remains. Special care was taken with the late, deep Shajn observation, which was affected strongly by bad seeing, where we used the PSF of a nearby, bright, unsaturated star as a model to subtract the two nearby stars. Still, residuals remain and no significant source could be found at the afterglow position. We place a conservative upper limit of $R_C>23$ on the afterglow magnitude, which is not constraining.}
\item{{\bf GRB 050908}: A single data point was obtained with the Maidanak 1.5m. Photometry was performed against the USNO A2.0 catalog.}
\item{{\bf GRB 050922C}: We obtained a very large data set, using the Zeiss-600 (0.6m) telescope of the Mt. Terskol observatory, Kabardino-Balkarija, Russia, the 1.3 m McGraw-Hill Telescope at the MDM Observatory, part of KPNO, the 8.2m European Southern Observatory Very Large Telescope (ESO VLT) at Paranal Observatory, Chile, the 4.2m William Herschel Telescope (WHT) on La Palma, the NOT, the D1.54m, and the INT. Data were calibrated using a Henden calibration.}
\item{{\bf GRB 051109A}: A single data point was obtained with the 0.38m K-380 telescope of CrAO. Photometry was performed against the USNO A2.0 catalog, and compares well with the data of \cite{Yost0511}.}
\item{{\bf GRB 051111}: We obtained data with the SARA and Maidanak 1.5m telescopes, and calibrated against the USNOB1.0 catalog. They agree well with other published data.}
\item{{\bf GRB 060206}: We obtained data with the SARA and Maidanak 1.5m telescopes, and calibrated against the USNOB1.0 catalog. They agree well with other published data.}
\item{{\bf GRB 060418}: Observations were obtained by the SMARTS 1.3m and the Maidanak 1.5m telescopes, yielding detections in seven filters ($BVR_CI_CJHK$). SMARTS observations were calibrated with Landolt stars, while for the Maidanak observations, a Henden calibration was used.}
\item{{\bf XRF 060512}: Observations were obtained with the NOT and NOTcam. They were calibrated against the single 2MASS star in the small FOV.}
\item{{\bf GRB 060607A}: Observations were obtained with the SMARTS 1.3m, yielding detections in all seven filters. Photometry was performed against Landolt standards.}
\item{{\bf GRB 060714}: Observations were obtained with the SMARTS 1.3m. Photometry was performed against 22 USNO-B1.0 stars in the $I_C$ band and six 2MASS stars in the $J$ band. Note that the $I_C$ magnitude is too faint compared with what is expected from the $R_C$ band, this may be due to a systematic offset in the USNO catalog.}
\item{{\bf GRB 060729}: Observations were obtained with the SMARTS 1.3m. Photometry was performed against Landolt standards.}
\item{{\bf GRB 060904B}: Observations were obtained with the RTT150, SNUCAM \citep{ImSNU} on the Maidanak 1.5m and the SMARTS 1.3m. The RTT150 observations were performed starting just eight minutes after the GRB, during dawn twilight, leading to large uncertainties especially in the $B$ and $V$ filters. The observations were calibrated against Landolt stars (RTT150, SMARTS) and the USNO A2.0 catalog, the latter agrees well.}
\item{{\bf GRB 060908}: A single observation was obtained with the Zeiss-2000 (2.0m) telescope of the Mt. Terskol observatory, under very bad seeing conditions, leading to a marginal detection which nonetheless agrees well with other data. Two observations obtained with the Maidanak telescope yielded upper limits only. We also present one UKIRT $K$-band detection which is not included in \cite{Covino060908}. Images were calibrated against comparison stars provided by S. Covino, and UKIRT data against 2MASS stars.}
\item{{\bf GRB 060927}: A single observation (upper limit only) was obtained with the 1.25m AZT-11 telescope of the Crimean Astrophysical Observatory. The observation were calibrated against the USNO A2.0 catalog.}
\item{{\bf GRB 061007}: Observations were obtained with the SMARTS 1.3m. They were calibrated against comparison stars taken from \cite{Mundell061007} in the optical and against 2MASS stars in the NIR.}
\item{{\bf GRB 061121}: Observations were obtained with the SMARTS 1.3m and the Shajn 2.6m telescope. SMARTS observations were calibrated against Landolt standards, while Shajn observations were calibrated against a set of stars provided by D.M.}
\item{{\bf GRB 070125}: Observations were obtained with the SMARTS 1.3m. In $BVRI$, we calibrated the afterglow against eight standard stars from \cite{Updike070125}, in $JHK$, we used four 2MASS stars.}
\item{{\bf GRB 070208}: A single data point was obtained with the AZT-33IK (1.5m) telescope of the Sayan Observatory, Mondy, Siberia, Russia. The observation was calibrated against the SDSS catalog and agrees well with other data. A nearby galaxy was removed via mask-subtraction. Analysis of the photometric redshift of this galaxy (from SDSS data) obtained with HyperZ code \citep{HyperZ} reveals two minima, at $z=0.52$ ($\chi^2_\nu=0.47$) and $z=0.07$ ($\chi^2_\nu=0.98$), indicating that this is clearly not the host galaxy but a foreground absorber. It did not fall in the slit during the spectroscopy reported by \cite{CucchiaraGCN070208}, and no absorption lines from this system are reported either.}
\item{{\bf GRB 070419A}: Two observations were obtained with the SARA 0.9m telescope, resulting in a shallow limit and a marginal detection, as well as an upper limit from the Sayan telescope. Data were calibrated against the USNOB1.0 catalog.}
\item{{\bf GRB 071020}: Observations were obtained with the SARA 0.9m, the 0.7m AZT-8 telescope of the Crimean Astrophysical Observatory, and the 0.6m Zeiss-600 telescope of the Maidanak observatory. The observations were calibrated against the SDSS.}
\item{{\bf XRF 071031}: Observations were obtained with the SMARTS 1.3m. They were calibrated against Landolt standards, and in some cases shifted slightly in zero-point to bring them into agreement with the data set of \cite{Kruehler071031}.}
\item{{\bf GRB 080129}: Observations were obtained with the SMARTS 1.3m. This GRB was highly extinct, and no detections were achieved. The observations were calibrated with Landolt standards.}
\item{{\bf GRB 080210}: A single data point was obtained with the Sayan 1.5m telescope. The observation was calibrated against the SDSS catalog. It implies a possible rebrightening in the light curve.}
\item{{\bf XRF 080310}: Observations were obtained with the SMARTS 1.3m. They were calibrated against Landolt standards.}
\item{{\bf XRF 080330}: Observations were obtained with the SMARTS 1.3m and the Terskol 2m. They were calibrated against Landolt standards (SMARTS) and SDSS standards transformed to the Johnson-Cousins $R_C$ band (Terskol).}
\item{{\bf GRB 080413A}: Observations were obtained with the SMARTS 1.3m. They were calibrated against Landolt standards.}
\item{{\bf GRB 080810}: Observations were obtained with the Maidanak, Sayan, AZT-8, Z-600 and RTT150 (TFOSC in the first and second epochs, and the Andor CCD in the third epoch) telescopes. They were calibrated against SDSS standards, except for the RTT150 data, which was calibrated against a Landolt standard star, and slightly adjusted to agree with the zero point of \cite{Page080810}.}
\item{{\bf GRB 081008}: Observations were obtained with the SMARTS 1.3m. They were calibrated against Landolt standards ($VI_C$), 2MASS ($JHK$) and the USNOB1.0 catalog ($BR_C$) Additionally, the $B$ data were made brighter by 0.25 mags to bring it into accordance with the zero point for \cite{Yuan081008}.}
\item{{\bf GRB 090323}: Observations were obtained with the RTT150, the Shajn telescope, the TLS 1.34m (upper limit not used in \citealt{McBreenLAT}) and the NOT, and calibrated against SDSS standards, either directly (RTT150) or transformed to the Johnson-Cousins bands.}
\item{{\bf GRB 090424}: Observations were obtained with the TLS 1.34m, the Maidanak 1.5m, and with the 1.23m Spanish telescope and the 2.2m telescope of the Centro Astron\'omico Hispano-Alem\'an (CAHA), Spain. They were calibrated against SDSS standards transformed to the Johnson-Cousins bands.}
\end{itemize}

\newpage\clearpage
\LongTables

\begin{deluxetable}{llcll}
\tablecolumns{5}
\tabletypesize{\scriptsize}
\tablefontsize{\tiny}
\tablewidth{0pt}
\tablecaption{Afterglow Photometry}
\tablehead{
\colhead{GRB}   &
\colhead{Days after trigger (midtime)} &
\colhead{Magnitude}  &
\colhead{Filter}  &
\colhead{Telescope}}
\startdata
040924	&	0.754109	& $ >	21.20			$ &	$B$	&	SMARTS	\\
	&	0.754109	& $ >	20.80			$ &	$V$	&	SMARTS	\\
	&	0.373410	& $	21.76	\pm	0.09	$ &	$R_C$	&	RTT150	\\
	&	0.390927	& $	21.96	\pm	0.10	$ &	$R_C$	&	RTT150	\\
	&	0.406554	& $	22.06	\pm	0.11	$ &	$R_C$	&	RTT150	\\
	&	0.422177	& $	21.93	\pm	0.09	$ &	$R_C$	&	RTT150	\\
	&	0.439080	& $	22.05	\pm	0.10	$ &	$R_C$	&	RTT150	\\
	&	0.454699	& $	22.12	\pm	0.10	$ &	$R_C$	&	RTT150	\\
\enddata
\end{deluxetable}

\newpage\clearpage

\section{Details on the GRB afterglow Samples}

In this appendix, we describe the GRBs and their afterglows in our different samples on a case-by-case basis. We give the GRB redshifts (references can be found in the caption of Table \ref{tabALL}) and cite the sources we took our data from. For many GRBs, we report additional light curve analysis results. We compare the results we derive from our SED analyses with those given by other sources in the literature. In some cases, we give further miscellaneous results and comments.

We followed \cite{ZKK} in terms of parameter designations. The parameter $m_k$ is the magnitude normalization, which is either the magnitude of the light curve fit at one day (if fit by a single power law), or the magnitude at the light curve break time $t_b$ assuming a break smoothness parameter $n=\infty$. In many cases, the break smoothness parameter needs to be fixed. By eye, most light curves (if they have breaks) show sharp breaks, here, we fix $n=10$ (or $-10$ in the case of steep-to-shallow transitions). In a few cases, the break is much smoother, and we choose $n=1$. $\alpha_1$ and $\alpha_2$ are the decay slopes of the pre-break and post-break power laws, respectively, and $m_h$ is the constant host galaxy magnitude. Often, no host galaxy has been reported, and the final data points show no indication of an upturn due to an emerging host. We fix the magnitude to a value significantly fainter than the the last data point.

\subsection{Details on the pre-\emph{Swift} Golden Sample extension}
\label{App0}

All in all, three GRBs described in K06 have been added to the Golden Sample due to additional data or improved analysis.

\paragraph{GRB 990510, $z=1.6187\pm0.0015$.}We constructed the light curve and the SED with data from the following sources: \cite{Israel990510}, \cite{Stanek990510}, \cite{Harrison990510}, \cite{Beuermann990510}, \cite{VreeswijkGCN990510}, \cite{GalamaGCN990510}, \cite{PietrzynskiGCN9905101}, \cite{PietrzynskiGCN9905102}, \cite{HjorthGCN320}, \cite{FruchterGCN990510}, \cite{BloomGCN990510}, and \cite{Curran990510}. This GRB has a densely sampled, very smooth light curve and is known for its very smooth ``rollover'' break which led to the introduction of the Beuermann equation \citep{Beuermann990510}. The GRB was not included in the Golden Sample of K06 because the short span of the SED ($BVR_CI_C$) led to large uncertainties. The addition of newly published $JHK$ data \citep{Curran990510} has strongly reduced these errors. We find a blue but curved SED, SMC dust is strongly preferred, and the extinction-corrected spectral slope, $\beta=0.17\pm0.15$, is very flat. From a combined fit with all colors, we find the following parameters, which update and supersede those presented in \cite{ZKK}: $\alpha_1=0.77\pm0.03$, $\alpha_2=2.45\pm0.12$, $t_b=1.80\pm0.18$ days, $n=0.81\pm0.16$, and $m_h=29$ was fixed.

\paragraph{GRB 011211, $z=2.1418\pm0.0018$.}We constructed the light curve and the SED with data from the following sources: \cite{Holland011211}, \cite{Jakobsson0112111}, \cite{Jakobsson0112112}, and \cite{CovinoGCN011211}. This afterglow shows early-time small-scale variability which has been attributed to a ``patchy shell'' model \citep{Jakobsson0112112}. A refined analysis has taken this variability into careful account, especially for the NIR data, and has yielded a significantly improved SED. We find that SMC dust is strongly preferred and there is a small, typical amount of extinction present.

\paragraph{GRB 030323, $z=3.3718\pm0.0005$.}We constructed the light curve and the SED with data from the following sources: \cite{Vreeswijk030323}, \cite{GilmoreGCN030323}, \cite{SmithGCN030323}, \cite{MasiGCN030323}, and \cite{Wood-VaseyGCN030323}. Similar to GRB 011211, this GRB contains small-scale variability which was not taken into account in K06. A refined analysis yields a significant improvement of the SED. The only band we do not add is the $K$ band which is too bright for unknown reasons. We find an SED best fit by SMC dust (MW dust is ruled out, and LMC dust yields an unrealistically flat intrinsic spectral slope) and a small amount of extinction.

\subsection{Details on the \emph{Swift}-era Golden Sample}
\label{AppA}

\paragraph{GRB 050319, $z=3.2425$.}We constructed the light curve and the SED with data from the following sources: \cite{Wozniak050319}, \cite{Mason050319}, \cite{Quimby050319}, \cite{Huang050319}, \cite{Kamble050319}, \cite{ToriiGCN050319}, as well as our own extensive data set (NOT, TNG, Maidanak, RTT150). We do not add the $J-$band data from \cite{George050319}. If this NIR flare is real, it is a quite mysterious event, as a contemporaneous $V-$band detection \citep{Huang050319} shows no sign of flaring activity. Initially, the light curve shows a transition from a moderately steep to a shallower decay: $m_k=18.80\pm0.14$ mag, $\alpha_1=0.93\pm0.026$, $\alpha_2=0.46\pm0.022$, $t_b=0.031\pm0.006$ days, and $m_h=28$ mag, $n=-10$ fixed. Another break is found in the late light curve, this is a good jet break candidate with a smooth transition: $m_k=21.11\pm0.12$ mag, $\alpha_1=0.46\pm0.020$, $\alpha_2=2.23\pm0.25$, $t_b=3.47\pm0.42$ days, and $m_h=28$ mag, $n=1$ fixed. The $B$ band is already affected by Lyman absorption \citep{Huang050319}, leaving us with a $VR_CI_CK$ SED. A fit without extinction yields $\beta_0=1.00\pm0.06$, in full agreement with \cite{Huang050319}. Due to the sparsity of colors and the large error in the $K$ band, we are not able to discern between extinction laws (the 2175 {\AA} feature lies in the $z$ band), and the derived errors of the SED parameters are quite large. Since it yields viable fits to most other GRB afterglows, we use the SMC values of $\beta=0.74\pm0.42$ and $\AV=0.05\pm0.09$, showing that extinction is negligible. \cite{OatesUVOT} find $\AV=0.06$ from UVOT data only (using an XRT-UVOT joint fit), in excellent agreement. \cite{Schady2010} find that a cooling break between optical and X-rays is preferred at low significance, but can only derive upper limits for this case, their upper limit of $\AV<0.09$ for SMC dust agrees very well with our result.

\paragraph{GRB 050408, $z=1.2357\pm0.0002$.}We constructed the light curve and the SED with data from the following sources:  \cite{Foley050408}, \cite{deUgarte050408}, \cite{WiersemaGCN050408}, \cite{MilneGCN050408}, \cite{KahharovGCN050408} and \cite{FlasherGCN050408}, as well as our own data set (TNG, Maidanak, AT-64 upper limit). From our $UBVR_CI_CZJHK$ SED, we derive results that are in full agreement with \cite{deUgarte050408}. We find that SMC dust is preferred and derive $\beta=0.28\pm0.27$ \citep[$\beta=0.28\pm0.33$ in][]{deUgarte050408} and $\AV=0.74\pm0.15$ \citep[$\AV=0.73\pm0.18$ in][]{deUgarte050408}. Thus, this is one of the highest line-of-sight extinctions found so far for a GRB with a detected afterglow (cf. the sample of K06). We concur that the rebrightening feature observed by \cite{deUgarte050408} is very unlikely to be a supernova rebrightening, as we derive a peak luminosity $k=1.65\pm0.55$ \citep[in units of the SN 1998bw peak luminosity in the same band at the same redshift,][]{Zeh2004} and a stretching factor $s=0.22\pm0.05$, which is much faster than any known GRB-SN \citep{Ferrero060218}, although we note that the shape of the rebrightening can be well approximated by a strongly compressed SN 1998bw light curve. Also, employing the extinction correction as described in \cite{Ferrero060218}, we derive $k=3.27^{+0.87}_{-0.68}$, which is also much brighter than any GRB-SN in the sample of \cite{Ferrero060218}. Excluding the rebrightening from the fit, we find the light curve parameters $m_k=21.96\pm0.44$ mag, $\alpha_1=0.48\pm0.10$, $\alpha_2=2.06\pm0.46$, $t_b=0.92\pm0.32$ days and $m_h=24.56\pm0.15$ mag. We fix $n=1$, as in this case there is a smooth transition between slopes.

\paragraph{XRF 050416A, $z=0.6528\pm0.0002$.}We constructed the light curve and the SED with data from the following sources: \cite{Holland050416A}, \cite{Soderberg050416A}, and \cite{PerleyHosts2009}, as well as our own data set (Maidanak, ANDICAM). Similar to GRB 050401, observations in different filters are taken with little overlap. The light curve evolution is not entirely clear, but we concur with \cite{Soderberg050416A} that there is an early break. We find the following results for a broken power law fit of the composite, host-subtracted light curve ($\chi^2=52.7$ for 47 degrees of freedom): $m_k=19.94\pm0.16$ mag, $\alpha_1=0.25\pm0.07$, $\alpha_2=0.97\pm0.05$, $t_b=0.025\pm0.005$ days, and $n=10$ fixed. The late decay slope is slightly steeper than what \cite{Soderberg050416A} find ($\alpha_{opt/NIR}\approx0.75$), possibly due to our use of a brighter host galaxy magnitude \citep{PerleyHosts2009}. We were not able to reproduce the SN results reported by \cite{Soderberg050416A}. We obtain a broad SED ($UVW2\;UVM2\;UBVR_CI_Cz^\prime K_S$), which has some scatter, but yields a small amount of dust for all extinction laws. While we can not rule out MW and LMC dust, we find no evidence for a 2175 {\AA} bump, which falls into the $U$ band.

\paragraph{GRB 050502A, $z=3.793$.}We constructed the light curve and the SED with data from the following sources: \cite{Guidorzi050502A}, \cite{Yost050502A}, and \cite{DurigGCN050502A}, as well as our own data (IAC80, INT, as well as an unused upper limit from the Maidanak telescope). We fit the $R$-band light curve (excepting the last two points), which also contains the \emph{ROTSE} $C_R$ data points, with a broken power-law. We find $m_k=19.55\pm0.39$ mag, $\alpha_1=1.06\pm0.043$, $\alpha_2=1.45\pm0.028$, $t_b=0.067\pm0.017$ days, $m_h=28$ mag is fixed. The data quality is high enough to leave the break smoothness parameter free, we find $n=149.3\pm6.1$, i.e., a very sharp break is required. This is an astonishing result, as no other afterglow light curve where it was possible to let $n$ be a free parameter of the fit showed $n>10$ \citep{ZKK}. Our results compare well with \cite{Yost050502A}, who find $\alpha_1=1.13\pm0.023$, $\alpha_2=1.44\pm0.022$ and $t_b=0.066\pm0.009$ days. \cite{Yost050502A} interpret the break as the passage of the cooling frequency $\nu_c$ through the $R$ band. We note that \cite{ZKK} found a similarly sharp break for the cooling frequency passage of the afterglow of GRB 030329 \citep{Sato2003}. Also, that break was a bit stronger ($\Delta\alpha=0.33\pm0.01$) than the theoretical prediction $\Delta\alpha=0.25$ \citep{PK2001a}, similar to this case ($\Delta\alpha=0.31\pm0.03$). The final deep INT data point shows that a second break must have occurred. Fitting all data from 0.07 days onward, we find $\alpha_2=1.39\pm0.031$, $\alpha_3=1.82\pm0.20$, $t_b=0.34\pm0.13$ days, and $n=10$, $m_h=28$ mag are fixed. The SED is well-fit by a small amount of SMC dust, but the preference is weak only. If the cooling break lies redward of the optical, $p$ derived from the intrinsic spectral slope ($p=1.52\pm0.32$) is in agreement with the value derived from the post-break decay slope ($p=1.82\pm0.20$), but no X-ray data exist to further examine this possibility.

\paragraph{GRB 050525A, $z=0.606$.}We constructed the light curve and the SED with data from the following sources: \cite{Klotz050525A}, \cite{Blustin050525A}, \cite{Dellavalle050525A}, \cite{Heng050525A}, \cite{RykoffROTSE}, \cite{YanagisawaGCN050525A}, \cite{KaplanGCN050525A}, and \cite{FlasherGCN050525A}, as well as our own data set (SARA, ANDICAM, RTT150, as well as upper limits from several Russian telescopes). Excluding the early data before the rebrightening \citep{Klotz050525A}, we find $m_k=18.87\pm0.13$ mag, $\alpha_1=1.09\pm0.032$, $\alpha_2=1.77\pm0.024$, $t_b=0.30\pm0.023$ days, $m_h=24.96\pm0.04$ mag, and $n=10$ was fixed. This agrees very well with \cite{Dellavalle050525A}, who find $\alpha_1=1.1$, $\alpha_2=1.8$ and $t_b=0.3$ days. The SN parameters we find are reported in \cite{Ferrero060218}. We find that SMC dust is preferred (although the SED shows scatter even for SMC dust, resulting in a high $\chi^2$/d.o.f.), $\AV=0.32\pm0.20$. This is in agreement with \cite{Blustin050525A}, who also find SMC dust and $\AV=0.23\pm0.15$. \cite{Schady2007} also find that SMC dust is preferred, and for a solution with a cooling break between the optical and X-rays, they derive $\AV=0.26\pm0.04$, also in agreement with our result. \cite{Heng050525A} use \emph{Spitzer Space Telescope} detections to extend the SED all the way to $24\micron$. Deriving the intrinsic SED from a fit to the X-ray and MIR data, they find $\AV\approx0.02-0.41$ for different models (e.g., amount of host contribution to the MIR data), with the most realistic model yielding $\AV=0.15\pm0.06$, smaller than our result due to the larger assumed intrinsic $\beta=0.87$. \cite{OatesUVOT} find $\AV=0.06$ from UVOT data only (using an XRT-UVOT joint fit), a significantly smaller value. \cite{Schady2010} find, similar to \cite{Schady2007}, the need for a cooling break and the preference of SMC dust, but derive $\AV=0.16\pm0.02$, comparable to \cite{Heng050525A} and still in agreement within errors with our value.

\paragraph{GRB 050730, $z=3.96855\pm0.00005$.}We constructed the light curve and the SED with data from the following sources: \cite{Pandey050730}, \cite{Perri050730}, \cite{BlustinGCN050730}, \cite{KlotzGCN050730}, \cite{DamerdjiGCN050730}, \cite{HolmanGCN050730}, \cite{HaislipGCN050730}, \cite{JacquesGCN050730}, and \cite{KannappanGCN050730}, as well as our own data set \citep[RTT150 and ANDICAM upper limits not given in][]{Pandey050730}. No host galaxy is found down to deep limits \citep{ChenHost}. The light curve of this burst is highly variable \citep{Pandey050730, Perri050730}, and the SED was evaluated carefully. We fit the light curve with a broken power law and find $m_k=18.20\pm0.10$ mag, $\alpha_1=0.56\pm0.008$, $\alpha_2=1.69\pm0.085$, $t_b=0.17\pm0.026$ days, and $m_h=27$ mag, $n=10$ were fixed. This is concurrent with the results of \cite{Pandey050730}: $\alpha_1=0.60\pm0.07$, $\alpha_2=1.71\pm0.06$ and $t_b=0.1$ days. The SED results, on the other hand, differ quite strongly between different works. \cite{Chen050730} find a steep $\beta=1.88\pm0.01$ from the continuum fitting of an echelle spectrum. \cite{Starling050730}, also fitting a spectrum, find $\beta=1.34\pm0.21$. \cite{Pandey050730}, using only $I$, $J$ and $K$ data, find $\beta=0.56\pm0.06$. Our result, for an $i^\prime I_CJK$ SED, is $\beta_0=0.82\pm0.041$ ($\beta_0$ being the spectral slope without any extinction correction) and $\beta=0.52\pm0.045$, $\AV=0.10\pm0.015$ for the preferred SMC dust. \cite{Starling050730} find $\AV\approx0.01$ from spectral fitting. \cite{OatesUVOT} find $\AV=0.15$ from UVOT data only (using an XRT-UVOT joint fit), a value comparable to ours. \cite{Schady2010} find that no dust model is able to fit the SED well, and derive $\AV\approx0.16-0.39$ for their different models, mostly in excess of our value.

\paragraph{GRB 050801, $z=1.56\pm0.1$.}We constructed the light curve and the SED with data from the following sources: \cite{Rykoff050801}, \cite{DePasquale050801}, \cite{MonardGCN050801}, as well as our own data set (ANDICAM, Danish 1.54m). No host was detected down to deep limits by \cite{OvaldsenHosts}. This GRB has a photometric redshift only, but it is precise enough for our analysis, $z=1.56\pm0.10$ \citep{DePasquale050801} (\citealt{OatesUVOT} find a slightly different value, $z=1.38\pm0.07$). The light curve shows an early, almost flat plateau phase before breaking sharply to a typical power law decay \citep{Rykoff050801}. Fitting the light curve with a broken power law, we find: $m_k=14.85\pm0.05$ mag, $\alpha_1=0.13\pm0.03$, $\alpha_2=1.19\pm0.01$, $t_b=247\pm11$ seconds, and $n=10$, $m_h=28$ were fixed. If we use all data, the SED of this afterglow does not yield a reasonable result, with the UVOT data \citep{DePasquale050801} being much brighter. Therefore, we use only the late-time ground-based data presented in this work, and derive a red SED which is well-fit by a moderate amount of SMC dust. This result is different from those based on UVOT data exclusively, \cite{DePasquale050801} find that the optical-to-X-ray SED is described by a single slope, and \cite{OatesUVOT} find no evidence for any extinction (using an XRT-UVOT joint fit).

\paragraph{GRB 050802, $z=1.7102$.}We constructed the light curve and the SED with data from the following sources: \cite{Oates050802} and our own extensive data set (OSN, NOT, TNG, Maidanak, Shajn). The two main data sets \citep[\emph{Swift} UVOT,][and ground-based, ours]{Oates050802}, hardly overlap. Creating a composite light curve, we find that the light curve is well described by a broken power law with the following parameters: $m_k=21.31\pm0.53$ mag, $\alpha_1=0.84\pm0.022$, $\alpha_2=1.43\pm0.25$, $t_b=0.63\pm0.31$ days, $m_h=25.36\pm0.46$ mag, and $n=10$ fixed. If this is a jet break, the post-break slope is very shallow, but this has been seen before \citep[e.g., GRB 010222 and GRB 041006,][]{ZKK}. We are able to derive a very smooth SED ($UBVR_CI_C$, the UV filters are all affected by Lyman damping), but the rather large errors of the data points lead to large uncertainties in the values of $\beta$ and $\AV$. LMC dust is marginally preferred, following the shape of the SED much better than MW or SMC dust, and we find $\beta=0.36\pm0.26$ and $\AV=0.21\pm0.13$. This is lower than the result of \cite{Oates050802}, who find $\EBV=0.18$ ($\AV=0.56$) for MW dust (they only fit with MW and SMC dust). A similar value is found by \cite{Nardini2009}. \cite{OatesUVOT}, on the other hand, find $\AV=0.11$ from UVOT data only (using an XRT-UVOT joint fit), which compares well with our result. \cite{Schady2010} find a marginal preference for MW dust and a cooling break between the optical and the X-rays, and derive $\AV\approx0.15$ for this case, in agreement with our value. At 210 days after the GRB, a faint host galaxy is detected.

\paragraph{GRB 050820A, $z=2.6147$.}We constructed the light curve and the SED with data from the following sources: \cite{Vestrand050820A}, \cite{Cenko050820A} and references therein, \cite{MacombGCN050820A}, as well as our own large data set (UAPhO, Maidanak, Shajn, RTT150, UKIRT, TNG, WIRO). Due to the complicated light curve structure \citep{Cenko050820A}, special care is taken in deriving the SED. Similar to \cite{Cenko050820A}, we find that the optical afterglow is composed of four parts, a moderately steep decay, a shallow rebrightening phase, then another phase with an identical decay slope as the first phase, and finally a steep decay that is probably due to a jet break. As our late data are taken almost exclusively from \cite{Cenko050820A}, our parameter results are identical within errors. Fixing the decay smoothness to $n=-10$ and the host galaxy magnitude $m_h=27$, we find $m_k=17.49\pm0.094$ mag, $\alpha_1=1.07\pm0.011$,  $\alpha_2=0.54\pm0.051$, $t_b=0.073\pm0.007$ days for the steep to shallow transition from 0.006 to 0.22 days. For the shallow to steep transition (0.09 to 7.7 days) we find (fixing $n=10$, $m_h=27$ mag): $m_k=18.22\pm0.088$ mag, $\alpha_1=0.50\pm0.051$,  $\alpha_2=1.04\pm0.016$, $t_b=0.27\pm0.025$ days. Clearly, the fit finds identical values for the shallow phase from both fits, and the steep phases are also almost identical within errors ($1.9\sigma$). Finally, fitting the data from 0.75 days onward, we follow \cite{Cenko050820A} and fix $\alpha_2=p=2.34$, and find $m_k=22.94\pm0.11$ mag, $\alpha_1=1.04\pm0.015$, $t_b=18.19\pm1.28$ days, again in agreement with \cite{Cenko050820A}. For the SED, we derive a steeper unextinct slope than \cite{Cenko050820A}, $\beta_0=0.96\pm0.028$, and find no conclusive evidence for color evolution. We find a very low host extinction $\AV=0.065\pm0.008$, in agreement with \cite{Cenko050820A}, who find $\AV=0$ for all dust extinction curves. \cite{Schady2010} find strong evidence for a cooling break between the optical and the X-rays, but are not able to distinguish the dust models beyond this, and find a higher value, $\AV\approx0.14-0.32$, in comparison to ours.

\paragraph{XRF 050824, $z=0.8281\pm0.004$.}We constructed the light curve and the SED with data from the following sources: \cite{Sollerman050824}, \cite{SchadyGCN050824}, and \cite{LipunovGCN050824}. The full light curve analysis (which also finds evidence for a rapid supernova) can be found in \cite{Sollerman050824}. The addition of \emph{Swift} UVOT data \citep{SchadyGCN050824} allows us to create a more detailed SED than that presented by \cite{Sollerman050824}. From a fit without extinction, we find $\beta_0=0.65\pm0.07$, in accordance, within errors, with the value found by \cite{Sollerman050824}, $\beta_0=0.56\pm0.04$. These authors are unable to discern MW, LMC and SMC extinction, and find $\AV=0.4\pm0.2$ mag for SMC dust, albeit with $\beta\approx0$. From our SED, we find a weak preference for SMC dust (with MW dust being ruled out, as the fit finds negative extinction), with $\AV=0.14\pm0.13$ mag and $\beta=0.45\pm0.18$, in accordance with the limit $\AV\leq0.5$ mag from \cite{Sollerman050824}. \cite{Schady2007} find that SMC dust is weakly preferred, and are not able to discern between a scenario with and one without a cooling break between the X-rays and the optical. In the former case, they find $\AV=0.12\pm0.04$, in the latter $\AV=0.16^{+0.06}_{-0.04}$, in excellent agreement with our value.

\paragraph{GRB 050904, $z=6.295\pm0.002$.}The complete analysis of this GRB is presented in \cite{Kann050904}. For Figure 1 in Paper II, we added the $dRc$ value derived in \cite{Kann050904} to the $z=1$ light curve and transformed the time back to the observer frame. This is what the observed light curve in the $R$ band would have looked like in a hypothetical fully ionized universe (no Lyman dropout). \cite{LiangLi} report higher extinction than we found using the ``Drude'' model of \cite{LiDust}. They claim to find a 2175 {\AA} bump as well as an evolution in the characteristics of the extinction. \cite{Stratta050904} claim the existence of non-standard ``high-$z$ SN dust'' \citep{MaiolinoDust} and a larger extinction value than we derive. Recently, these results have been refuted by a careful analysis of \cite{Zafar050904}, who concur with the analysis of \cite{Kann050904}, finding no evidence for dust or evolution.

\paragraph{GRB 050922C, $z=2.1992\pm0.0005$.}We constructed the light curve and the SED with data from the following sources: \cite{RykoffROTSE}, \cite{OfekGCN050922C}, \cite{DurigGCN050922C}, \cite{HenychGCN050922C}, \cite{NovakGCN050922C}, \cite{HunsbergerGCN050922C}, and \cite{LiGCN050922C}, as well as our own extensive data set (Terskol, NOT, MDM, Danish 1.54, INT, WHT, VLT). No host galaxy is found down to deep limits \citep{OvaldsenHosts, ChenHost}. The composite light curve shows some small variations in the early UVOT data \citep{LiGCN050922C, HunsbergerGCN050922C} but is otherwise a smooth broken power law with a soft rollover. We find $m_k=17.70\pm0.089$ mag, $\alpha_1=0.76\pm0.016$, $\alpha_2=1.39\pm0.023$, $t_b=0.11\pm0.008$ days, $n=3.36\pm0.69$ and $m_h=28$ mag was fixed. If this is due to a jet break, it is very early and the post-break slope is very shallow \citep[see][for pre-\emph{Swift} values]{ZKK}. The SED shows essentially no signs of dust extinction, and no dust model can be preferred. \cite{OatesUVOT} find $\AV=0.08$ from UVOT data only (using an XRT-UVOT joint fit), a comparably small value. \cite{Schady2010} are also unable to prefer any dust model (or make a statement concerning the location of the cooling break), but find a comparably low $\AV=0.07\pm0.02$ for SMC dust and no cooling break (but higher values up to $\AV\approx0.28$ for other models).

\paragraph{GRB 060124, $z=2.297\pm0.001$.}We constructed the light curve and the SED with data from the following sources: \cite{Misra060124}, \cite{Deng060124} and D. A. Kann et al., in preparation (paper on the GRB 060124 afterglow). Preliminary analyses of the data set of Kann et al. have been published by \cite{Curran060124} and \cite{Covino060124}. The afterglow of this extremely long GRB \citep{Romano060124} is very luminous and well-described by an achromatic broken power law, with a contemporary break seen in the X-rays. The SED is best fit with a MW dust law, as there is evidence for a small 2175 {\AA} bump. \cite{Schady2010} also find marginal evidence for a preference of MW dust, but derive an extinction value ($\AV=0.52\pm0.13$) much in excess of our own ($\AV=0.17\pm0.03$).

\paragraph{GRB 060206, $z=4.04795\pm0.0002$.}We constructed the light curve and the SED with data from the following sources: \cite{Wozniak060206}, \cite{Stanek060206}, \cite{Monfardini060206}, \cite{Curran060206}, \cite{OfekGCN060206}, \cite{LinGCN060206}, \cite{MilneGCN060206}, \cite{AlataloGCN060206}, \cite{TeradaGCN060206}, \cite{BoydGCN060206}, \cite{GrecoGCN060206}, \cite{LacluyzeGCN060206}, \cite{ReichartGCN060206}, and \cite{Thoene060206} (host galaxy detection) as well as our own data set (SARA, Maidanak). The light curve is very complex, showing an extreme rebrightening feature \citep{Wozniak060206, Monfardini060206}, and special care was taken in determining the SED. Excluding the early rebrightening feature ($t<0.09$ days), we fit the light curve with a broken power law and find $m_k=17.79\pm0.049$ mag, $\alpha_1=0.80\pm0.031$, $\alpha_2=1.42\pm0.021$, $t_b=0.26\pm0.008$ days, $n=7.73\pm2.48$ and $m_h=24.91\pm0.26$ mag \citep[note that the actual host is several magnitudes fainter, but can only be separated from a very nearby galaxy by the HST,][]{Thoene060206, ChenHost}. These results differ quite strongly from \cite{Stanek060206}, who find $\alpha_1=0.7$, $\alpha_2=2.0$, and $t_b=0.6$ days. There are small-scale variations present in the densely sampled data of \cite{Stanek060206}, and it is possible that the light curve parameters and especially the break time are very sensitive to which data are actually included in the fit. As the final deep data points \citep{Curran060206} are overestimated in our fit, this may be indicative of another break and a following, steeper decay. These authors use only data after 0.2 days, and we are able to exactly reproduce their results if we use the same interval, we find $m_k=18.87\pm0.11$ mag, $\alpha_1=1.13\pm0.016$, $\alpha_2=1.64\pm0.035$, $t_b=0.57\pm0.044$ days, $n=10$, $m_h=24.9$ fixed. This afterglow is a good example of why dense sampling can be very important. From the SED, we find that MW dust is ruled out, but the preference for SMC dust is only weak. We find almost negligible extinction, in concurrence with \cite{OatesUVOT}, who use UVOT data only (using an XRT-UVOT joint fit). \cite{Schady2010} also find (low) upper limits only on the extinction.

\paragraph{GRB 060418, $z=1.49010\pm0.0001$.}We constructed the light curve and the SED with data from the following sources: \cite{Molinari060418}, \cite{MelandriROBONET}, \cite{Cenko060418}, \cite{FalconeGCN060418}, \cite{NysewanderGCN060418}, \cite{JelinekGCN060418}, \cite{SchadyGCN060418}, \cite{ChenGCN060418}, \cite{HuangGCN060418}, D. Malesani et al., in preparation (paper on the GRB 060418 afterglow), as well as our own data set (ANDICAM, Maidanak). As reported by several groups, the light curve shows a strong initial rise, which then rolls over into a power-law decay \citep{Molinari060418, NysewanderGCN060418, JelinekGCN060418}. Shifting all data to the $R_C$ zero point, we find no evidence for color evolution, in accordance with \cite{Molinari060418}. The composite light curve is well fit with a broken power law and we find $m_k=11.92\pm0.18$ mag, $\alpha_R=-5.89\pm1.38$, $\alpha_1=1.19\pm0.003$, $t_b=80.7\pm10.8$ seconds, $n=0.42\pm0.11$ and $m_h=24.97\pm0.14$ mag \citep[][find no host underlying the direct afterglow position, but a very nearby galaxy complex would influence any ground-based imaging]{Cenko060418}. We find no evidence for a jet break out to at least 10 days, in accordance with \cite{Cenko060418}, who detect a break only in late, very deep HST imaging. A host-corrected light curve confirms that the HST data lies more than a magnitude under the extrapolation of the earlier decay. These fit values are in similar to those found by \cite{Molinari060418}, though we find a smoother rollover and a steeper early rise (which is sparsely sampled). The SED has been derived from a joint fit to all bands. We derive a very broad SED in 11 filters, and find that while no dust model can be strongly preferred, LMC dust yields the best result, with $\AV=0.20\pm0.08$. This is interesting, as \cite{Ellison060418} find clear evidence of a 2175 {\AA} bump deriving from a dusty foreground absorber at $z=1.1$ in an echelle spectrum of the afterglow. While we fit at the redshift of the burst \cite[$z=1.49$,][]{ProchaskaGCN060418}, the low resolution of the SED may lead to the bump also affecting our SED. The extinction we derive is therefore probably a superposition of extinction in the host frame and in the foreground absorbers. Our extinction results are marginally consistent to those derived by \cite{Schady2007}, who derive higher values, but also prefer LMC dust. \cite{OatesUVOT} find $\AV=0.08$ from UVOT data only (using an XRT-UVOT joint fit), a smaller value. \cite{Schady2010} find marginal preference for SMC dust and a value similar to \cite{OatesUVOT}, $\AV\approx0.09$.

\paragraph{GRB 060526, $z=3.211\pm0.001$.}The complete analysis of this GRB is presented in \cite{Thoene060526}. Those authors find that the complex, highly variable light curve can be modeled with multiple energy injections, as well as jet break \citep[see also][]{Dai060526}. The SED is fit well by SMC dust and a small amount of extinction. \cite{OatesUVOT} find no evidence for dust from UVOT data only (using an XRT-UVOT joint fit). \cite{Schady2010} derive mostly low upper limits and two low detections and are not able to discern between dust models.

\paragraph{GRB 060607A, $z=3.0749$.}We constructed the light curve and the SED with data from the following sources: \cite{Molinari060418},  \cite{Nysewander060607A}, \cite{Ziaeepour060607A}, and \cite{FynboSpectra} as well as our own ANDICAM data set. No host is detected down to deep levels \citep{ChenHost}. The afterglow shows an early rise and several achromatic bumps \citep{Molinari060418, Nysewander060607A, Ziaeepour060607A}. As the last detection is at 0.2 days in the observer frame, no statement can be made about the magnitude at one day. The SED shows very little dust, the SMC fit is weakly preferred, but the extinction is 0 within errors. This is comparable to \cite{OatesUVOT}, who find $\AV=0.05$ from UVOT data only (using an XRT-UVOT joint fit). \cite{Schady2010} similarly derive low upper limits only and are not able to discern between any models.

\paragraph{GRB 060904B, $z=0.7029$.}We constructed the light curve and the SED with data from the following sources: \cite{Klotz060904B}, \cite{RykoffROTSE}, \cite{FynboSpectra}, \cite{deUgartePostigoGCN060904B}, \cite{SkvarcGCN060904B}, \cite{OatesGCN060904B},  \cite{GrecoGCN060904B}, \cite{PrymakGCN060904B}, \cite{SoyanoGCN060904B}, and \cite{HuangGCN060904B}, as well as our own data set (ANDICAM, RTT150, Maidanak). This GRB features strong early variability during the prompt emission \citep{RykoffROTSE}. In agreement with \cite{Klotz060904B}, we find a peaking afterglow from 0.001 to 0.012 days, with parameters $m_k=16.347\pm0.031$ mag, $\alpha_R=-0.665\pm0.085$ (rising slope), $\alpha_F=1.143\pm0.069$ (falling slope), $t_b=546\pm20$ s and $n=10$ fixed. At $t_{b2}=1213\pm110$ s, the decay goes over into a slow decay/plateau phase with $\alpha_P=0.145\pm0.077$, before resuming, at $t_{b3}=4281\pm314$ s, a smooth decay with $\alpha=1.170\pm0.026$. As the slopes before and after the plateau are identical, this may indicate a ``step'' due to an energy injection \citep[see][]{Klotz060904B}. The SED is very broad, and includes 10 filters from $UVW2$ to $K$. The SED is relatively steep but also flat, and we only find a small amount of dust. This is indicative of a cooling break redward of the optical. SMC dust is (weakly) preferred, and the bright detections in $UB$ clearly rule out a 2175 {\AA} bump. \cite{Schady2010} find no good fit for any dust model (with fits without a cooling break between optical and X-rays being even less likely), but derive low extinction in all cases, in agreement with our result.

\paragraph{GRB 060908, $z=1.8836$.}We constructed the light curve and the SED with data from the following sources: \cite{Covino060908}, \cite{CenkoDark}, \cite{FynboSpectra}, \cite{NysewanderGCN060908}, as well as a single data point published here. The light curve is adequately fit with a single power law: $m_k=22.66\pm0.03$ mag and $\alpha=1.08\pm0.007$ (for our fits, we use host-subtracted data). But there are significant deviations from the power law, $\chi^2=192$ for 117 degrees of freedom. The earliest data points show a steeper decay, removing these yields a significant improvement ($\chi^2=115$ for 109 degrees of freedom), and the parameters $m_k=22.52\pm0.03$ and $\alpha=1.04\pm0.007$. Two data points at one day lie significantly ($\approx0.5$ mag) below the extrapolation of the early decay, indicating that a break may have occurred (we note that a further P60 data point at four days is overluminous and has not been included). Due to the sparsity of the data, the break time and post-break decay slope are unconstrained in a free fit, but fixing $\alpha_2=1.6$ (from the SED, see below), we find $t_b=0.44\pm0.12$ days. The SED is broad, from $B$ to $K$. We find a very flat SED ($\beta_0=0.30\pm0.06$, which implies $p=1.6$ for $\nu_c$ blueward of the optical bands, in excellent agreement with the hard X-ray slope $\Gamma=1.798$) and no evidence for dust. MW and LMC fits yield negative extinction and are ruled out. The SMC fit yields a small amount of dust (albeit 0 within errors) but makes the SED even more blue. Therefore, we use $\beta_0$. The flat SED strongly resembles that of GRB 021211 \citep{PaperIII}. \cite{OatesUVOT} also find very low extinction, $\AV=0.02$ from UVOT data only (using an XRT-UVOT joint fit). Similarly, \cite{Schady2010} derive low upper limits for all models only, they marginally prefer a cooling break between the optical and the X-rays.

\paragraph{GRB 061007, $z=1.2622$.}We constructed the light curve and the SED with data from the following sources: \cite{Mundell061007}, \cite{Schady061007}, and \cite{RykoffROTSE}, as well as our own ANDICAM data, which includes the only NIR detections of this GRB that we are aware of. We confirm the rebrightening feature \citep{Mundell061007}. Excluding this feature, we can fit the data with a single power law with $m_k=22.32\pm0.033$ (note that in this case this is the $R$ magnitude at 1 day after the GRB) and $\alpha=1.71\pm0.009$. We fix $m_h=26$ mag. This is fully consistent with \cite{Mundell061007}, who find $\alpha=1.72\pm0.10$, and also consistent with \cite{Schady061007}, who find $\alpha=1.64\pm0.01$. Furthermore, consistent with \cite{Mundell061007} and \cite{Schady061007}, we find that the afterglow, notwithstanding its extreme brightness, is moderately extinct. While there is only weak evidence for the 2175 {\AA} feature, we find that LMC is the preferred dust model, since both MW and SMC dust lead to much smaller extinction and intrinsic spectral slopes that are too steep for the standard fireball model. We find $\beta=1.07\pm0.19$ and $\AV=0.48\pm0.10$, which is among the highest values in comparison with the sample of K06. \cite{Mundell061007} derive $\beta=1.02\pm0.05$ and $\AV=0.48\pm0.19$ for SMC dust and a joint optical to X-ray fit, in excellent agreement with our value. From a similar fit, \cite{Schady061007} find $\beta=0.90\pm0.005$ and $\AV=0.39\pm0.01$ for SMC dust but they also prefer LMC dust, where they find $\beta=0.98\pm0.007$ and $\AV=0.66\pm0.02$. \cite{OatesUVOT} also find a high value, $\AV=0.66$, from UVOT data only (using an XRT-UVOT joint fit). \cite{Schady2010} also find high extinction values, they strongly rule out MW dust and prefer LMC dust, where, for a fit with no cooling break between the optical and the X-rays, they find $\AV=0.75\pm0.02$. Similar to the SED of GRB 050525A, there is scatter while the errors themselves are small, leading to a high $\chi^2$/d.o.f. 

\paragraph{GRB 061126, $z=1.1588\pm0.0006$.}We constructed the light curve and the SED with data from the following sources: \cite{Perley061126}, and \cite{Gomboc061126}. The light curve analysis is presented in the latter paper. We confirm the early color evolution found by \cite{Perley061126}, and thus exclude this data from the SED construction. We find a rather steep but very straight broad SED ($U$ to $K_S$, the \emph{Swift} UVOT UV filters lie beyond Lyman $\alpha$ and are not included.). Fitting with dust, we find that SMC dust gives the best fit, with a very low amount of line-of-sight extinction, $\AV=0.095\pm0.055$. This is fully in agreement with \cite{Perley061126}, who find no ``classical'' dust but imply the need for strong gray extinction to explain the optical subluminosity in contrast with the bright X-ray afterglow \citep[see also][]{LiLiWei}. \cite{Gomboc061126} find higher extinction values from joint X-ray-to-optical fits, as well as variable extinction, which stems from the fact that the X-ray afterglow decays more rapidly than the optical afterglow, reducing the offset \cite{Perley061126} first detected. At 40 ks, \cite{Gomboc061126} find $\AV\lesssim0.13$ mag, in agreement with our result. \cite{Nardini2009} interpret the light curve as being dominated at late times by ``late prompt'' emission, and also find no need for any extinction. \cite{Schady2010} find marginal preference for SMC dust and a strong preference for a cooling break between the optical and the X-rays, and find $\AV=0.10\pm0.04$, in excellent agreement with our result.

\paragraph{GRB 070125, $z=1.5477\pm0.0001$.}We constructed the light curve and the SED with data from the following sources: \cite{Updike070125}, \cite{Chandra070125}, \cite{Dai070125}, \cite{YoshidaGCN070125}, \cite{UemuraGCN070125}, \cite{DurigGCN070125}, and \cite{SposettiGCN070125}, as well as our own ANDICAM data set. By shifting data to the $R$-band zero point, we find an achromatic evolution (except for very early UVOT observations, which have a different $B-V$ color than later on), and construct a cleaned composite light curve. In accordance with \cite{Updike070125}, we find an early rebrightening. Fitting the data up to 0.95 days (before the plateau and rebrightening), we find $m_k=18.806\pm0.014$ mag and $\alpha_1=1.460\pm0.063$. The following rebrightening episodes are described in detail in \cite{Updike070125}. Using all data from beyond 1.5 days, we find clear evidence for a break in the light curve, in accordance with \cite{Updike070125}, \cite{Chandra070125}, and \cite{Dai070125}. We derive the following parameters: $m_k=21.442\pm0.303$ mag, $\alpha_1=1.793\pm0.014$, $\alpha_2=2.875\pm0.200$, $t_b=5.309\pm0.810$ days, and $n=10$ fixed. We followed \cite{Cenko070125} and assume an extremely faint ``host galaxy'' (e.g., tidal tail starburst). The SED, reaching from $UVW2$ to $K$, consists of 14 filters (the UVOT UV filters are affected by Lyman damping and are not included in the fit). We find a very good fit with SMC dust, which is strongly preferred.

\paragraph{GRB 070419A, $z=0.9705$.}We constructed the light curve and the SED with data from the following sources: \cite{Melandri070419A}, \cite{CenkoDark}, \cite{Dai070125}, \cite{FynboSpectra}, \cite{IizukaGCN070419A}, \cite{WrenGCN070419A}, \cite{SwanGCN070419A}, and \cite{LandsmanGCN070419A}, and we also publish SARA data (one upper limit and one marginal detection) and an unused upper limit from the 1.5m Sayan telescope. The complex light curve, featuring an early rise, a broken power-law decay, and a late flattening, is analyzed in \cite{Melandri070419A}. At very late times, there is a possible supernova bump \citep{Dai070125}. We derive the SED from a joint fit to the broken power-law portion, and find $\alpha_1=0.54\pm0.08$, $\alpha_2=1.42\pm0.04$, $t_b=0.0158\pm0.0014$ days and $n=10$ fixed (no host galaxy). These values are in good agreement with \cite{Melandri070419A}. For the SED, we use UVOT $v$ data, as the $V$ data from \cite{Melandri070419A} is too bright, probably due to the NOMAD calibration (A. Melandri, priv. comm.). The SED shows some scatter. We find the best fit is achieved with MW dust, though the preference is weak, our value ($\AV=0.32\pm0.27$) is in excellent agreement with that of \cite{Melandri070419A} ($\AV=0.37\pm0.19$ from a joint optical-X-ray fit), but only marginally in agreement with \cite{CenkoDark}, who find $\AV=0.70^{+0.31}_{-0.10}$.

\paragraph{GRB 070802, $z=2.4541$.}We constructed the light curve and the SED with data from the following sources: \cite{Kruehler070802}, and \cite{Eliasdottir070802}. The afterglow of this faint GRB features an early peak/flare peaking at 0.027 days \citep{Kruehler070802}. Late detections are sparse, and we need to extrapolate to determine the magnitude at one day after the GRB. The afterglow is highly reddened ($\beta_0\approx3$) and shows a clear 2175 {\AA} dust feature, the most distant detected so far except for the also highly extinct GRB 080607 \citep{Prochaska080607}. In agreement with \cite{Kruehler070802} and \cite{Eliasdottir070802}, we find that the SED is best fit (among the three dust laws we use) by a large amount of LMC dust. We find $\AV=1.18\pm0.19$, in combination with an intrinsic spectral slope $\beta=1.07\pm0.31$, which is in perfect agreement with the X-ray spectrum \citep[$\beta_X=1.02^{+0.17}_{-0.15},$][]{Eliasdottir070802}, implying that both lie on the same slope and the derived extinction is actually the lowest possible amount \citep[from an optical/NIR only spectrum, using the complete broadband spectrum can actually lead to lower values, see][]{Eliasdottir070802}. The high $\chi^2_\nu=2.84$ of the fit indicates that even LMC dust is not able to model the SED perfectly. This is a good candidate for the advanced dust model (``Drude'' model) of \cite{LiDust}, and \cite{Eliasdottir070802} find that the best fits are obtained from the spectrum and a Fitzpatrick-Massa parametrization. \cite{LiangLi} have used the ``Drude'' model and find that the extinction curve consists of a 2175 {\AA} bump similar to that of the MW combined with a UV extinction similar to the LMC. We use our SED (excluding the $H$ band, so we have eleven filters) and derive a very peculiar result: $\chi^2=6.62$ for 4 d.o.f., $F_0=5022677\pm30347844$, $\beta=-11.83\pm10.95$, $\AV=12.22\pm7.61$, $c_1=0.41\pm3.78$, $c_2=-0.82\pm4.46$, $c_3=-1.68\pm2.75$, $c_4=0.0031\pm0.0019$ ($F_0$ is the normalization. $\beta$ and $\AV$ are identical in usage to this paper, and the constants $c_{1\cdots4}$ determine the specific shape of the ``Drude'' functions, see \citealt{LiDust} for details). By eye, the fit is excellent, but clearly this result is unphysical, and reveals two fundamental problems of the ``Drude'' approach: Firstly, the function is too flexible, and can achieve good fitting results which are nonsensical in terms of the resulting parameters. Secondly, and related, the parameters are almost all unconstrained. This is mainly due to parts of the function being ``anchored'' by only very few data points at the blue and red ends of the SED. Fixing $\beta=1.02$, we find $\chi^2=10.15$ for 5 d.o.f., $F_0=590\pm1112$, $\AV=1.35\pm2.04$, $c_1=0.22\pm22.6$, $c_2=-0.60\pm31.8$, $c_3=-1.94\pm6.14$, $c_4=0.033\pm0.054$. These values are roughly comparable with those of \cite{LiangLi}, but they are all unconstrained. \cite{LiangLi} do not publish the errors of their values. This is, next to the more unsure extinction of GRB 060210, the largest extinction found in the sample, clearly exceeding all values from the pre-\emph{Swift} era (K06), though still much less than some very dark GRBs described in \cite{PerleyHosts2009} or the case of GRB 080607 \citep{Prochaska080607}.

\paragraph{GRB 071003, $z=1.60435$.}We constructed the light curve and the SED with data from the following sources: \cite{Perley071003}, and \cite{GuidorziGCN071003}. This very energetic GRB is discussed in detail in \cite{Perley071003}. The early light curve features a small chromatic flare, and at 0.04 days, the afterglow decay turns over into a large (but only sparsely sampled) rebrightening. The late decay is probably smooth but most measurements are low S/N due to a nearby bright star. Very late NIR AO observations reveal no underlying galaxy to deep levels \citep{Perley071003}. Construction of the SED of this GRB was complicated, as observations in different filters often do not occur simultaneously. We assume an achromatic evolution (excepting the small early flare) but can not conclusively show that this is the case. The SED is well-fit by SMC dust and moderately high extinction, in agreement with \cite{Perley071003}.

\paragraph{XRF 071010A, $z=0.985\pm0.005$.}We constructed the light curve and the SED with data from the following sources: \cite{Covino071010A}, and \cite{CenkoDark}. \cite{Covino071010A} present a detailed analysis of this XRF and its optical/NIR afterglow, which features an early rise, an energy injection at about 0.6 days and a late, steep decay, very probably post-jet break. We find the following parameters from a compound light curve: $\alpha_{rise}=-0.94\pm0.10$, $t_{peak}=0.005\pm0.0002$ days, $\alpha_{1}=0.73\pm0.01$. After the energy injection, we find: $\alpha_{EI}=1.01\pm0.22$, $t_b=1.08\pm0.10$ days, and $\alpha_{2}=2.14\pm0.04$. These values are in full agreement with \cite{Covino071010A}. We furthermore confirm that this GRB afterglow is moderately highly extinct, for the strongly preferred SMC dust, we find $\AV=0.64\pm0.09$, one of the highest values and clearest detections ($7.5\sigma$) in the sample.

\paragraph{XRF 071031, $z=2.6918$.}We constructed the light curve and the SED with data from the following sources: \cite{Kruehler071031}, \cite{FynboSpectra}, \cite{HaislipGCN071031}, \cite{StrohGCNR071031} as well as our own ANDICAM data set. The early afterglow of this XRF shows a complex evolution, with an early rise and several superposed flares which show color evolution \citep{Kruehler071031}. We derive the SED from data after 0.07 days, where the afterglow is mostly achromatic. We find an intrinsically blue SED with a small amount of SMC dust, which is strongly preferred. The magnitude at one day can only be derived via extrapolation and thus has a large error associated with it. ANDICAM limits at two days are not deep enough to allow any assertions about the existence of a break in the light curve.

\paragraph{GRB 071112C, $z=0.8227$.}We constructed the light curve and the SED with data from the following sources: \cite{FynboSpectra}, \cite{YuanGCN071112C}, \cite{WangGCN071112C}, \cite{KlotzGCN071112C}, \cite{BureninGCN071112C}, \cite{ChenGCN071112C1, ChenGCN071112C2}, \cite{DintinjanaGCN071112C}, \cite{OatesGCN071112C}, \cite{IshimuraGCN071112C}, \cite{GrecoGCN071112C}, \cite{SposettiGCN071112C}, \cite{YoshidaGCN071112C}, \cite{UemuraGCN071112C}, \cite{MinezakiGCN071112C}, and \cite{HuangGCN071112C}. This GRB afterglow was rapidly observed by many observatories on the western Pacific rim. It shows a plateau phase early on \citep{YuanGCN071112C, HuangGCN071112C}, after 0.003 days, we find a decay slope of $\alpha=0.93\pm0.02$ in the $R_C$ band. The SED is broad ($UVW1$ to $K$), shows slight curvature, is well fit by a small amount of SMC dust and shows a strong dropout in $UVM2$ and $UVW2$, in agreement with the low redshift \citep[z=0.8227,][]{FynboSpectra}.

\paragraph{GRB 080129, $z=4.349\pm0.002$.}We constructed the light curve and the SED with data from the following sources: \cite{Greiner080129}, and \cite{PerleyGCN080129} (we also present ANDICAM upper limits). This faint high-$z$ event had a remarkable light curve, the most striking feature being a huge optical/NIR flare that peaks at about 0.006 days after the GRB, a phenomenon that has not been seen in any other optical/NIR light curve. This flare is followed by a slow rise and a very long plateau phase extending up to 2 days after the GRB, where a steep decay sets in \citep{Greiner080129}. Even at late times, the afterglow exhibits small-scale variations, e.g. a flare at 0.82 days for which we find a clear spectral hardening. The early flare shows clear color evolution \citep{Greiner080129}, and we derive the SED from a joint fit of data after 0.78 days, where the afterglow is mostly achromatic. From this fit, we find parameters $\alpha_1=-0.15\pm0.06$, $\alpha_2=2.15\pm0.23$, $t_b=2.03\pm0.13$ days, and $n=10$ fixed (no host galaxy). The burst lay in the Galactic plane, with high foreground extinction ($\EBV\approx1$), after correcting for this, we find a straight SED with no sign of dust extinction. The very long plateau phase combined with the high redshift leads to this afterglow being the most luminous ever discovered at 0.5 days after the GRB in the host frame, even though especially the early afterglow is significantly fainter than most afterglows in the sample.

\paragraph{GRB 080210, $z=2.6419$.}We constructed the light curve and the SED with data from the following sources: \cite{FynboSpectra}, \cite{KlotzGCN080210}, \cite{KupcuYoldasGCN080210}, \cite{BrennanGCN080210}, \cite{UpdikeGCN080210}, \cite{PerleyGCN080210}, \cite{GrupeGCNR080210}, as well as one data point from the 1.5m Sayan telescope. Data on this GRB is rather sparse. It seems to show an early peak, after which the light curve decays with $\alpha=1.06\pm0.02$ (though the Sayan observation indicates a rebrightening feature). The SED is broad, encompassing ten filters from GROND $g$ to GROND $K$, very red and shows a spectral ``plateau'' in the GROND $i$ and GROND $z$ bands which may indicate the presence of a 2175 {\AA} bump. Indeed, while there is still some deviation from the model, the best fit is achieved by a quite large ($\AV=0.7$) amount of LMC dust. SMC dust shows strong residuals, and MW dust undercorrects the extinction. We caution that this fit is based on preliminary data from GROND only. The VLT spectrum also indicates significant reddening \citep{FynboSpectra}. The large extinction leads to a large $dRc$, making it this a very bright afterglow at early times.

\paragraph{XRF 080310, $z=2.4274$.}We constructed the light curve and the SED with data from the following sources: \cite{CenkoDark},  \cite{ChornockGCN080310},  \cite{CovinoGCN0803101, CovinoGCN0803102},  \cite{MilneGCN080310},  \cite{ChenGCN080310},  \cite{HoverstenGCN080310},  \cite{WozniakGCN080310},  \cite{PerleyGCN080310},  \cite{GarnavichGCN0803101, GarnavichGCN0803102},  \cite{YoshidaGCN080310},  \cite{YuanGCN080310},  \cite{UrataGCN080310},  \cite{WegnerGCN080310}, and \cite{HillGCN080310} as well as our own ANDICAM data set. This XRF has a complex optical light curve, with an early plateau phase which smoothly rolls over into a steeper decay, and a second flat phase from 1 to 2 days after the trigger, after which the light curve decays steeply. From a joint fit of the data up to 0.5 days, we find $\alpha_1=-0.12\pm0.05$, $\alpha_2=1.64\pm0.11$, $t_b=0.049\pm0.0041$ days, $n=0.83\pm0.18$, and $m_h=27$ fixed. The SED is broad and well-sampled, from $B$ to $K$ bands, and is well-fit with SMC dust and a small amount of extinction.

\paragraph{GRB 080319B, $z=0.9371$.}We constructed the light curve and the SED with data from the following sources: \cite{Bloom080319B},  \cite{Racusin080319B},  \cite{Wozniak080319B},  \cite{Tanvir080319B},  \cite{CenkoDark}, \cite{Pandey080319B}, \cite{SwanGCN080319B},  \cite{HentunenGCN080319B}, and \cite{Cenko060418}. GRB 080319B is optically the most extreme GRB ever observed. Observationally, the prompt optical flash \citep{Racusin080319B, Bloom080319B, SwanGCN080319B} reached 5th magnitude, earning it the moniker ``naked-eye GRB''. Due to the proximity of GRB 080319A which happened just 27 minutes earlier and 14 degrees away on the sky, it was observed even before the trigger by wide-field sky monitors \citep{Racusin080319B, Wozniak080319B}, allowing, for the first time, a sub-second sampling of the prompt optical light curve \citep{Beskin080319B}. The event also had the highest fluence of any GRB in the \emph{Swift} era including IPN GRBs which were not localized and followed up. In the last 20 years, only GRB 021206 has had a higher fluence \citep{Wigger021206}. The extremely bright early afterglow combined with favorable observing conditions lead to the second-best afterglow sampling ever (after GRB 030329), but at one day already, the afterglow is not especially luminous anymore. Concurrent with \cite{Bloom080319B} and \cite{Racusin080319B}, we find a very flat spectrum \citep[determined from 0.04 to 0.4 days after the GRB, there is strong color evolution earlier on,][]{Bloom080319B, Racusin080319B, Wozniak080319B} and no sign of any dust extinction. The early prompt flash is the most luminous source ever observed, and exceeds GRB 050904 \citep{Kann050904} by over a magnitude \citep{Bloom080319B}. The afterglow evolution is complex, with evidence for a late jet break \citep{Racusin080319B} as well as an underlying supernova \citep{Bloom080319B, Tanvir080319B}.

\paragraph{GRB 080319C, $z=1.9492$.}We constructed the light curve and the SED with data from the following sources: \cite{CenkoDark},  \cite{FynboSpectra}, \cite{WilliamsGCN080319C},  \cite{LiGCN080319C},  \cite{WrenGCN080319C},  \cite{PaganiGCNR080319C}, and \cite{PerleyHosts2009}. Due to GRB 080319B, this burst does not have extensive follow-up. At 0.004 days, it shows an optical flare also seen in the X-rays \citep{PerleyHosts2009}. These authors also report on an intrinsically luminous host galaxy. Starting at 0.01 days, the afterglow decays quite steeply ($\alpha=1.47\pm0.08$). As the data around one day at $z=1$ \citep{CenkoDark} are already strongly dominated by the host galaxy, we are unable to firmly deduce a magnitude at one day. We find moderately high extinction and a strong preference for SMC dust, in accordance with \cite{CenkoDark} and \cite{PerleyHosts2009}.

\paragraph{XRF 080330, $z=1.5115$.}We constructed the light curve and the SED with data from the following sources: \cite{Guidorzi080330},  \cite{YuanNanjing},  \cite{SchaeferGCN080330},  \cite{BloomGCN080330},  \cite{ImGCN080330},  \cite{WangGCN080330},  \cite{SchubelGCN080330},  \cite{WrenGCN080330},  \cite{BlockGCN080330}, and additional Seeing In The Dark Internet Telescope data (A. Block, priv. comm.) as well as our own data set (ANDICAM, Terskol 2m). The light curve of this X-ray flash is complex (and similar to that of XRF 080310), with initial rapid variability \citep{YuanNanjing, WrenGCN080330}, followed by a plateau phase, and then a gradual steepening of the decay. Finally, another break (possibly a jet break) may occur \citep{Guidorzi080330}. Fitting the data from 0.04 to 1 days, we find: $\alpha_1=0.09\pm0.02$, $\alpha_2=1.127\pm0.005$, $t_b=0.021\pm0.0005$ days, $\n=3.60\pm0.36$ (no host galaxy). We note a small amplitude optical flare from 0.2 to 1 day. The SED spans from $U$ to $K$ and consists of 15 data points. We find weak evidence for a preference of Milky Way type dust, but all three dust models fit well with a small amount of dust. Intrinsically, the afterglow is not very luminous.

\paragraph{GRB 080413A, $z=2.4330$.}We constructed the light curve and the SED with data from the following sources: \cite{YuanNanjing},  \cite{KlotzGCN080413A},  \cite{AntonelliGCN080413A},  \cite{FukuiGCN080413A}, as well as our own ANDICAM data set. We are able to jointly fit all afterglow data with a broken power-law with parameters $\alpha_1=0.64\pm0.03$, $\alpha_2=1.58\pm0.04$, $t_b=0.013\pm0.002$ days, and $n=2.5$ fixed, and no host galaxy. We find a strong preference for SMC dust (MW and LMC dust yield negative extinction) and low extinction (equal to 0 within errors).

\paragraph{GRB 080710, $z=0.8454$.}We constructed the light curve and the SED with data from the following sources: \cite{Kruehler080710}, \cite{LiGCN080710}, \cite{D'AvanzoGCN080710}, \cite{BersierGCN080710}, \cite{LandsmanGCN080710}, \cite{WeaverGCN080710}, \cite{PerleyGCN0807102}, and \cite{YoshidaGCN080710}. This afterglow showed a slow rise and multiple breaks \citep{Kruehler080710} and is one of the observationally brightest afterglows ever observed around 0.05 days. The low redshift of $z=0.8454$ \citep{FynboSpectra} and the small amount of dust extinction imply that it is much fainter intrinsically, though. Using data beyond the light curve peak at 0.03 days, a joint fit finds $\alpha_1=0.54\pm0.04$, $\alpha_2=1.60\pm0.01$, $t_b=0.110\pm0.003$ days, $n=3.68\pm0.82$, and we assumed no host galaxy. These values are in agreement with \cite{Kruehler080710}. A fit to the broad SED ($UVM2$ to $K_{G}$, 14 filters) shows that only a small amount of extinction is needed. While we can not discern between SMC and LMC dust, there is no evidence of a 2175 {\AA} bump.

\paragraph{GRB 080913, $z=6.733$.}We constructed the light curve and the SED with data from the following sources: \cite{Greiner080913}. This very high redshift GRB \citep[the record holder until the recent GRB 090423 at $z=8.2$, ][]{Tanvir090423, Salvaterra090423} has been extensively analyzed in \cite{Greiner080913}. The light curve shows a strong late-time rebrightening. In accordance with these authors, we find no evidence for dust. Intrinsically, the afterglow is fainter than the mean at early times, but the rebrightening causes it to become more luminous than the mean at late times.

\paragraph{GRB 080916C, $z=4.35\pm0.15$.}We constructed the light curve and the SED with data from the following sources: \cite{Greiner080916C}. This extremely energetic GRB was detected by Fermi GBM/LAT and extensively discussed by \cite{Abdo080916C}. The optical observations are sparse and only a photo-$z$ (albeit with good precision) is known \citep{Greiner080916C}. We find no evidence for any dust along the sightline. The afterglow decays with $\alpha=1.40\pm0.10$. Both results are in perfect agreement with \cite{Greiner080916C}. The afterglow, despite the extremely energetic prompt emission, is not overly luminous and comparable to most afterglows in the sample. We note that if the LAT emission of this GRB is dominated by external shock emission \citep{Kumar080916C}, the isotropic energy release may be overestimated, on the other hand, the bolometric bandpass should be dominated by the true prompt emission, implying that the correction, if any, is small. This changes if the energetics are computed up to 10 GeV in the rest frame, though \citep{Amati080916C}.

\paragraph{GRB 080928, $z=1.6919$.}We constructed the light curve and the SED with data from the following sources: \cite{Rossi080928}, \cite{FynboSpectra}, and \cite{FerreroGCN080928} (see \citealt{Rossi080928} for further details). The light curve evolution and the SED analysis are reported in \cite{Rossi080928}, who find a strongly variable early light curve with multiple peaks and a steep late decay, as well as a rather red SED which is best fit by a small amount of MW dust; the underlying spectral slope is still steep, and in full agreement with the X-ray spectral slope.

\paragraph{GRB 081008, $z=1.9685$.}We constructed the light curve and the SED with data from the following sources: \cite{Yuan081008}, as well as our own ANDICAM data set. \cite{Yuan081008} report an initially rising afterglow with small-scale variability (probably linked to central engine activity) superposed, which then goes over into a smooth decay which later shows a break. From a multicolor joint fit, we find $\alpha_1=0.931\pm0.006$, $\alpha_2=1.132\pm0.008$, $t_b=0.048\pm0.004$ days, $n=10$ fixed, and a faint host galaxy assumed. Our decay slopes agree very well with \cite{Yuan081008}, but we find an earlier break. Late upper limits hint that a further break may have occurred. The SED shows a bit of scatter, we find it is best fit by a small amount of SMC dust and an intrinsically blue slope. \cite{Yuan081008} find a break in the optical SED, with a larger amount of dust and $\beta\approx0$, we can not confirm this.

\paragraph{GRB 090102, $z=1.547$.}We constructed the light curve and the SED with data from the following sources: \cite{Gendre090102}, \cite{deUgarteGCN0901021, deUgarteGCN0901022}, \cite{CurranGCN090102}, \cite{CenkoGCN090102}, \cite{MalesaniGCN090102}, and \cite{LevanGCN090102}. We set $T_0$ as 14 seconds before the Swift trigger. This GRB shows an early steep decay, which may be due to a reverse shock flash, caught in high time resolution by TAROT \citep{Gendre090102}. Note that this refined analysis comes to a very different result from \cite{KlotzGCN090102}, who found a rapid rise to a peak. The reverse shock flash interpretation is further supported by the recent report of significant polarization in the early optical afterglow \citep{Steele090102}. \cite{Gendre090102} fit the complete, well-sampled light curve with a steep-to-shallow broken power law. Using a similar fit, we find $\alpha_1=1.79\pm0.09$, $\alpha_2=1.10\pm0.02$, $t_b=0.004\pm0.00075$ days, $n=-10$ fixed, in decent agreement with the values from \cite{Gendre090102}: $\alpha_1=1.50\pm0.06$, $\alpha_2=0.97\pm0.03$, $t_b\approx0.012$ days. But there are clear residuals seen between 0.01 and 0.1 days, and we find the light curve can be fit by a total of four power law segments: $\alpha_1=1.92\pm0.03$, $t_{b1}=0.0060\pm0.0016$ days, $\alpha_2=1.54\pm0.13$, $t_{b2}=0.0167\pm0.0064$ days, $\alpha_3=0.86\pm0.05$, $t_{b3}=0.200\pm0.027$ days, $\alpha_4=1.38\pm0.04$ ($n=-10$ or $n=10$ fixed), $m_h=24.2\pm0.12$ mags. \cite{Gendre090102} have already noted that the afterglow behavior does not agree with the standard fireball model, if this more complex evolution is correct, the problem becomes even more persistent. We furthermore confirm no (jet) break out to several days. Both afterglow behavior and the SED are comparable to those of GRB 061126, according to \cite{Gendre090102}, and we also confirm this in terms of the SED. The moderately red, well-sampled SED is best fit by a small amount of SMC dust (with LMC dust yielding similar results), with an intrinsic spectral slope in good agreement with the blue X-ray slope \citep{Gendre090102}. But \cite{Gendre090102} find a $\beta_{OX}=0.53$ which makes this almost a dark burst, implying either very gray extinction or an additional emissive component in the X-rays. They find higher extinction values for MW and LMC dust, but their SMC result is in agreement with our value within errors. Note that the redshift of this GRB is almost identical to that of GRB 070125.

\paragraph{GRB 090313, $z=3.3736\pm0.0004$.}We constructed the light curve and the SED with data from the following sources: \cite{UpdikeGCN090313},  \cite{GuidorziGCN090313},  \cite{NissinenGCN090313},  \cite{deUgarteGCN0903131, deUgarteGCN0903132},  \cite{MorganGCN090313},  \cite{VaalstaGCN090313},  \cite{PerleyGCN0903131},  \cite{KlotzGCN090313},  \cite{PerleyGCN0903132},  \cite{CobbGCN090313}, and \cite{MaioranoGCN090313}. This moderately high redshift \emph{Swift} GRB has an exceptionally luminous afterglow. The afterglow evolution is complex, with an early rise and a possible long plateau phase \citep{PerleyGCN0903133, deUgarteGCN0903131} and a late steep decay \citep{deUgarteGCN0903132, CobbGCN090313}. We find that SMC dust is strongly preferred and the extinction is quite high for a burst at $z>3$, with $\AV=0.34\pm0.15$. From about 0.02 to 0.5 days (at $z=1$), this is the most luminous afterglow ever detected.

\paragraph{GRB 090323, $z=3.568\pm0.004$.}We constructed the light curve and the SED with data from the following sources: \cite{McBreenLAT}, \cite{CenkoLAT}, \cite{WangGCN090323}, \cite{PerleyGCN0903231, PerleyGCN0903232}, \cite{GuidorziGCN090323} as well as our own data set (RTT150, NOT, Shajn). We find that until 5 days after the GRB, the decay is described well with a power-law $\alpha=1.93\pm0.02$, before a plateau phase sets in \citep{deUgarteGCN090323, McBreenLAT}, before returning to a similar decay rate as before. \cite{McBreenLAT} give evidence that this is a post-jet-break decay slope, this is contested in \cite{CenkoLAT}. This burst was extremely energetic \citep{Bissaldi090323090328} \citep[within the rest-frame 1 - 10000 keV band, it has the highest isotropic energy in the entire sample, larger even than that of GRB 080916C, see also][]{Amati080916C} and was detected by Fermi LAT up to several kiloseconds after the event \citep{OhnoGCN090323}. We find that the SED is well-fit by a small amount of SMC dust, which is strongly preferred. The afterglow is intrinsically highly luminous, among the brightest in the sample.

\paragraph{GRB 090328, $z=0.7354\pm0.0003$.}We constructed the light curve and the SED with data from the following sources: \cite{McBreenLAT}, \cite{CenkoLAT}, and \cite{AllenGCN090328}. \cite{McBreenLAT} report the afterglow can be fit with a steep decay and a late-time, achromatic bump before going over into a bright host galaxy. We find that it can be fit equally well with a steep-to-shallow transition, with the following parameters: $\alpha_1=2.36\pm0.50$, $\alpha_2=1.36\pm0.26$, $t_b=2.33\pm0.61$ days, $n=-10$ fixed, and host galaxy magnitudes left free for each filter except for $K_{G}$ and $i^\prime$. The steep decay is consistent with a post-jet-break decay \citep[but see][]{CenkoLAT}, and there is marginal evidence for a more shallow decay at one day and earlier. The SED is red but straight, and we find that we can not discern between MW, LMC and SMC dust, in full agreement with what was found by \cite{McBreenLAT}. For SMC dust, we find $\beta=1.17\pm0.17$, $\AV=0.18\pm0.13$, indicating that the cooling break lies redwards of the optical band.

\paragraph{GRB 090423, $z={8.23}^{+0.06}_{-0.07}$.}We constructed the light curve and the SED with data from the following sources: \cite{Tanvir090423}, and \cite{YoshidaGCN090423}. With a redshift of $z={8.23}^{+0.06}_{-0.07}$ \citep{Tanvir090423}, this is the most distant spectroscopically confirmed source in the universe \citep[see also][]{Salvaterra090423}. It is only detected in the NIR $JHK$ bands (therefore, we can not do free dust model fits, and the GRB is not listed in Table \ref{tabALL}). From a joint fit of all data \citep[excepting the last detections, which show a flare,][]{Tanvir090423}, we find, as \cite{Tanvir090423}, an early plateau. It is $\alpha_1=0.063\pm0.052$, $\alpha_2=1.76\pm0.31$, $t_b=0.38\pm0.06$ days ($t_b=0.041\pm0.0066$ days in the rest frame), and $n=10$ was fixed, no host galaxy was assumed. These values are broadly in agreement with those of \cite{Tanvir090423}. From the joint fit, we find a flat SED, $\beta_0=0.45\pm0.13$, also in broad agreement with \cite{Tanvir090423}. \cite{CharyGCN090423} report a $2\sigma$ detection at 46 days from an ultra-deep $3.6 \mu m$ Spitzer observation. We find, from extrapolation of our SED, that the color $J_{Vega}-3.6\mu m_{AB}=-0.3$ mag, and $J_{Vega}-3.6\mu m_{AB}=-0.18$ mag from a (long!) extrapolation of the light curve decay. Both values are in good agreement, as \cite{CharyGCN090423} also report, indicating an as yet negligible contribution by the host galaxy, and no jet break until at least 5 days post-burst in the rest frame. There is no evidence for dust. While the latter fact is similar to the two other $z\geq6$ GRBs, GRB 050904 and GRB 080913, both of those showed straight but red SEDs, with $\beta_0\approx1.1$. The flat spectral slope leads to a blue $R_C-J$ color, we find that the composite $J$-band light curve needs to be dimmed by 0.98 magnitudes to derive the $R_C$ light curve for the case of a hypothetically fully ionized universe \citep[see][]{Kann050904}. The $dRc$ correction is also smaller than for the two closer GRBs mentioned above, and at $z=1$, we find the light curve to be quite unremarkable,, though it is brighter than most GRBs at one day due to the optical flare. A later deep upper limit \citep{Tanvir090423} indicates that this flare is not similar to the strong rebrightenings experienced by the afterglow of GRB 080913 \citep{Greiner080913}, though. All in all, in terms of duration, isotropic energy release and optical luminosity, GRB 080913 and GRB 090423 are similar, in stark contrast to the very powerful GRB 050904.

\paragraph{GRB 090424, $z=0.544$.}We constructed the light curve and the SED with data from the following sources: \cite{YuanGCN090424},  \cite{XinGCN090424},  \cite{SchadyGCN090424},  \cite{GuidorziGCN090424},  \cite{UrataGCN090424},  \cite{OlivaresGCN090424},  \cite{NissinenGCN090424},  \cite{ImGCN090424},  \cite{RoyGCN090424},  \cite{MaoGCN090424},  \cite{CobbGCN090424}, and \cite{RumyantsevGCN090424}, as well as our own TLS, CAHA and Maidanak data. This was, in terms of peak photon flux \citep{SakamotoGCN090424}, the brightest burst \emph{Swift} has detected so far, mainly due to it lying at a comparatively low redshift \citep{ChornockGCN090424}, with a host galaxy that has been pre-imaged in the SDSS \citep{EvansGCN090424}. We find from fitting a composite light curve that the early afterglow may show the signature of a reverse shock flash: $\alpha_1=2.12\pm0.16$, $\alpha_2=0.92\pm0.006$, $t_b=0.002\pm0.0001$ days, $m_h=22.77\pm0.12$, $n=-10$ was fixed. From the available data, we find no evidence for a jet break, which was suggested by \cite{ImGCN090424}. We derive a very broad ($UVW1$ to $K$) SED which is clearly reddened ($\beta_0=1.58\pm0.03$). The best fit is achieved with LMC dust and a moderate extinction of $\AV=0.50\pm0.12$ mag. The intrinsic SED is still red ($\beta=0.97\pm0.15$) and in excellent agreement with the X-ray spectrum \citep[from the \emph{Swift} XRT repository,][]{EvansXRT1, EvansXRT2}, indicating that the cooling break lies redward of the optical.

\paragraph{GRB 090902B, $z=1.8229\pm0.0004$.}We constructed the light curve and the SED with data from the following sources: \cite{McBreenLAT}, \cite{Pandey090902B}, and \cite{CenkoLAT}. This extremely energetic GRB has the third-highest fluence of all GRBs in the \emph{Swift} era, featured the highest observed energy ever for a GeV photon, and shows a high-energy spectrum consisting of a superposition of a Band function and a power-law rising to GeV energies which also dominates below 50 keV \citep{Abdo090902B}. It is the first \emph{Fermi} GRB for which the afterglow was detected (by ROTSE) before the LAT position (or a \emph{Swift} XRT position) was published, by mosaicing the GBM ground position. This detection shows the afterglow was probably dominated in the first hours by a reverse shock flash \citep{Pandey090902B}. We find $\alpha_{RS}>1.8$, $\alpha_1=0.96\pm0.02$, $t_b<0.46$ days. Fitting only data after 0.5 days, we find $\alpha_1=0.95\pm0.03$, $\alpha_2=1.37\pm0.17$, $t_b=10.97\pm3.91$ days, $n=10$ fixed, and a faint host galaxy assumed. This late break is probably a jet break \citep{McBreenLAT, CenkoLAT}; while the post-break decay slope is very shallow, it is derived from only a few data points (the GRB became unobservable), so it is entirely possible that we are seeing a smooth break which has not reached its asymptotic decay value yet. This very late break makes the GRB hyper-energetic according to \cite{McBreenLAT}, though note that \cite{CenkoLAT} find evidence for a very low circumburst density which reduces the energy requirements. The SED disfavors MW dust, but we can not discern between LMC and SMC dust (we work with the latter). The very slow decay puts the afterglow among the most luminous ones at late times \citep{McBreenLAT}.

\paragraph{GRB 090926A, $z=2.1062\pm0.0004$.}We constructed the light curve and the SED with data from the following sources: \cite{Rau090926A} and \cite{CenkoLAT}. This was another high-fluence \emph{Fermi} LAT/GBM GRB. The afterglow is observationally very bright, and began a powerful rebrightening phase \citep{Rau090926A, CenkoLAT} just as ground-based observations started. In this way, it strongly resembles GRB 070125 \citep{Updike070125}. Fitting data from 0.8 to 2 days, we find $\alpha_1=-2.27\pm0.13$, $\alpha_2=1.64\pm0.01$, $t_b=0.947\pm0.003$ days, $n=10$ fixed. The afterglow then undergoes two further rebrightenings, the first which is also found by \cite{Rau090926A}. As the decay slopes after each rebrightening are identical within errors, this is a classical example of multiple energy injections into the forward shock. Joining the two large data sets from \cite{Rau090926A} and \cite{CenkoLAT}, we concur with \cite{CenkoLAT} \citep[in disagreement with][]{Rau090926A} that the late afterglow shows another break; using only data after the last rebrightening (4.8 days after the GRB), we find: $\alpha_3=1.53\pm0.32$, $\alpha_4=2.09\pm0.17$, $t_b=7.96\pm4.20$ days, $n=10$ fixed. This is probably a jet break, together with that of GRB 090902B one of the latest discovered, again implying that this is a hyper-energetic event \citep{Rau090926A, CenkoLAT}. Concerning the SED, we note that there seems to be a discrepancy between the calibrations from \cite{CenkoLAT} (unextincted slope for their data only: $\beta_0=1.53\pm0.08$) and \cite{Rau090926A} (unextincted slope for their data only: $\beta_0=0.89\pm0.05$, in agreement with their result). Joining all data yields a usable SED, but a high $\chi^2$ due to the scatter, we find it is best fit by a small amount of SMC dust, in general agreement with \cite{Rau090926A}. After shifting it to $z=1$, the afterglow is seen to be among the most luminous ever detected at any time during its observations.

\subsection{Details on the \emph{Swift}-era Silver Sample}
\label{AppB}

\paragraph{GRB 050401, $z=2.8992\pm0.0004$.}We constructed the light curve and the SED with data from the following sources: \cite{Rykoff050401}, \cite{DePasquale050401}, \cite{Watson050401}, \cite{Kamble050401} and \cite{ChenHost} (we also publish upper limits from ANDICAM in this work). We note that the three magnitude values given in \cite{Kamble050401} are about one magnitude too bright. For the SED, we derive the $V-R_C$ and $R_C-I_C$ colors from the \cite{Kamble050401} data only, as the colors are in agreement with the rest of the SED. Construction of the SED is hampered by the fact that the NIR data ($JHK$) are for the most part not contemporaneous with the densely sampled part of the $R_C$ light curve. The SED we derive yields negative $\beta$ for a free fit (LMC and SMC), or strongly unphysical results (MW). While being very red ($\beta_0=1.88\pm0.11$), the curvature is too strong to be explained even by SMC dust. As already noted earlier \citep{Rykoff050401, Watson050401}, the light curve is reasonably well fitted ($\chi^2=35.2$ for 28 degrees of freedom) with a simple power law, we find $\alpha=0.84\pm0.02$. We point out that if one fits only the later $R_C$ data, without the early ROTSE points \citep{Rykoff050401}, a broken power law gives a significantly improved fit ($\chi^2=12.5$ for 21 degrees of freedom). We find the following results: $m_k=22.43\pm0.52$ mag, $\alpha_1=0.58\pm0.11$, $\alpha_2=0.96\pm0.04$, $t_b=0.32\pm0.17$ days, and $n=10$, $m_h=27$ were fixed \citep[the host galaxy magnitude is given in][]{ChenHost}. The early time (2 hours in the rest frame) and the small $\Delta\alpha=0.38\pm0.12$, close to the expected value $\Delta\alpha=0.25$ may indicate that this is a cooling break. This would also mean that the early light curve shows a more complicated evolution than a single power law. We find no evidence for a jet break at late times, in agreement with \cite{Watson050401}. Using this $\alpha$ value (what is labeled $\alpha_2$ in our fit represents the pre-break decay slope $\alpha_1$), and assuming a wind environment and a cooling frequency above the optical band, we derive $\beta=0.30$. While this is untypically shallow, the X-ray spectrum is also quite hard \citep[$\Gamma=1.85-1.89$,][]{Watson050401}, yielding a similar spectral slope ($0.35-0.39$, from $\beta=\Gamma-1-0.5$) to what we derive from the decay slope. Using this fixed value, we find $\AV=0.69\pm0.02$ (we caution that this strongly underestimates the error, see K06). This is close to the result \cite{Watson050401} derive, they fix $\beta=0.39$ and find $\AV=0.67$ from the spectrum. This is a rare GRB at high redshift where significant extinction has been detected.

\paragraph{GRB 051109A, $z=2.346$.}We constructed the light curve and the SED with data from the following sources: \cite{Yost0511}, \cite{HollandGCN051109A}, \cite{BloomGCN051109A}, \cite{MilneGCN051109A}, \cite{HaislipGCN051109A}, \cite{HuangGCN051109A}, \cite{WozniakGCN051109A}, \cite{LiGCN051109A}, \cite{MisraGCN051109A}, and \cite{KinugasaGCN051109A}, as well as our own data (one point from the 0.38m K-380 telescope). While this GRB had one of the densest early multi-color follow-ups ever obtained, most of this data are as yet unpublished, therefore our analysis is preliminary. The $R_C$ light curve is well fit with a broken power law with the following parameters: $m_k=19.99\pm0.099$ mag, $\alpha_1=0.64\pm0.006$, $\alpha_2=1.88\pm0.17$, $t_b=1.44\pm0.16$ days, $m_h=23.37\pm0.15$ mag, and $n=1$ (smooth rollover) fixed. This is one of the best examples of an optical light curve with a probable jet break in the \emph{Swift} era. Data in other colors is sparse. We derive a broad SED ($VR_CI_CJHK_S$), but it is strongly curved, free fits yield negative $\beta$ (LMC, SMC) or unphysical results (MW). A fit without extinction yields $\beta_0=0.61\pm0.057$. Deriving $\beta$ from $\alpha$, we find that the only physical solution is found for an ISM-BLUE model, with $\beta=0.42$. For this case, $p=1.85\pm0.012$, identical to $p=1.88\pm0.17$ derived from $p=\alpha_2$. SMC dust gives the significantly best fit, we find $\AV=0.09\pm0.03$. \cite{OatesUVOT} find $\AV=0.01$ from UVOT data only (using an XRT-UVOT joint fit), a smaller value. \cite{Schady2010} find strong evidence for a cooling break between the optical and the X-rays, but are not able to discern between the extinction models beyond that, they derive low upper limits on any extinction, in agreement with our result.

\paragraph{GRB 051111, $z=1.54948\pm0.00001$.}We constructed the light curve and the SED with data from the following sources: \cite{Yost0511}, \cite{Butler051111}, \cite{Guidorzi051111}, \cite{MilneGCN051111}, \cite{PooleGCN051111}, \cite{NanniGCN051111}, and \citet[][host galaxy]{Penprase051111}, as well as our own data set (SARA, Maidanak). Similar to GRB 051109A, this GRB was observed in multiple colors at early times, and the $R_C$ light curve is very well sampled. Basically, it can be described by a broken power law with the following parameters: $m_k=19.64\pm0.34$ mag, $\alpha_1=0.91\pm0.006$, $\alpha_2=2.32\pm0.80$, $t_b=0.55\pm0.18$ days, $m_h=25.6$ mag and $n=10$ fixed. The late slope indicates the existence of a jet break, but there are few late data points, and the error on $\alpha_2$ is very large. Furthermore, the light curve residuals show significant variability ($\chi^2=542$ for 118 degrees of freedom), e.g., the earliest data points follow a steeper decay \citep[cf.][]{Yost0511}. \cite{Butler051111} also present an early, shallow break. Another similarity to GRB 051109A is that this SED is also strongly curved. Only SMC dust yields an acceptable result, but the fit finds negative $\beta$. Assuming a constant density environment and a cooling break bluewards of the optical yields $\beta=0.61$, fixing this value, we find $\AV=0.19\pm0.02$. This is fully in agreement with the result of \cite{Butler051111}, who find $\AV=0.23\pm0.07$ for SMC dust, which they find is clearly preferred.

\paragraph{GRB 060210, $z=3.9133$.}We constructed the light curve and the SED with data from the following sources: \cite{Stanek060206}, \cite{Curran060210}, \cite{CenkoDark}, \cite{PerleyHosts2009}, \cite{FynboSpectra}, \cite{LiGCN060210}, \cite{WilliamsGCN060210}, and \cite{HeartyGCN060210}. This GRB at a moderately high redshift was rapidly observed by several ground-based telescopes, yielding a dense early light curve coverage. At early times, the afterglow shows a plateau, which is followed by a small rebrightening \citep{LiGCN060210}, before it decays achromatically with $\alpha=1.29\pm0.02$ \citep[fully in agreement with, e.g.,][]{Stanek060206}. Note that this GRB shows rather strong pre-trigger activity \citep{Curran060210}, and we shifted all afterglow data by 230 seconds. With the exception of a quite bright host galaxy \citep{PerleyHosts2009}, a bright detection which is is in conflict with surrounding upper limits \citep{MisraGCN060210} and the late $K_S$ observation \citep{HeartyGCN060210}, there are no reported detections after $\approx0.1$ days. The $K_S$ data point can only be added to the SED if one does a long extrapolation of the initial light curve decay, assuming achromaticity (we also assume $K_S=22$ for the host galaxy, but note that \citealt{HeartyGCN060210} report possible blending with an extended source). In such a case, one finds a large $R_C-K_S=5$ mag, but this is still $\approx2$ magnitudes lower than what is expected from the fit only to the optical data (see the following). \cite{DaiGCN060210} report the detection of a light curve break at 7.9 hours in the XRT light curve, which, if achromatic, would imply a larger $R_C-K_S$ color. Due to these insecurities, we therefore choose not to use the $K_S$ point in the SED, and point out that there may also be a problem with the photometric calibration of the P60 data of \cite{CenkoDark}, as their $i^\prime$ magnitudes are already significantly brighter (instead of slightly fainter, as expected) than $I_C$ data from \cite{Curran060210} and \cite{LiGCN060210}. As already found by \cite{CenkoDark}, the SED is extremely red \citep[$\beta_0=7.09\pm0.52$, in agreement with][]{CenkoDark}, this cannot be attributed to the redshift. Spanning only from $R_C$ to $z^\prime$, a free fit results in negative $\beta$ for the SMC or too little correction for LMC and MW dust. We can rule out MW dust from the bright $z^\prime$ detection, but can not discern between SMC and LMC dust as no NIR data has been published. We choose SMC dust as this is most often found in GRB hosts (and also needs less extinction). From the optical decay slope, and assuming an ISM environment with the cooling break blueward of the optical, we infer $\beta=0.76$ and a very high extinction $\AV=1.18\pm0.10$. This is fully in agreement with the result of \cite{CenkoDark}, $\AV=1.21^{+0.16}_{-0.12}$. This result is remarkable, not only is there only one other GRB in the sample for which a similarly high extinction was found (GRB 070802), but the high redshift leads to the highest correction in the whole sample, $dRc\approx10$. If this value is correct, then GRB 060210 has not only one of the most luminous early afterglows ever discovered (Fig. \ref{Bigfig2}), it also becomes, by 0.005 days after the GRB, the most luminous of all afterglows, roughly one magnitude brighter than the upper boundary now given by GRBs 090313 and 080129. This either implies that afterglows, even at times when prompt flashes like those of GRBs 050904 and 080319B have significantly decayed, can be much more luminous than is known now, or that there is a problem with the data. Indeed, \cite{CenkoDark} state that they did not derive an independent photometric calibration of the field, which may lead to especially the $z^\prime$ magnitudes to be too bright. Though as long as no additional data are published, especially NIR data, there is no way to discern between the two scenarios.

\paragraph{GRB 060502A, $z=1.5026$.}We constructed the light curve and the SED with data from the following sources: \cite{CenkoDark}, \cite{FynboSpectra}, \cite{PooleGCN060502A}, and \cite{JakobssonGCN060502A}. Data on this afterglow is sparse. From a joint fit, we find a shallow decay, $\alpha=0.56\pm0.17$. The SED is red \citep[$\beta_0=1.99\pm0.35$, in agreement with][]{CenkoDark}. Free fits result in negative $\beta$. We take $\beta=1.01$ from the X-ray spectrum \citep[from the \emph{Swift} XRT repository,][]{EvansXRT1, EvansXRT2} and find, with SMC dust, $\AV=0.38\pm0.15$. If we assume a cooling break redward of the optical, we derive $\beta=0.71$ and $\AV=0.50\pm0.15$, comparable to \cite{CenkoDark}, who derive $\AV=0.53\pm0.13$. \cite{Schady2010} strongly rule out MW dust, and marginally prefer SMC dust and a cooling break between the optical and the X-rays, for this model, they derive an extinction identical to our result.

\paragraph{GRB 060906, $z=3.685\pm0.002$.}We constructed the light curve and the SED with data from the following sources: \cite{CenkoDark}, \cite{FynboSpectra}, and \cite{LiGCN060906}. This moderately high redshift GRB has a peculiar early afterglow featuring a strong optical rebrightening \citep{CenkoDark}. Only three colors are given, as the Gunn $g$ band is affected by Lyman absorption. From the X-ray spectrum \citep[from the \emph{Swift} XRT repository,][]{EvansXRT1, EvansXRT2} and under the assumption of a cooling break between the X-rays and the optical (the uncorrected optical slope is already bluer than the X-ray slope), we use $\beta=0.61$ and find only a very small amount ($\AV=0.05\pm0.05$) of SMC dust. This value is smaller than what was found by \cite{CenkoDark}, $\AV=0.20^{+0.01}_{-0.12}$.

\paragraph{GRB 060927, $z=5.4636$.}We constructed the light curve and the SED with data from the following sources: \cite{Ruiz-Velasco060927}, \cite{ToriiGCN060927}, and \cite{Zheng060927} (we also publish an unused upper limit). This GRB has the fourth highest spectroscopic redshift determined to date, and incidentally, it also had the most rapid follow-up observation ever \citep{Ruiz-Velasco060927}, excepting the coincidental observations of GRB 080319B. Similar to GRB 050904, observations in most colors yielded only upper limits due to Lyman dropout. After a strong rebrightening at 0.01 days \citep{Ruiz-Velasco060927}, the light curve follows a simple power law, we find $m_k=23.394\pm0.118$ and $\alpha=1.235\pm0.033$ ($m_h=28$ fixed), in agreement with \cite{Ruiz-Velasco060927}. We construct the $R_C$ light curve following the method \cite{Kann050904} used for GRB 050904. Only three colors are unaffected by Lyman damping. From the decay slope, we derive $\beta=0.823$ from a model where the cooling break lies redward of the optical (the unextinct slope is quite steep, $\beta=1.297$), and find $\AV=0.209\pm0.084$ for SMC dust. This is also in agreement with the simultaneous optical/X-ray fits presented in \cite{Ruiz-Velasco060927}. The additional extinction correction makes this GRB afterglow have an even larger $dRc$ shift than found for GRB 050904 \citep{Kann050904}.

\paragraph{GRB 070208, $z=1.165$.}We constructed the light curve and the SED with data from the following sources: \cite{MelandriROBONET},  \cite{CenkoDark},  \cite{BloomGCN070208},  \cite{HalpernGCN070802},  \cite{WrenGCN070208},  \cite{SatoGCNR070208},  \cite{SwanGCN070208}, as well as one data point from the Sayan telescope. This GRB had a faint afterglow which exhibited an early rise \citep{WrenGCN070208, MelandriROBONET} and a slow decay, we find $\alpha=0.55\pm0.02$ from a joint fit. The SED is red \citep[$\beta_0=2.27\pm0.12$, in agreement with][]{CenkoDark}, and, under the assumption of a cooling break between the X-rays and the optical, the X-ray spectrum \citep[from the \emph{Swift} XRT repository,][]{EvansXRT1, EvansXRT2} yields $\beta=0.66$ and we derive $\AV=0.74\pm0.03$ with SMC dust. Similar values are found if we assume a cooling break redward of the optical and derive the spectral index from the optical decay slope. \cite{CenkoDark} derive a slightly higher value, $\AV=0.96\pm0.09$.

\paragraph{GRB 071020, $z=2.1462$.}We constructed the light curve and the SED with data from the following sources: \cite{CenkoDark}, \cite{FynboSpectra}, \cite{SchaeferGCN071020},  \cite{YuanGCN071020},  \cite{BloomGCN071020},  \cite{XinGCN071020},  \cite{ImGCN071020},  \cite{HentunenGCN071020}, and \cite{IshimuraGCN071020}, as well as our own data set (SARA, Z-600 and AZT-8). This very bright GRB had a bright early afterglow \citep{SchaeferGCN071020} and featured a strong rebrightening or plateau phase after about 0.2 days \citep{ImGCN071020}. Excluding the very first ROTSE point, we find the afterglow decays with $\alpha=1.11\pm0.07$, and, after $\approx0.2$ days, rises with $\alpha_R\approx-0.4$. \cite{YuanGCN071020} report a steeper early decay slope of $\alpha=1.52$, which may point to an additional component from a reverse shock flash. The X-ray spectrum \citep[from the \emph{Swift} XRT repository,][]{EvansXRT1, EvansXRT2} is hard, with $\beta\approx0.8$. Assuming no cooling break between the X-rays and the optical, we find, with SMC dust, $\AV=0.28\pm0.09$. The uncorrected SED is moderately reddened, it is $\beta_0=1.47\pm0.21$.

\paragraph{GRB 071025, $z=4.8\pm0.4$.}We constructed the light curve and the SED with data from the following sources: \cite{Perley071025}, and \cite{JiangGCN071025}. This afterglow has been extensively discussed in \cite{Perley071025}. No spectroscopic redshift is know, but photometric constraints as well as the trace of a very low-S/N Keck HIRES spectrum \citep{FynboSpectra} indicate $z=4.4-5.2$, and \cite{Perley071025} use $z=5$ in their discussion, which we adopt. They find a complex, double-peaked light curve. Fitting the two peaks separately (but using all available filters in each case), we find $\alpha_1=-1.39\pm0.18$, $\alpha_2=1.89\pm0.30$, $t_b=0.0073\pm0.0008$ days, $n=1$ fixed, and $\alpha_3=-5.10\pm0.58$, $\alpha_4=1.42\pm0.02$, $t_b=0.0138\pm0.0002$ days, $n=1$ fixed, for the first and second peak, respectively. These values are in rough agreement with those found by \cite{Perley071025}. Those authors find that the SED, which is well-determined in the NIR, shows a plateau phase between the $J$ and $H$ bands, which we fully confirm from both our early and late SED (which also show a different color, like \citealt{Perley071025} find). They are unable to fit the SED with any typical dust model (which we also confirm), but find that it can be fit well with SN-synthesized dust \citep{MaiolinoDust}. To correctly derive the intrinsic luminosity of the afterglow, we used the dust extinction in the individual filters $A_\lambda$ (it is $A_{I_C}=1.98$ mag, $A_Y=1.29$ mag, $A_J=1.14$ mag, $A_H=1.18$ mag, and $A_K=0.95$ mag, D. A. Perley, priv. com.). Correcting for these values, we find a straight SED with $\beta=0.93\pm0.05$, in accordance with \cite{Perley071025}, who find $\beta=0.96\pm0.14$. Extrapolating this to the $R_C$ band, we find $A_{R_C}=3.74$ mag, note that this is a combined effect of Lyman $\alpha$ absorption as well as dust reddening, the two can not be disentangled due to insecurities in the dust model (D. A. Perley, priv. com.). We find $dR_C=4.20$ mag from $\beta=0.93\pm0.05$, $z=5$ and no dust, for a total shift of $dR_c=7.94$. This large correction makes the afterglow of GRB 071025 one of the most luminous known at early times.

\paragraph{GRB 071122, $z=1.14$.}We constructed the light curve and the SED with data from the following sources: \cite{CenkoDark}, and \cite{BrownGCN071122}. This GRB had a faint afterglow which exhibited a slow rise and long plateau phase \citep{CenkoDark}. There are not enough filters to freely fit the SED. The X-ray spectrum \citep[from the \emph{Swift} XRT repository,][]{EvansXRT1, EvansXRT2} is hard \citep[though this may be influenced by a flare,][]{StamatikosGCNR071122}, with $\beta=0.66$. Assuming no cooling break between the X-rays and the optical, we find, with SMC dust, $\AV=0.22\pm0.23$, significantly less than \cite{CenkoDark}, who derive $\AV=0.58\pm0.05$. This is puzzling, as our uncorrected slope ($\beta_0=0.99\pm0.43$) is in agreement with their value ($\beta_0=1.3\pm0.6$).

\paragraph{GRB 080721, $z=2.591\pm0.001$.}We constructed the light curve and the SED with data from the following sources: \cite{Starling080721}, and \cite{HuangGCN080721}. This very energetic GRB, which also had a very bright afterglow, is discussed in detail in \cite{Starling080721}. We confirm that the data up to 30 days can be fit with a single power-law with neither a (jet) break nor host galaxy contribution. Our value ($\alpha=1.239\pm0.005$) agrees with that of \cite{Starling080721} ($\alpha=1.256\pm0.010$). There is some scatter involved in the data, leading to a high $\chi^2$ (97 for 25 degrees of freedom). We note that all data after three days lies systematically beneath the fit. Fixing $n=10$ and fitting with a broken power law, we find $\alpha_1=1.217\pm0.008$, $\alpha_2=1.446\pm0.071$ and $\tb=1.32\pm0.88$ days, and an improvement in the fit ($\chi^2=69$ for 23 degrees of freedom, $\Delta\chi^2=28$ for 2 additional parameters). If real, this is a very shallow break, and $\Delta\alpha=0.23\pm0.07$ may indicate that it is a cooling break. While the SED data has large errors \citep[see][]{Starling080721}, we find a red slope ($\beta_0=2.36\pm0.30$), and, using $\beta=0.86$ from the X-rays \citep[no cooling break between X-rays and optical,][]{Starling080721}, we find $\AV=0.35\pm0.07$. Correcting for this extinction, the afterglow of GRB 080721 is seen to be one of the most luminous known at early times, in accordance with its extreme energetics \citep{Starling080721}.

\paragraph{GRB 080810, $z=3.35104$.}We constructed the light curve and the SED with data from the following sources: \cite{Page080810} \cite{IkejiriGCN080810}, \cite{UemuraGCN080810}, \cite{OkumaGCN080810}, and \cite{YoshidaGCN080810}, as well as our own data set (RTT150 and several Russian telescopes). This complex, moderate redshift GRB had a very luminous afterglow featuring early variability contemporaneous with the $\gamma$-ray emission \citep{Page080810}. After about 0.004 days, the afterglow can be fit by a single power-law decay, with $\alpha=1.153\pm0.003$, in agreement with \cite{Page080810}, who find $\alpha=1.22\pm0.09$. There is marginal evidence for late steepening, which was already noted by \cite{GaleevGCN0808102} and \cite{ThoeneGCN080810}, this may be a candidate for an optical jet break, but the post-break decay is not measured long enough to make a significant case for a break \citep[see also][]{Page080810}. The SED shows a bit of scatter, and free fits do not yield good results. The uncorrected SED is moderately red ($\beta_0=1.24\pm0.08$), indicating possible additional dust along the line of sight. The spectral slope of the X-ray spectrum \citep[$\beta=1.00\pm0.09$, for the late data,][]{Page080810} is in agreement with the spectral slope ($\beta=0.44$) derived from the optical decay assuming an ISM environment and a cooling break blueward of the optical \citep[as found by][]{Page080810}, so we use $\beta=0.5$ and find a typical amount of SMC dust, $\AV=0.16\pm0.02$. Note that \cite{Page080810} find the same slope, but intrinsic, with no need for extinction. The dust models can not be distinguished yet, this will need further NIR data. After correction, the afterglow of GRB 080810 is found to be one of the most luminous ever observed, comparable to GRB 061007 at early times and to GRB 090313 at later times.

\paragraph{GRB 081203A, $z=2.05\pm0.01$.}We constructed the light curve and the SED with data from the following sources: \cite{AndreevGCN081203A1, AndreevGCN081203A2}, \cite{DePasqualeGCN081203A}, \cite{VolkovGCN081203A}, \cite{WestGCN081203A}, \cite{LiuGCN081203A}, \cite{MoriGCN081203A}, \cite{IsogaiGCN081203A}, \cite{RumyantsevGCN081203A}, and \cite{FatkhullinGCN081203A}. We set $T_0$ as 69 seconds before the Swift trigger. This GRB hat an very bright early afterglow \citep{Kuin081203A, WestGCN081203A} observationally and shows a strong early rise and peak. Most of the filters it was detected in lie beyond Lyman $\alpha$. The SED comprises only five filters ($Bg^\prime VR_CI_C$) and shows strong curvature, which leads to unphysical results in free fits (especially for SMC dust). The X-ray spectrum \citep[from the \emph{Swift} XRT repository,][]{EvansXRT1, EvansXRT2} gives $\beta=1.096$, and as the uncorrected spectral slope in the optical is already shallower than this, we assume a cooling break between the two bands and fix $\beta=0.596$, and the best fit is found with a small amount of SMC dust, $\AV=0.09\pm0.04$. At $z=1$, this would have been one of the brightest afterglows ever seen at very early times, reaching 10th magnitude.

\subsection{Details on the \emph{Swift}-era Bronze Sample}
\label{AppC}

\paragraph{GRB 050315, $z=1.9500\pm0.0008$.}Data taken from \cite{BergerSecondSwift}, \cite{BersierGCN050315}, \cite{GorosabelGCN050315} (host galaxy) as well as our own ANDICAM data set.

\paragraph{GRB 050318, $z=1.4436\pm0.0009$.}Data taken from \cite{BergerSecondSwift} and \cite{Still050318}. \cite{Schady2010} find SMC dust is preferred for this GRB, and derive a rather high $\AV\approx0.53$.

\paragraph{GRB 050603, $z=2.821$.}Data taken from \cite{BergerGCN050603}, \cite{Grupe050603}, and \cite{Li050603}.

\paragraph{GRB 050908, $z=3.3467$.}Data taken from \cite{CenkoDark}, \cite{ToriiGCN050908}, \cite{LiGCN050908}, \cite{KirschbrownGCN050908}, and \cite{DurigGCN3950}, as well as a single point from Maidanak.

\paragraph{GRB 060512, $z=0.4428$.}Data taken from \cite{MelandriROBONET}, \cite{CenkoGCN060512}, \cite{HeartyGCN060512}, \cite{MilneGCN060512}, \cite{TanakaGCN060512}, \cite{DePasqualeGCN060512}, and \cite{KlotzGCN060512}, as well as NOT $J$ band data we publish here. \cite{Schady2010} find that no dust model is capable of fitting the SED well, and derive moderate-to-high extinction in all cases ($\AV\approx0.47-0.66$ for the ``least bad'' SMC dust.)

\paragraph{GRB 060605, $z=3.773\pm0.001$.}Data taken from \cite{Ferrero060605}, \cite{RykoffROTSE}, \cite{ZhaiGCN060605}, \cite{KhamitovGCN0606051}, \cite{KhamitovGCN0606052}, \cite{MalesaniGCN060605}, \cite{SharapovGCN060605}, and \cite{KarskaGCN060605}. We follow \cite{Ferrero060605} and assume $\beta=1.06$ in this case. \cite{Schady2010} find that no dust model can be preferred, and find moderate extinction ($\AV\approx0.25-0.35$) for all cases.

\paragraph{GRB 060707, $z=3.425\pm0.002$.}Data taken from \cite{deUgartePostigoGCN060707}, \cite{StefanescuGCN060707}, \cite{SchadyGCN060707}, and \cite{JakobssonGCN060707}.

\paragraph{GRB 060714, $z=2.711\pm0.001$.}Data taken from \cite{StefanescuGCN060714}, \cite{JakobssonGCN060714A}, \cite{MelandriGCN060714}, A. Pozanenko, in preparation (paper on the GRB 060714 afterglow), as well as our own ANDICAM data set. \cite{Schady2010} find that no dust model can be preferred significantly, and find moderate extinction ($\AV\approx0.46$) for the SMC single power-law case.

\paragraph{GRB 060729, $z=0.5428$.}Data taken from \cite{Grupe060729}, \cite{RykoffROTSE}, and \cite{QuimbyGCN060729}, as well as our own ANDICAM data set. \cite{Schady2010} find that no dust model can be preferred significantly, and find low extinction ($\AV\approx0.03-0.18$) for all cases.

\paragraph{GRB 061121, $z=1.3145$.}Data taken from \cite{Page061121}, \cite{MelandriROBONET}, \cite{YostGCN061121}, \cite{UemuraGCN061121}, \cite{CenkoGCN061121}, \cite{HalpernGCN061121A}, \cite{HalpernGCN061121B, HalpernGCN061121C}, as well as our own data set (ANDICAM, Shajn). \cite{Schady2010} find that a cooling break between the X-ray and the optical bands is preferred, but can not distinguish the dust models significantly beyond this, finding moderate extinction ($\AV\approx0.28-0.55$) for all cases.

\paragraph{GRB 070110, $z=2.3521$.}Data taken from \cite{Troja070110} and \cite{MalesaniGCN070110A, MalesaniGCN070110B}. \cite{Schady2010} find that a cooling break between the X-ray and the optical bands is preferred at low significance, but can not distinguish the dust models significantly beyond this, finding moderate extinction ($\AV\approx0.23-0.49$) for all cases.

\paragraph{GRB 070411, $z=2.9538$.}Data taken from \cite{Ferrero070411}, \cite{MelandriROBONET}, \cite{JelinekGCN070411}, \cite{BergerGCN070411}, \cite{PrietoGCN070411}, \cite{PerleyGCN070411}, and D. Malesani et al., in preparation (paper on the GRB 070411 afterglow). \cite{Schady2010} are only able to rule out high extinction toward this GRB, with $\AV<0.21-0.47$ depending on the dust different models.

\paragraph{GRB 070612A, $z=0.617$.}Data taken from \cite{UpdikeGCN070612AA, UpdikeGCN070612AB}, \cite{CenkoGCN070612AA}, \cite{MirabalGCN070612AA, MirabalGCN070612AB}, \cite{YoshidaGCN070612A}, \cite{MalesaniGCN070612A}, \cite{D'AvanzoGCN070612}, and \cite{TaubenbergerGCN070612A}.

\paragraph{GRB 070810A, $z=2.17$.}Data taken from \cite{ChesterGCN070810A}, \cite{ThoneGCN070810A}, \cite{PerleyGCN070810A}, and \cite{YuanGCN070810A}.


\newpage\clearpage

\LongTables

\begin{deluxetable}{lccccccccl}
\tablecolumns{9}
\tablefontsize{\tiny}
\tablewidth{0pt}
\tablecaption{Energetics of the Type II GRB Sample}
\tablehead{
\colhead{GRB} &
\colhead{Redshift $z$} &
\colhead{Fluence ($10^{-7}\;$} &
\colhead{Band} &
\colhead{Low-energy} &
\colhead{High-energy} &
\colhead{$E_{\tn{p,rest}}$} &
\colhead{log $E_{\tn{iso,bol}}$} &
\colhead{References} \\
\colhead{} &
\colhead{} &
\colhead{$\rm erg\;\rm cm^{-2}$)} &
\colhead{(keV)} &
\colhead{index $\alpha_B$} &
\colhead{index $\beta_B$} &
\colhead{(keV)} &
\colhead{(erg)} &
\colhead{(for Energetics)}
}
\startdata
970508& $0.8349\pm0.0003$ & $26\pm3$ & $20-2000$ & $-1.71\pm0.1$ & $-2.2\pm0.25$ & $168.8\pm47.7$ & ${51.87}^{+0.06}_{-0.07}$ &1, 2\\
990123& $1.6004\pm0.0008$ & $5090\pm20$ & $20-2000$ & $-0.89\pm0.08$ & $-2.45\pm0.97$ & $2031\pm161.0$ & ${54.60\pm0.05}$ &1, 3\\
990510 & $1.6187\pm0.0015$ & $270\pm50$ & $20-2000$ & $-1.23\pm0.05$ & $-2.7\pm0.4$ & $423.0\pm42.0$ & ${53.34}^{+0.07}_{-0.09}$ &1, 2\\
991208& $0.7063\pm0.0017$ & $1630\pm50$ & $20-2000$ & $-1.1\pm0.4$ & $-2.2\pm0.4$ & $312.3\pm5.1$ & ${53.43}^{+0.05}_{-0.06}$ &1, 2, 4\\
991216& $1.02\pm0.01$ & $2480\pm120$ & $20-2000$ & $-1.23\pm0.25$ & $-2.18\pm0.39$ & $763.6\pm20.2$ & ${53.93}^{+0.04}_{-0.05}$ &1, 2\\
000131& $4.5\pm0.015$ & $440\pm60$ & $20-2000$ & $-1.2\pm0.1$ & $-2.4\pm0.1$ & $1293\pm110.0$ & ${54.28}^{+0.06}_{-0.07}$ &1, 2\\
000301C& $2.0404\pm0.0008$ & $83\pm6$ & $20-2000$ & $-1.0\pm0.3$ & $-2.3\pm0.3$ & $1213\pm303.0$ & ${52.99}^{+0.03}_{-0.04}$ &2\\
000911& $1.0585\pm0.0001$ & $1920\pm340$ & $20-2000$ & $-1.11\pm0.12$ & $-2.32\pm0.41$ & $2106\pm177.0$ & ${53.86}^{+0.08}_{-0.09}$ &1, 2\\
\enddata
\end{deluxetable}

\newpage\clearpage

\begin{deluxetable}{lcrcrcr}
\tablecolumns{7}
\tabletypesize{\tiny}
\tablewidth{0pt}
\tablecaption{Results of the SED Fitting for the Golden Sample -- Part I}
\tablehead{
\colhead{GRB}   &
\colhead{Filters} &
\multicolumn{2}{c}{No Extinction} &
\multicolumn{3}{c}{MW Dust}\\
\multicolumn{2}{c}{} &
\colhead{$\chi^2_{dof}$}  &
\colhead{$\beta_0$}  &
\colhead{$\chi^2_{dof}$}  &
\colhead{$\beta$}  &
\colhead{$\AV$}
}
\startdata
990510	&	$BVR_ci^{G}I_cJHK_S$	&	1.83	&	$0.61\pm0.04$	&	1.95	&	$0.69\pm0.09$	&	$-0.06\pm0.06$	\\
011211	&	$BVR_cI_cJK$	&	1.57	&	$0.72\pm0.04$	&	1.3	&	$0.84\pm0.09$	&	$-0.09\pm0.06$	\\
030323	&	$R_cI_cJH$	&	0.94	&	$0.85\pm0.03$	&	0	&	$1.22\pm0.27$	&	$-0.35\pm0.25$	\\
050319	&	$VR_CI_CK$	&	0.38	&	$1.00\pm0.06$	&	0.71	&	$0.94\pm0.33$	&	$0.06\pm0.31$	\\
050408	&	$UBVR_cI_cZJHK$	&	3.12	&	$1.61\pm0.14$	&	3.27	&	$1.17\pm0.60$	&	$0.33\pm0.44$	\\
050416A	&	$U_3U_2UBVR_CI_Cz^\prime K_S$	&	0.39	&	$1.15\pm0.17$	&	0.91	&	$1.20\pm0.45$	&	$0.11\pm0.38$	\\
050502A	&	$R_ci^\prime JHK_s$	&	0.76	&	$0.95\pm0.06$	&	0.39	&	$0.46\pm0.22$	&	$0.40\pm0.18$	\\
050525A	&	$U_1UBVR_cJHKS_{3.6}S_{4.5}S_{8.0}S_{24.0}$	&	12.7	&	$1.02\pm0.03$	&	13.7	&	$1.12\pm0.07$	&	$-0.11\pm0.06$	\\
\enddata
\tablecomments{Filters that are not used for the fit (e.g, due to Lyman $\alpha$ dampening) are not included. The degrees of freedom of the fit are always number of filters minus three for the fits with extinction, and minus two for the fit without extinction. $\beta_0$ is the spectral slope without extinction correction. $U_{1,2,3}$ denote the \emph{Swift} UVOT UVW1, UVM2 and UVW2 filters, respectively. $S_{3.6, 4.5, 8.0, 24.0}$ denote \emph{Spitzer Space Telescope} IRAC $3.6\micron$, $4.5\micron$, $8.0\micron$ and MIPS $24.0\micron$ filters, respectively. A superscript \emph{G} denotes a Gunn filter. A subscript $G$ denotes \emph{GROND} dichroic filters. The results on GRB 050904 are taken from \cite{Kann050904}, the results on GRB 060526 are from \cite{Thoene060526}.}
\label{tabALL}
\end{deluxetable}

\newpage\clearpage

\begin{deluxetable}{lrrrrcr}
\tablecolumns{7}
\tabletypesize{\tiny}
\tablewidth{0pt}
\tablecaption{Results of the SED Fitting for the Golden Sample -- Part II}
\tablehead{
\colhead{GRB}   &
\multicolumn{3}{c}{LMC Dust} &
\multicolumn{3}{c}{SMC Dust}\\
\colhead{}  &
\colhead{$\chi^2_{dof}$}  &
\colhead{$\beta$}  &
\colhead{$\AV$} &
\colhead{$\chi^2_{dof}$}  &
\colhead{$\beta$}  &
\colhead{$\AV$}}
\startdata
990510	&	1.81	&	$0.36\pm0.18$	&	$0.16\pm0.11$	&	0.21	&	$0.17\pm0.15$	&	$0.22\pm0.07$\\	
011211	&	1.82	&	$0.51\pm0.24$	&	$0.13\pm0.14$	&	0.45	&	$0.41\pm0.15$	&	$0.14\pm0.06$\\	
030323	&	0.03	&	$0.15\pm0.51$	&	$0.35\pm0.26$	&	0.03	&	$0.46\pm0.28$	&	$0.13\pm0.09$\\	\hline\hline
050319	&	0.17	&	$0.64\pm0.52$	&	$0.12\pm0.17$	&	0.31	&	$0.74\pm0.42$	&	$0.05\pm0.09$	\\
050408	&	1.1	&	$0.01\pm0.47$	&	$1.06\pm0.31$	&	0.59	&	$0.28\pm0.27$	&	$0.74\pm0.15$\\	
050416A	&	0.56	&	$0.84\pm0.38$	&	$0.33\pm0.24$	&	0.46	&	$0.92\pm0.30$	&	$0.21\pm0.14$	\\
050502A	&	0.35	&	$0.43\pm0.22$	&	$0.24\pm0.10$	&	0.16	&	$0.76\pm0.16$	&	$0.05\pm0.05$\\	
050525A	&	11.27	&	$0.53\pm0.10$	&	$0.42\pm0.08$	&	8.42	&	$0.52\pm0.08$	&	$0.36\pm0.05$\\
\enddata
\tablecomments{Filters that are not used for the fit (e.g, due to Lyman $\alpha$ dampening) are not included. The degrees of freedom of the fit are always number of filters minus three for the fits with extinction, and minus two for the fit without extinction. $\beta_0$ is the spectral slope without extinction correction. $U_{1,2,3}$ denote the \emph{Swift} UVOT UVW1, UVM2 and UVW2 filters, respectively. $S_{3.6, 4.5, 8.0, 24.0}$ denote \emph{Spitzer Space Telescope} IRAC $3.6\micron$, $4.5\micron$, $8.0\micron$ and MIPS $24.0\micron$ filters, respectively. A superscript \emph{G} denotes a Gunn filter. A subscript $G$ denotes \emph{GROND} dichroic filters. The results on GRB 050904 are taken from \cite{Kann050904}, the results on GRB 060526 are from \cite{Thoene060526}.}
\label{tabALL2}
\end{deluxetable}

\newpage\clearpage

\begin{deluxetable}{lcccccc}
\tablecolumns{7}
\tabletypesize{\scriptsize}
\tablewidth{0pt}
\tablecaption{Corrected Apparent and Absolute Magnitudes of Type II Afterglows}
\tablehead{
\colhead{GRB\tablenotemark{a}} &
\colhead{$dRc$\tablenotemark{b}} &
\colhead{$R_C$ (at 1 day\tablenotemark{c})} &
\colhead{$\MB$ (at 1 day\tablenotemark{c})} &
\colhead{$R_C$ (at 4 days\tablenotemark{c})} &
\colhead{$\MB$ (at 4 days\tablenotemark{c})} &
\colhead{Type\tablenotemark{d}}
}
\startdata
990510  & $-1.60$ & $17.95\pm0.15$ & $-24.73\pm0.16$  & $20.83\pm0.16$ & $-21.85\pm0.16$  & B \\
011211  & $-2.18$ & $19.22^{+0.14}_{-0.16}$ & $	-23.53^{+0.15}_{-0.16}$  & $21.92\pm0.32$ & $-20.83\pm0.32$  & B \\
030323  & $-3.38$ & $17.80^{+0.19}_{-0.20}$ & $	-24.96^{+0.20}_{-0.22}$  & $20.80\pm0.24$ & $	-21.96\pm0.25$  & B \\\hline\hline
050319  & $-3.20$ & $18.09^{+0.08}_{-0.12}$ & $-24.77^{+0.14}_{-0.16}$  & $20.41^{+0.18}_{-0.20}$ & $-22.43^{+0.21}_{-0.23}$  & B \\
050408	& $	-2.00	$ & $	20.73	^{+	0.27	}_{-	0.28	}$ & $-21.99^{+0.28}_{-0.29}$ & $	22.40	^{+	0.29	}_{-	0.30	}$ & $	-20.32	^{+	0.30	}_{-	0.31	}$	&	A	\\
050416A & $+0.83$ & $24.35\pm0.48$ & $-18.54\pm0.49$  & $\cdots$ & $\cdots$  & A \\
050502A	& $	-3.56	$ & $	22.30	^{+	0.31	}_{-	0.32	}$ & $-20.55^{+0.31}_{-0.32}$ & $	\cdots$					&	$\cdots$						&	B	\\
050525A	& $+0.73	$ & $	20.40	\pm	0.36			$ & $	-22.39\pm0.37$ & $	23.15	\pm	0.40			$ & $	-19.64	\pm	0.41			$	&	A	\\
\enddata
\tablenotetext{a}{First block: Additional pre-\emph{Swift} Golden Sample; second block: Golden Sample; third block: Silver Sample; fourth block: Bronze Sample.}
\tablenotetext{b}{Magnitude shift, see K06.}
\tablenotetext{c}{After the GRB, assuming $z$=1.}
\tablenotetext{d}{Type A: $z<1.4$; Type B: $z\geq1.4$. See K06 for more information on the two types.}
\label{tabMB}
\end{deluxetable}

\newpage\clearpage

\begin{deluxetable}{llcr}
\tablecolumns{4}
\tabletypesize{\scriptsize}
\tablewidth{0pt}
\tablecaption{Early Afterglow Peak or Limit Magnitudes}
\tablehead{
\colhead{Type\tablenotemark{a}} &
\colhead{GRB} &
\colhead{Time\tablenotemark{b}} &
\colhead{$R_C$ Magnitude\tablenotemark{c}}
}
\startdata
Peak + Fast Decay	&	990123		&	0.0004440	& $	7.60	\pm	0.02	$ \\
	&	080319B\tablenotemark{d}	&	0.0006359	& $	5.18	\pm	0.02	$ \\
	&	061121\tablenotemark{e}	&	0.0007647	& $	14.19	\pm	0.16	$ \\
	&	060526\tablenotemark{f}	&	0.0014088	& $	12.18	\pm	0.14	$ \\
	&	060729\tablenotemark{g}	&	0.0014365	& $	17.20	\pm	0.14	$ \\
	&	050904\tablenotemark{h}	&	0.0016318	& $	6.48	\pm	0.24	$ \\
	&	080129\tablenotemark{i}	&	0.0022666	& $	13.51	\pm	0.10	$ \\
	&	070802\tablenotemark{j}	&	0.0155854	& $	15.34	\pm	0.07	$ \\
\enddata
\end{deluxetable}

\newpage\clearpage

\begin{deluxetable}{lccccccccl}
\tablecolumns{9}
\tabletypesize{\tiny}
\tablewidth{0pt}
\tablecaption{Energetics of the Low-Redshift Type II GRB Sample}
\tablehead{
\colhead{GRB} &
\colhead{Redshift $z$} &
\colhead{Fluence} &
\colhead{Band (keV)} &
\colhead{Low-energy} &
\colhead{High-energy} &
\colhead{$E_{\tn{p,rest}}$} &
\colhead{log $E_{\tn{iso,bol}}$} &
\colhead{References} \\
\colhead{} &
\colhead{} &
\colhead{($10^{-7}\;\rm erg\;\rm cm^{-2}$)} &
\colhead{} &
\colhead{index $\alpha$} &
\colhead{index $\beta$} &
\colhead{(keV)} &
\colhead{(erg)} &
\colhead{}
}
\startdata
980425	& $	0.00867	\pm	0.00004	$ & $	38.7	\pm	3.9			$ & $	20	-	2000	$ & $	-1.27	\pm	0.25	$ & $	-2.3	\pm	0.46	$ & ${	119.0	}^{+	24.2	}_{-	24.2	}$ & ${	48.03	}^{+	0.07	}_{-	0.08	}$ &	1	\\
020903	& $	0.2506	\pm	0.0003	$ & ${	0.98	}^{+	0.67	}_{-	0.36	}$ & $	2	-	400	$ & $	-1	\pm	0.3	$ & $	-2.6		^{-0.4}_{+0.5}	$ & ${	3.25	}^{+	1.75	}_{-	1.00	}$ & ${	49.29	}^{+	0.23	}_{-	0.21	}$ &	2	\\
031203	& $	0.10536	\pm	0.00007	$ & $	20	\pm	4			$ & $	20	-	200	$ & $	-1.63	\pm	0.06	$ & $	\cdots			$ & $	159.2	\pm	50.8	$ & ${	50.06	}^{+	0.08	}_{-	0.10	}$ &	3, 4	\\
050223	& $	0.584	\pm	0.005	$ & ${	8.8	}^{+	0.9	}_{-	3	}$ & $	15	-	350	$ & $	-1.7	\pm	0.2	$ & $	\cdots			$ & ${	106.1	}^{+	179.0	}_{-	36.4	}$ & ${	50.85	}^{+	0.23	}_{-	0.07	}$ &	5	\\
050826	& $	0.296	\pm	0.001	$ & ${	10	}^{+	2	}_{-	3	}$ & $	15	-	350	$ & $	-1.2	\pm	0.3	$ & $	\cdots			$ & ${	440.6	}^{+	1023.8	}_{-	272.2	}$ & ${	50.48	}^{+	0.37	}_{-	0.48	}$ &	5	\\
051016B	& $	0.9364			$ & ${	1.4	}^{+	0.2	}_{-	0.3	}$ & $	15	-	350	$ & $	-1	\pm	0.3	$ & $	-2.1	\pm	0.3	$ & ${	71.6	}^{+	36.8	}_{-	69.7	}$ & ${	50.57	}^{+	0.40	}_{-	0.08	}$ &	5	\\
060202	& $	0.783			$ & ${	38	}^{+	3	}_{-	4	}$ & $	15	-	350	$ & $	-1.7	\pm	0.1	$ & $	\cdots			$ & ${	156.9	}^{+	613.4	}_{-	46.4	}$ & ${	51.85	}^{+	0.27	}_{-	0.07	}$ &	5	\\
060218	& $	0.03342	\pm	0.00002	$ & ${	173	}^{+	18	}_{-	78	}$ & $	0.5	-	150	$ & $	-1.44	\pm	0.06	$ & $	-2.54	\pm	0.07	$ & $	4.83	\pm	1.19			$ & ${	49.50	}^{+	0.05	}_{-	0.26	}$ &	6, 7	\\
\enddata
\tablecomments{References for $z$:
GRB	980425: \cite{Foley980425},
XRF	020903: \cite{Bersier020903},
GRB	031203: \cite{Margutti031203},
GRB	050223: \cite{Pellizza050223},
GRB	050826: \cite{Mirabal050826},
XRF	051016B: \cite{SoderbergGCN051016B},
GRB	060202: \cite{Butler060202},
XRF	060218: \cite{Pian060218},
XRF	060319: \cite{Perley060319},
GRB	060602A: \cite{JakobssonGCN060602A},
GRB	061028: \cite{Kocevski061028},
GRB	061110A: \cite{FynboSpectra}.
References for energetics (Fluence, Band, $E_p$, Band function parameters): (1) \cite{FB2005}, (2) \cite{Sakamoto020903}, (3) \cite{Sazonov031203}, (4) \cite{UlanovKonus}, (5) \cite{ButlerBAT}, (6) \cite{Kaneko2007}, (7) \cite{Campana060218}, (8) this work.}
\label{tabTypeIIlowzSample}
\end{deluxetable}


\end{document}